\documentclass[twocolumn]{aastex62}
\shorttitle{Neutrino-process in Core-Collapsing Supernovae }
\shortauthors{Ko et al.}

\usepackage{mathrsfs}
\usepackage{amsmath,amssymb}
\usepackage{slashed} 
\usepackage{bbm} 
\usepackage{accents} 
\usepackage{soul}

\begin{document}

\title{Comprehensive Analyses of the Neutrino-Process in the Core-collapsing Supernova}

\correspondingauthor{Myung-Ki Cheoun}
\email{cheoun@ssu.ac.kr}

\author{Heamin Ko}
\affiliation{Department of Physics and OMEG institute, Soongsil University, Seoul 07040, Korea}

\author{Dukjae Jang}
\affiliation{Center for Relativistic Laser Science, Institute for Basic Science (IBS), Gwangju 61005, Korea}

\author{Myung-Ki Cheoun}
\affiliation{Department of Physics and OMEG institute, Soongsil University, Seoul 07040, Korea}
\affiliation{School of Physics and International Research Center for Big-Bang Cosmology and Element Genesis, Beihang University, Beijing 100083, China}
\affiliation{National Astronomical Observatory of Japan, Mitaka, Tokyo 181-8588, Japan}

\author{Motohiko Kusakabe}
\affiliation{School of Physics and International Research Center for Big-Bang Cosmology and Element Genesis, Beihang University, Beijing 100083, China}
\affiliation{National Astronomical Observatory of Japan, Mitaka, Tokyo 181-8588, Japan}

\author{Hirokazu Sasaki}
\affiliation{National Astronomical Observatory of Japan, Mitaka, Tokyo 181-8588, Japan}
\affiliation{Graduate School of Science, The University of Tokyo, Bunkyo-ku, Tokyo 113-0033, Japan}
\affiliation{Theoretical Division, Los Alamos National Laboratory, Los Alamos, NM 87545, USA}

\author{Xingqun Yao}
\affiliation{School of Physics and International Research Center for Big-Bang Cosmology and Element Genesis, Beihang University, Beijing 100083, China}

\author{Toshitaka Kajino}
\affiliation{School of Physics and International Research Center for Big-Bang Cosmology and Element Genesis, Beihang University, Beijing 100083, China}
\affiliation{National Astronomical Observatory of Japan, Mitaka, Tokyo 181-8588, Japan}
\affiliation{Graduate School of Science, The University of Tokyo, Bunkyo-ku, Tokyo 113-0033, Japan}

\author{Takehito Hayakawa}
\affiliation{National Institutes for Quantum and Radiological Science and Technology, 2-4 Shirakata, Tokai, Naka, Ibaraki 319-1106, Japan}
\affiliation{Institute of Laser Engineering, Osaka University, Suita, Osaka 565-0871, Japan}

\author{Masaomi Ono}
\affiliation{Kyushu University, Hakozaki, Fukuoka 812-8581, Japan}

\author{Toshihiko Kawano}
\affiliation{Theoretical Division, Los Alamos National Laboratory, Los Alamos, NM 87545, USA}

\author{Grant J. Mathews}
\affiliation{School of Physics and International Research Center for Big-Bang Cosmology and Element Genesis, Beihang University, Beijing 100083, China}
\affiliation{National Astronomical Observatory of Japan, Mitaka, Tokyo 181-8588, Japan}
\affiliation{Department of Physics, Center for Astrophysics, University of Notre Dame, Notre Dame, IN 46556, USA}

\nocollaboration

\affiliation{IOP Senior Publisher for the AAS Journals}
\affiliation{IOP Publishing, Washington, DC 20005}



\begin{abstract}
We investigate the neutrino flavor change effects due to neutrino self-interaction, shock wave propagation as well as matter effect on the neutrino-process of the core-collapsing supernova (CCSN). For the hydrodynamics, we use two models: a simple thermal bomb model and a specified hydrodynamic model for SN1987A. As a pre-supernova model, we take an updated model adjusted to explain the SN1987A employing recent development of the $(n,\gamma)$ reaction rates for nuclei near the stability line $(A \sim 100)$. As for the neutrino luminosity, we adopt two different models: equivalent neutrino luminosity and non-equivalent luminosity models. The latter is taken from the synthetic analyses of the CCSN simulation data which involved quantitatively the results obtained by various neutrino transport models. Relevant neutrino-induced reaction rates are calculated by a shell model for light nuclei and a quasi-particle random phase approximation model for heavy nuclei. For each model, we present abundances of the light nuclei ($^7$Li, $^7$Be, $^{11}$B and $^{11}$C) and heavy nuclei ($^{92}$Nb, $^{98}$Tc, $^{138}$La and $^{180}$Ta) produced by the neutrino-process. The light nuclei abundances turn out to be sensitive to the Mikheyev-Smirnov-Wolfenstein (MSW) region around O-Ne-Mg region while the heavy nuclei are mainly produced prior to the MSW region. Through the detailed analyses, we find that neutrino self-interaction becomes a key ingredient in addition to the MSW effect for understanding the neutrino-process and the relevant nuclear abundances. The normal mass hierarchy is shown to be more compatible with the meteorite data. Main nuclear reactions for each nucleus are also investigated in detail.
\end{abstract}




\section{Introduction} \label{sec:intro}
The observation of a supernova in 1987 (SN1987A) has been considered as the brightest supernova (SN) with naked eyes at the nearest space from the earth. A few hours before the optical observation, it was predicted by the detection of neutrinos, which is the first record of the neutrino detection from extrasolar objects \citep{1987Natur.330..142S}. The Kamiokande and Irvine-Michigan-Brookhaven detectors measured 8-11 neutrino events with Cherenkov detectors \citep{1987PhRvL..58.1490H,1987PhRvL..58.1494B}, and Mont Blanc Underground Neutrino Observatory found 5 events from the neutrino burst  using a liquid scintillation detector \citep{1987EL......3.1315A}. The detection made it possible to point out that the location of the SN1987A event is in our satellite galaxy, Large Magellanic Cloud.

Ever since SN1987A was observed, the explosion mechanism in massive stars has been extensively studied \citep{2012ARNPS..62..407J}. The development of simulations for SN1987A enabled evaluations of the SN mass and light curve \citep{1988ApJ...330..218W,1990ApJ...360..242S}, and subsequently various pre-supernova (pre-SN) models and explosive nucleosynthesis have been investigated \citep{1995PThPh..94..663H}. In particular, by the neutrino detection from the core collapsing SN (CCSN), the neutrino process ($\nu$-process) in explosive nucleosynthesis has been considered to trace the origin of several elements unexplained by the traditional nuclear processes \citep{1990ApJ...356..272W,2014JPhG...41d4007K}. Table \ref{tab:ele} tabulates nuclides thought to be produced mainly in the $\nu$-process, which we closely examine in this paper.

\begin{table}[h]
    \begin{tabular}{c | c}
        \hline
        \hline
        \textbf{Element} & \textbf{Related references.} \\
        \hline
            $^{7}$Li & \citep{1990ApJ...356..272W}, \citep{2004ApJ...600..204Y}, \\
            & \citep{2005PhRvL..94w1101Y}, \citep{2006PhRvL..96i1101Y}, \\
            & \citep{2008ApJ...686..448Y}, \citep{2013JPhG...40h3101S}, \\
            & \citep{2019ApJ...872..164K} \\
        \hline
            $^{11}$B & \citep{1990ApJ...356..272W}, \citep{2004ApJ...600..204Y},\\
            & \citep{2005PhRvL..94w1101Y}, \citep{2006PhRvL..96i1101Y}, \\
            & \citep{2008ApJ...686..448Y}, \citep{2010ApJ...718L.137N}, \\
            & \citep{2011PhRvL.106o2501A}, \citep{2013JPhG...40h3101S}, \\
            & \citep{2019ApJ...872..164K}\\
        \hline
            $^{19}$F & \citep{1990ApJ...356..272W}, \citep{2018ApJ...865..143S}, \\
            & \citep{2019MNRAS.490.4307O}\\
        \hline
            $^{93}$Nb & \citep{2013ApJ...779L...9H}  \\
        \hline
            $^{98}$Tc &\citep{2018PhRvL.121j2701H} \\
        \hline
            $^{138}$La &\citep{1990ApJ...356..272W}, \citep{2018ApJ...865..143S}, \\
            & \citep{2005PhLB..606..258H}, \citep{2008PhRvC..77f5802H}, \\
            & \citep{2007PhRvL..98h2501B}, \citep{2013RPPh...76f6201R}, \\
            & \citep{2014JPhG...41d4007K}, \citep{2015PhLB..744..268K}\\
            & \citep{2017JApA...38....8L} \\
        \hline
           $^{180}$Ta & \citep{1990ApJ...356..272W}, \citep{2005PhLB..606..258H}, \\
           & \citep{2010PhRvC..81e2801H}, \citep{2007PhRvL..98h2501B}, \\
           & \citep{2013RPPh...76f6201R}, \citep{2014JPhG...41d4007K} \\
           & \citep{2017JApA...38....8L}, \citep{2019PhLB..791..403M}, \\
        \hline \hline
    \end{tabular}
     \caption{Main elements produced by the $\nu$-process and related references}
    \label{tab:ele}
\end{table}

In general, it is known that the heavy elements within $70 \lesssim A \lesssim 209$ are produced by the rapid neutron capture process ($r$-process) \citep{1957RvMP...29..547B,2014JPhG...41d4007K}.
However, since the heavy elements at Table \ref{tab:ele} are surrounded by stable nuclei in the nuclear chart, they are blocked by the relevant nuclear reactions such as neutron capture or $\beta^\pm$-decay. Consequently, the origin of those nuclei cannot be sufficiently explained by the $r$-process. However, the stable nuclei at Table \ref{tab:ele} exist in  our solar system being contained in the primitive meteorites \citep{2009LanB...4B..712L}. For example, the existence of the short-lived unstable isotope $^{92}$Nb at the solar system formation has been verified from the isotropic anomaly in $^{92}$Zr to which $^{92}$Nb decays \citep{1996ApJ...466..437H, 2000Sci...289.1538M,2000E&PSL.184...75S,2021PNAS..11820177H}. To explain the meteorite data, another production mechanism for the nuclides---such as the $\nu$-process---has been required to be necessary.
In the study of the $\nu$-process, neutrino properties and stellar environments play vital roles in determining the neutrino oscillation behavior, which is critical for studying the process, as argued below.

First, neutrino mixing parameters such as squared mass differences and mixing angles significantly affect the neutrino oscillation in CCSN environments. The neutrino oscillation experiments do not measure absolute masses of neutrinos but the squared mass differences defined by $\Delta m^2_{ij}$, where $i,j=1,2,$ and $3$ in the mass eigenstates. Also, the mixing angles of $\theta_{12}, \theta_{23}$, and $ \theta_{13}$, which are deeply related to the neutrino oscillation behavior, have been measured by various experiments \citep{2014ChPhC..38i0001O}. However, in spite of the challenging experiments, the problem regarding the neutrino mass hierarchy (MH) remained unsolved. Although it is reported that the inverted hierarchy is disfavored with $93\%$ confidence level \citep{2017PhRvL.118w1801A}, it still requires the precise verification whether the neutrino mass follows the normal hierarchy ($m_1 < m_2 < m_3 $, NH) or the inverted hierarchy ($m_3 < m_1 < m_2$, IH).

Second, during the CCSN explosion, neutrinos pass through the matter composed of protons, neutrons, charged leptons, neutrinos, and nuclides. In the propagation, the neutrinos scatter with the background particles through the weak interaction \citep{1988NuPhB.307..924N} classified to the charged current (CC) and neutral current (NC) interactions. Since both interactions would affect the time evolution of the neutrino flavors, the neutrino oscillation behavior differs from the case in the free space. The effective potential describing the interaction of the propagating neutrinos depends on the matter density, so that the hydrodynamics models become important in the CCSN $\nu$-process. In this paper, we compare the results of the $\nu$-process by two different hydrodynamics models and discuss how the models change the behavior of the neutrino oscillation and affect the nucleosynthesis.

Third, the neutrino emission mechanism in the CCSN significantly impacts on the neutrino flux. The neutrinos in the CCSN are trapped for a few seconds until the dynamical time scale of the core collapsing becomes longer than the neutrino diffusion time scale \citep{2017hsn..book.1575J}. After the trapping, the explosion creates the emission of a huge number of neutrinos. Near the proto-neutron star, due to the emitted high-density neutrino gas, the self-interaction (SI) among neutrinos should be considered \citep{1993NuPhB.406..423S,1992PhLB..287..128P,1993PhRvD..48.1462S,1995PhRvD..51.1479Q, 2006PhRvD..74j5014D, 2006PhRvD..73b3004F}. By the SI, the neutrino flux in the CCSN could be changed and would impact on the $\nu$-process \citep{2020ApJ...891L..24K} as well as $\nu p$-process \citep{2017PhRvD..96d3013S}, and also neutrino signals from the CCSN \citep{2015PhRvD..91f5016W}. Such a collective neutrino flavor conversion derived by the neutrino SI ($\nu$-SI) is distinct from the matter effects in that the background neutrinos provide the non-linear contribution due to mixed neutrino states represented by an off-diagonal term. The SI effects can be suppressed when the electron background is more dominant than the neutrino background near the proto-neutron star \citep{2010PhRvD..81i3008D, 2011PhRvL.107o1101C,2011PhRvL.106i1101D}. This suppression depends on the neutrino decoupling model \citep{2019PhRvD.100d3004A}.

Finally, one of the main concerns in the neutrino physics for the CCSN is the shape of the statistical distribution of neutrinos. In the neutrino decoupling from a proto-neutron star, various collision processes---such as neutrino elastic scattering on nucleon ($\nu N-N\nu$), nucleon-nucleon bremsstrahlung ($NN-NN \nu \bar{\nu}$), and leptonic processes ($\nu e - e \nu,~e^+e^- - \nu\bar{\nu}$)---make the neutrino spectra deviate from the Fermi-Dirac distribution \citep{2001ApJ...561..890R,2003ApJ...590..971K}. Because muon- and tau-type neutrinos ($\nu_\mu$ and $\nu_\tau$) interact only through the NC interaction being weaker than CC interactions, they are decoupled earlier than electron-type neutrino ($\nu_e$), and consequently they have different temperature. The decoupling temperature and neutrino distribution function are also crucial to determine their flux which affects the neutrino-induced reaction rates. In principle, the exact decoupling process should be investigated by the neutrino transport equation because it has significant effects on the CCSN $\nu$-process and the CCSN explosion. As a cornerstone, \citet{2018JPhG...45j4001O} showed the consistent agreement of the results by six different neutrino transport models for the SN simulation. For the further step, we expect that multi-dimensional neutrino transport simulations are to be performed in the near future.

Based on the neutrino properties and related explosive models, the CCSN $\nu$-process has been investigated in detail. Furthermore, by comparing the calculated results with the observed solar abundance, one may find the contribution of the SN $\nu$-process to the solar system material. For instance, it is known that the stable nuclides such as $^{7}$Li and $^{11}$B are meaningfully produced by the $\nu$-process, while $^{6}$Li and $^{10}$B productions are insignificant amounts \citep{2000ApJ...540..930F,2012PhRvD..85j5023M}. For the short-lived unstable nuclide such as $^{98}$Tc, one can use the estimated abundance as a cosmic chronometer predicting the epoch when the supernovae (SNe) flows in our solar system \citep{2013ApJ...779L...9H,2018PhRvL.121j2701H}. Besides, the comparison between estimated abundances and observed meteoritic data may give a clue for the neutrino temperature in the CCSN explosion \citep{2005PhRvL..94w1101Y}.

In this paper, we investigate how the CCSN $\nu$-process is affected by the various neutrino properties and CCSN models currently available, and discuss some questions remaining in the CCSN $\nu$-process and future works.
For this purpose, we organize this paper as follows. In Section \ref{sec:nu_osc}, we explain the effects of different SN hydrodynamics models on the CCSN $\nu$-process. After establishing the hydrodynamics models, in Section \ref{sec:nu_SI}, we review the neutrino oscillation feature in the CCSN environment. Section \ref{sec:nusyn} covers the formulae of nuclear reaction rates for the nucleosynthesis in CCSN explosion. The results of synthesized elements are presented in Section \ref{sec:abun}. Finally, discussion and summary are done in Section \ref{sec:sum_con}. Detailed formulae for the calculation and numerical results of some updated nuclear reactions are summarized at Appendix A, B and C, respectively. Each contribution of main nuclear reactions to each nuclear abundance is provided in detail at Appendix D. All equations in this paper follow the natural unit, i.e., $\hbar = c =1$.

\section{Neutrino Oscillation in Core Collapse Supernova}  \label{sec:nu_osc}
In CCSN explosion, the propagating neutrinos interact with the leptonic background through the CC or NC reactions. Although various lepton components are interacting with neutrinos in the CCSN environments, because of their too high threshold energies or small numbers of the background leptons, only the net electron composed by electrons and positrons {---}as a background component{---}contributes to the neutrino oscillation through the CC reaction. Hence, we consider the matter effect only with the varying electron density, which is known as Mikheyev-Smirnov-Wolfenstein (MSW) effect \citep{1986NCimC...9...17M}. In dealing with the electron density as the background, it is important which of the hydrodynamical models of SN explosion we choose. In this section, with two kinds of SN hydrodynamical models, we discuss the MSW effects on the neutrino oscillation in the CCSN environments.

\subsection{The MSW effect in SN1987A}
The typical CCSN environment constrains the particle momenta scale within the MeV range, while the masses of gauge bosons in the standard weak interaction have 100 GeV scale mass. Because of the low energy environments, we treat the weak interaction of the neutrinos with a point-like coupling scheme. By taking the thermal average of the electron background and the charge neutrality condition, we obtain the following total Hamiltonian for CC interaction, which is decomposed into the vacuum term and the MSW matter potential,
 	\begin{eqnarray}			
		H_{\rm total} &=& H_{\rm vacuum} + \mathcal{V}_{\nu_e e}  \nonumber \\
		                    &=& U {\rm diag} \left(0,\frac{\Delta m^2_{21}}{2E_\nu},\frac{|\Delta m^2_{31}|}{2E_\nu} \right) U^\dagger \nonumber \\
		                      && + {\rm diag} \left(\sqrt{2}G_F n_e,0,0 \right), \label{Ham_tot}
	\end{eqnarray}
where $U$ is the Pontecorvo-Maki-Nakagawa-Sakata (PMNS) unitary mixing matrix, the values of $\Delta m_{ij}^2$ are adopted from \citet{2014ChPhC..38i0001O}, and $E_{\nu}$ denotes the neutrino energy. ${n_e}$ stands for the net electron density. Detailed derivation of the Hamiltonian is succinctly summarized at Appendix \ref{app_Basic formula}. Note that the effective potential for the NC interactions contributing to all flavors is not included in the matter Hamiltonian because it is absorbed into a phase shift \citep{2007fnpa.book.....G}.
By solving the Schr\"odinger-like equation with the total Hamiltonian, we obtain the neutrino flavor change probability. As seen in the Hamiltonian, the neutrino oscillation probability depends on $n_e$, which implies the importance of chosen hydrodynamics models. In the next subsection, we introduce two kinds of hydrodynamics models to describe the time evolution of the neutrino flavors.

\subsubsection{Profiles of SN1987A model}
The SN1987A is verified to be an explosion of a blue supergiant star Sk--69 202 in the Large Magellanic Cloud which is estimated to have had a (19$\pm$3) solar mass ($M_\odot$) in the main-sequence by the analysis of the light curve, and the metallicity is given as $Z\sim Z_\odot/4$ \citep{1988ApJ...330..218W}. Among various explosive models satisfying the given condition \citep{2012ARNPS..62..407J}, as a pre-SN model, we adopt the initial density and temperature profiles from \citet{2015PTEP.2015f3E01K} whose results are similar to \citet{1990ApJ...360..242S}. For the hydrodynamics models, we exploit the model in \citet{2019ApJ...872..164K} based on the blcode (\url{https://stellarcollapse.org/index.php/SNEC.html}) with an explosion energy of $10^{51}$ erg. To discuss the effects of hydrodynamical models, we introduce another model used in \cite{2018PhRvL.121j2701H}, which was gleaned from the pre-SN model of \citet{2000ApJ...532.1132B} and also used in \citet{2020ApJ...891L..24K, 2013ApJ...779L...9H,2018PhRvL.121j2701H}. We call the former and the latter model as `{\it KCK19}' \citep{2019ApJ...872..164K} and `{\it HKC18}' \citep{2000ApJ...532.1132B} model, respectively.
The {\it HKC18} model turns out to have some inconsistency with the adopted pre-SN model. Detailed explanation about this inconsistency between the hydrodynamics and pre-SN model is explained in Section \ref{sec:abun}.

\begin{figure}[t]
		\begin{center}
			\includegraphics[width=8.4cm]{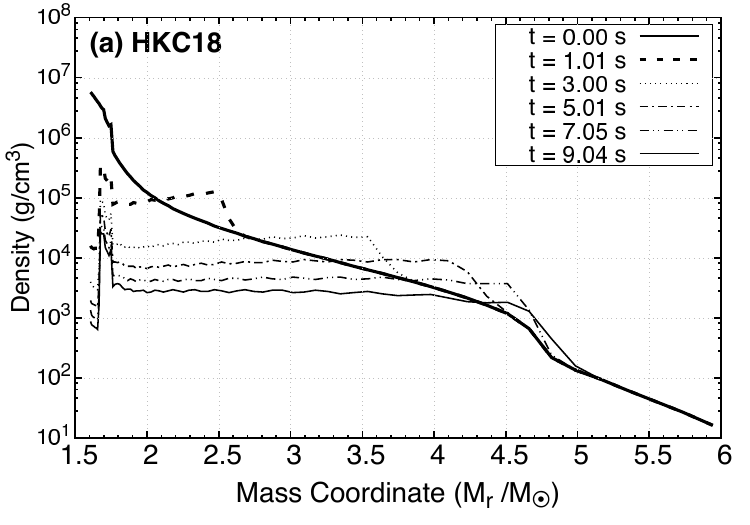}
			\includegraphics[width=8.4cm]{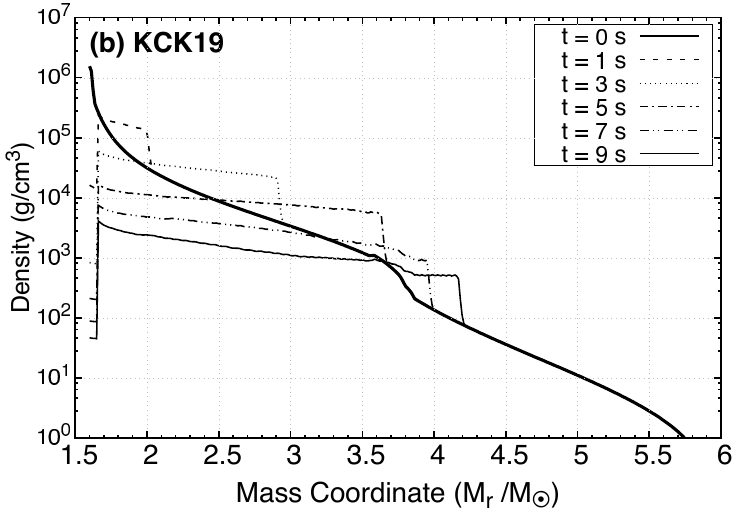}
		\end{center}
		\caption{The time-evolving density profiles in the Lagrange mass coordinate. The upper and lower panels show the hydrodynamics models used in {\it HKC18}  \citep{2000ApJ...532.1132B} and {\it KCK19} \citep{2019ApJ...872..164K}, respectively. The time range is taken from about 0 to 7 seconds.}
		\label{hydro_den}
	\end{figure}

\begin{figure}[t]
		\begin{center}
			\includegraphics[width=8.4cm]{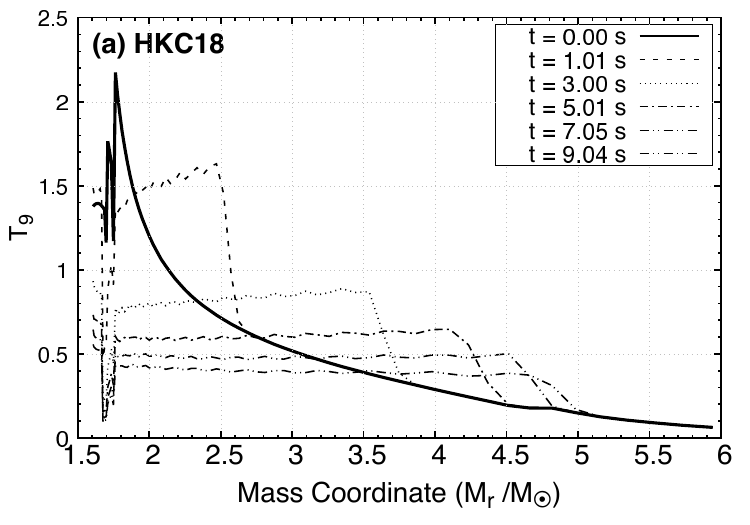}
			\includegraphics[width=8.4cm]{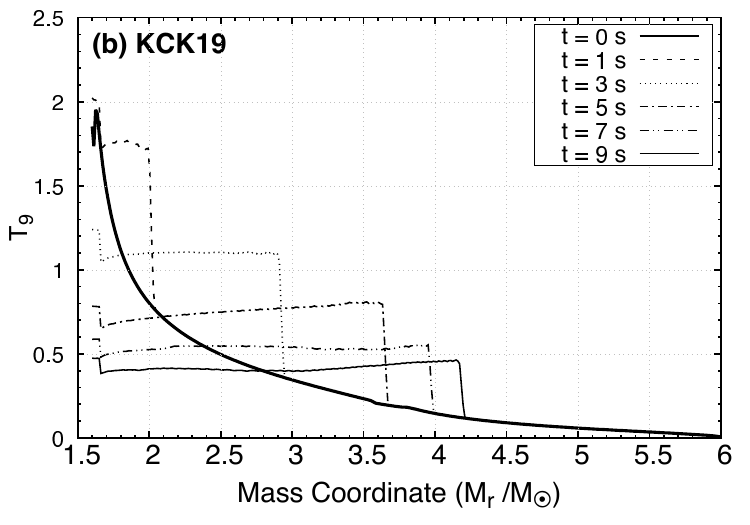}
		\end{center}
		\caption{The time-evolving temperature profiles as a function of the Lagrange mass coordinate. The upper and lower panels adopt the same models in Figure \ref{hydro_den}, respectively.
					  The temperature unit is taken as $T_9=T/(10^{9}\,{\rm K})$.}
		\label{hydro_tem}
	\end{figure}

\begin{figure}[t]
		\begin{center}
			\includegraphics[width=8.4cm]{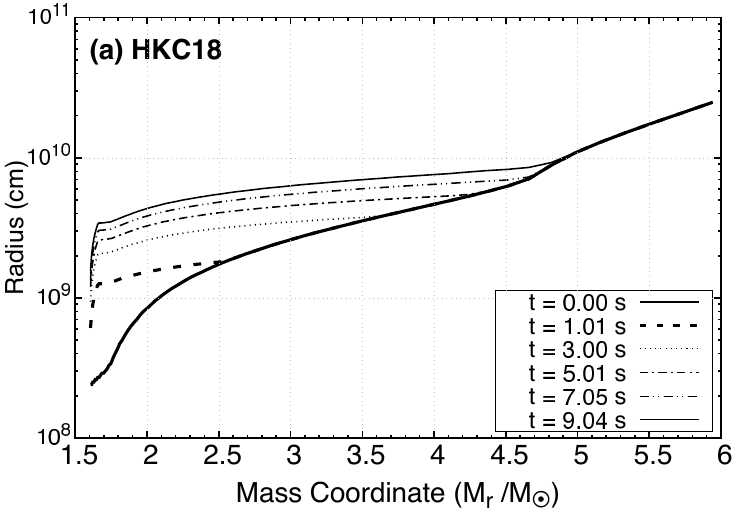}
			\includegraphics[width=8.4cm]{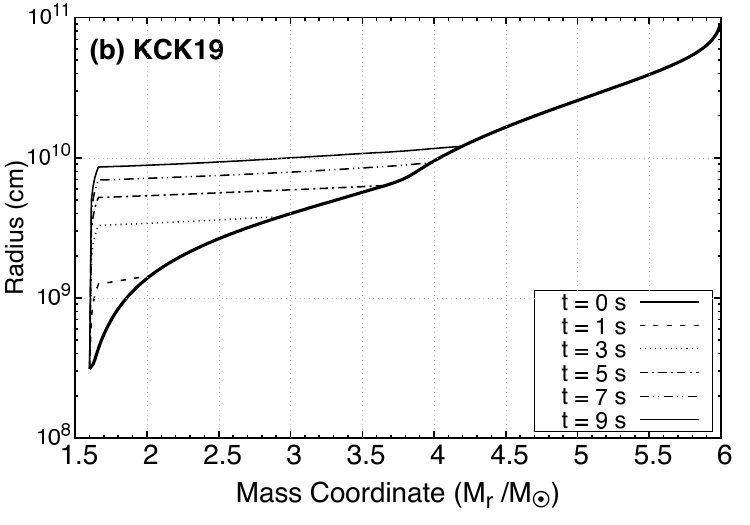}
		\end{center}
		\caption{The time-evolving radius profiles as a function of the Lagrange mass coordinate. The upper and lower panels adopt the same models in Figure \ref{hydro_den}, respectively. }
		\label{hydro_radi}
	\end{figure}

In the Lagrange mass coordinate, Figures \ref{hydro_den}, \ref{hydro_tem}, and \ref{hydro_radi} show the results of the time-evolving density, temperature, and radius from about 0 to 7 seconds, respectively, by the two different hydrodynamics. The upper and lower panels in the figures illustrate the results by {\it HKC18} and {\it KCK19} models, respectively. The different evolutions of the density affect the neutrino oscillation probability through the change of the effective potential in $\mathcal{V}_{\nu_ee}$. Also, the change of temperature is deeply involved in the thermonuclear reaction rates affecting the explosive nucleosynthesis as explained in Sections \ref{sec:nusyn} and \ref{sec:abun}.

\subsubsection{Neutrino flavor change probability}
\begin{figure*}[h!]
		\begin{center}
			\includegraphics[width=8cm]{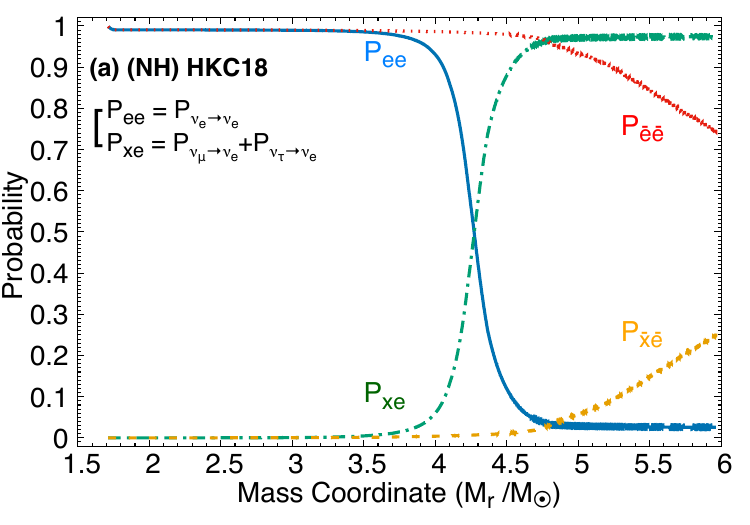}
		    \includegraphics[width=8cm]{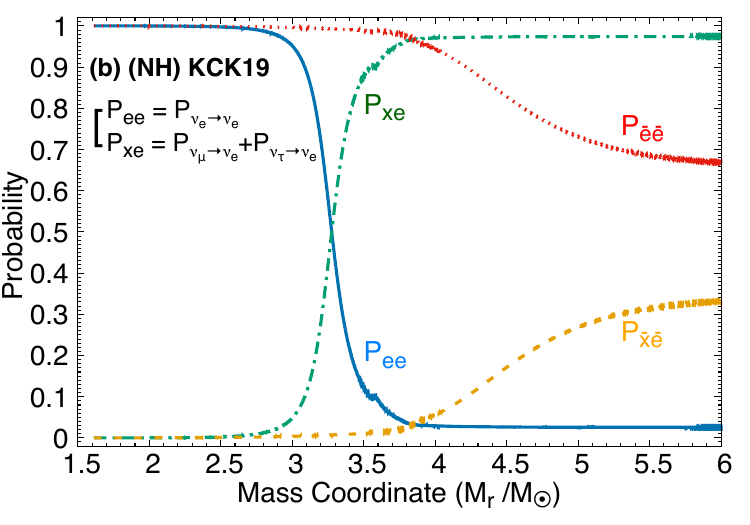}
		    \includegraphics[width=8cm]{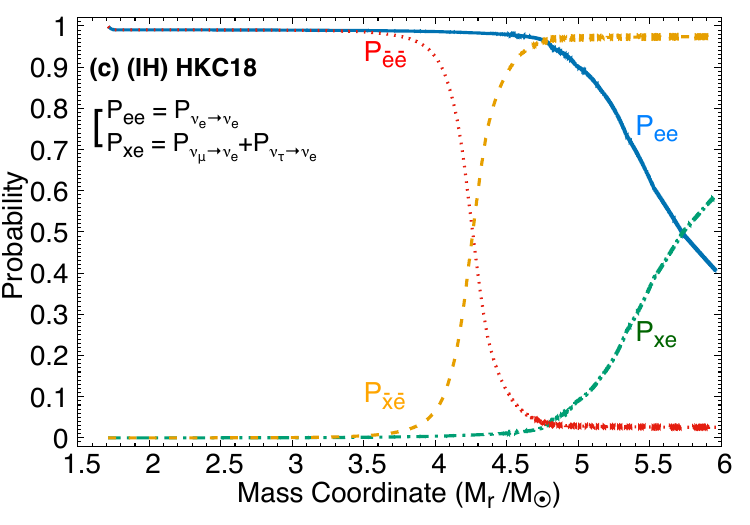}
		    \includegraphics[width=8cm]{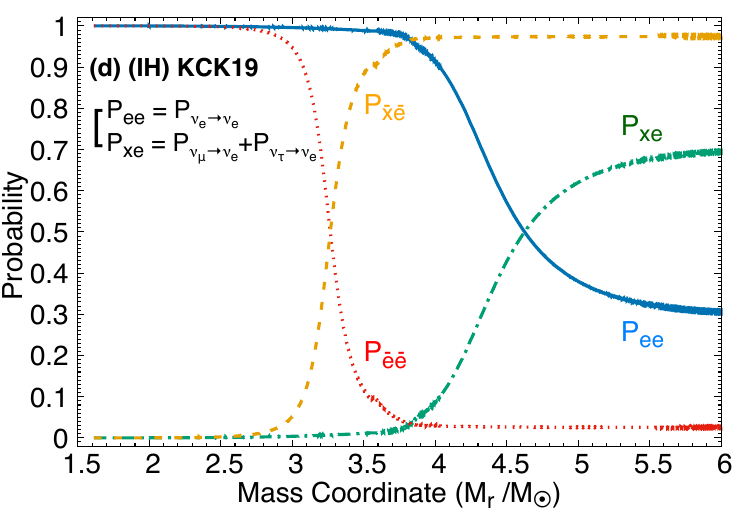}
		\end{center}
			\caption{The flavor change probability for $\nu_e$ with neutrino energy $E_\nu=15$ MeV calculated at $t=0$.
			Left and right panels adopt the hydrodynamics model of {\it HKC18} \citep{2000ApJ...532.1132B} and {\it KCK19} \citep{2019ApJ...872..164K}, respectively.
			Upper and lower panels correspond to the NH and IH, respectively. }
		 \label{MSW}
	 \end{figure*}

For the given density profile, we solve the Schr\"odinger-like equation with the total Hamiltonian in Equation (\ref{Ham_tot}). As a result, we obtain the neutrino flavor change probability from $\alpha$ to $\beta$ denoted by $P_{\alpha\beta}$ in the CCSN environments. Figure \ref{MSW} shows the survival probability of $\nu_e$ and $\bar{\nu}_e$ at $t=0$ as a function of the Lagrange mass coordinate for $E_\nu = 15\,{\rm MeV}$. The four panels show that the $P_{\alpha\beta}$ depends on the neutrino mass hierarchy and hydrodynamics models. The difference between the hydrodynamical models is shown at the left and right panels, while that by the mass hierarchy is distinguished by the top and bottom panels.

In all the panels of Figure \ref{MSW}, there are regions where the neutrino flavor probability changes drastically. During the neutrino propagation from the neutrino sphere, as the matter density decreases, the value of the vacuum oscillation term becomes comparable to the matter potential term in a specific region. As a result, the maximal neutrino flavor change is induced, which is called MSW resonance. Precisely, when the vacuum term is the same as the matter potential term, i.e., $\sqrt{2}G_F n_e = \cos 2\theta \Delta m^2/(2E_\nu) $, the MSW resonance occurs at the density
	\begin{eqnarray}
		\rho_{res} = \frac{\cos 2\theta_{ij} |\Delta m^2_{ji}| }{2\sqrt{2}G_F Y_e E_\nu N_A}.
	\label{eq-res}
	\end{eqnarray}
Here we approximately take the electron fraction as $Y_e=0.5$.

If we compare the left and right panels in Figure \ref{MSW}, we can note that the resonance region depends on the hydrodynamics models. The left panels adopting the {\it HKC18} model indicate the resonance region as $M_r \sim 4.6 M_{\odot}$, while the right panels for the {\it KCK19} model indicate a different resonance region at $M_r \sim 3.7 M_{\odot}$. This is because the density satisfying the resonance $(\sim 10^3\,{\rm g/cm^3})$ appears in the different regions. (See Figure \ref{hydro_den}).

The upper and lower panels in Figure \ref{MSW} show the different resonance patterns due to the neutrino MH. The difference stems from the value of density-dependent $\Delta m_{31}^2$ involved in the resonance density. As a result, for the NH case, the resonance leads the average of the initial $\nu_\mu$ and $\nu_\tau$ spectra to the final spectra of $\nu_e$. On the other hand, for the IH case, the resonance for the anti-neutrinos occurring in the inner side drives the average of the initial $\bar{\nu}_\mu$ and $\bar{\nu}_\tau$ spectra to be the final spectra of $\bar{\nu}_e$.

\subsection{The shock wave propagation effect}
Another noteworthy point is that the temperature and density are increased when the shock passes the stellar region. The shock propagation changes drastically the density in a specific region and time, which is related to the matter Hamiltonian. However, in the adiabatic approximation usually adopted in the quantum mechanics, quantum states are gradually varying with the external condition acting slowly enough. Therefore, we should weigh the adiabatic condition for the neutrino oscillation---to verify whether the quantum states of neutrinos react with the drastic change of the background by the shock propagation or not.

\begin{figure}[h!]
		\begin{center}
			\includegraphics[width=8cm]{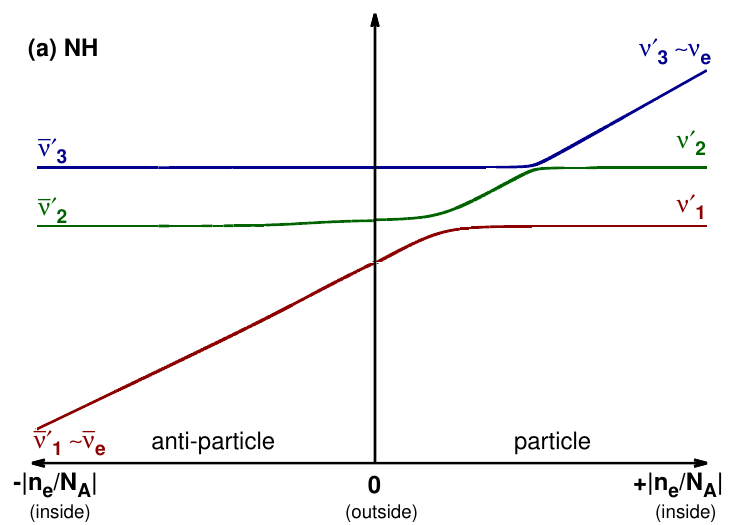}
			\includegraphics[width=8cm]{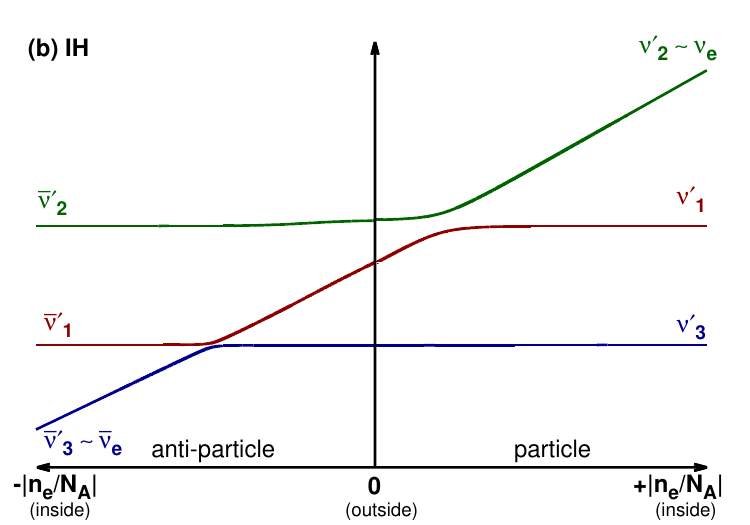}
		\end{center}
		\caption{The eigenvalues of matter states according to the electron number density per $N_A$ for $E_{\nu}=15$\,MeV.
		Upper and lower panels are the results calculated in the NH and IH scheme, respectively. }
		\label{lev_cross}
	\end{figure}

For the time evolution of the neutrino flavors, if the energy gap between mass eigenstates is given as $\Delta E$, it would satisfy the following condition:
	\begin{eqnarray}
		 \Delta E \times \Delta t \ll \hbar,
	\end{eqnarray}
where $\Delta t = \Delta r = \frac{\delta n_e}{n_e} ({n_e}/{\frac{d n_e}{d r}})$ is the time gap for the transition. Figure \ref{lev_cross} shows the eigenstates for 15\,MeV neutrino energy. In the adiabatic process, the diagonalized states are gradually changed and the resonance occurs when the mass eigenvalue of each state is close to each other, while the rapid external variation such as a shock can make the transition between the states as non-adiabatic process.

The flip probability for linear density case is given as \citep{2000PhRvD..62c3007D}
	\begin{eqnarray}
		P_f = \exp \left( -\frac{\pi}{2} \gamma \right),
	\end{eqnarray}
where the adiabatic parameter $\gamma$ is defined as \citep{1989PhRvD..39.1930K},
	\begin{eqnarray}
		\gamma = \frac{\Delta m^2_{ij}}{2E_\nu}\frac{\sin ^2 2 \theta_{ij}}{\cos 2 \theta_{ij}} \frac{n_e}{dn_e/dr}.
	\end{eqnarray}
For $\gamma \gg 1$, $P_f$ goes to zero, indicating that there is no transition between eigenstates, which corresponds to the adiabatic condition. On the other hand, for $\gamma \lesssim 1$, we expect a considerable flip probability, which implies the non-adiabatic condition.

We investigate the flavor-change probabilities for the three active neutrino flavors by the shock propagation effect. It was found that the first resonance occurs at $M_r \sim 1.6 M_{\odot}$ owing to the shock propagation. The resonance pattern is distinct from that in Figure \ref{MSW}. The adiabatic parameter has a value of $\gamma \approx 1.63 - 4.8$ at the resonance region where the flip probability is approximately evaluated as 8$\times 10^{-2} - 5 \times 10^{-4}$.

\begin{figure*}[h!]
		\begin{center}
			 \includegraphics[width=15cm]{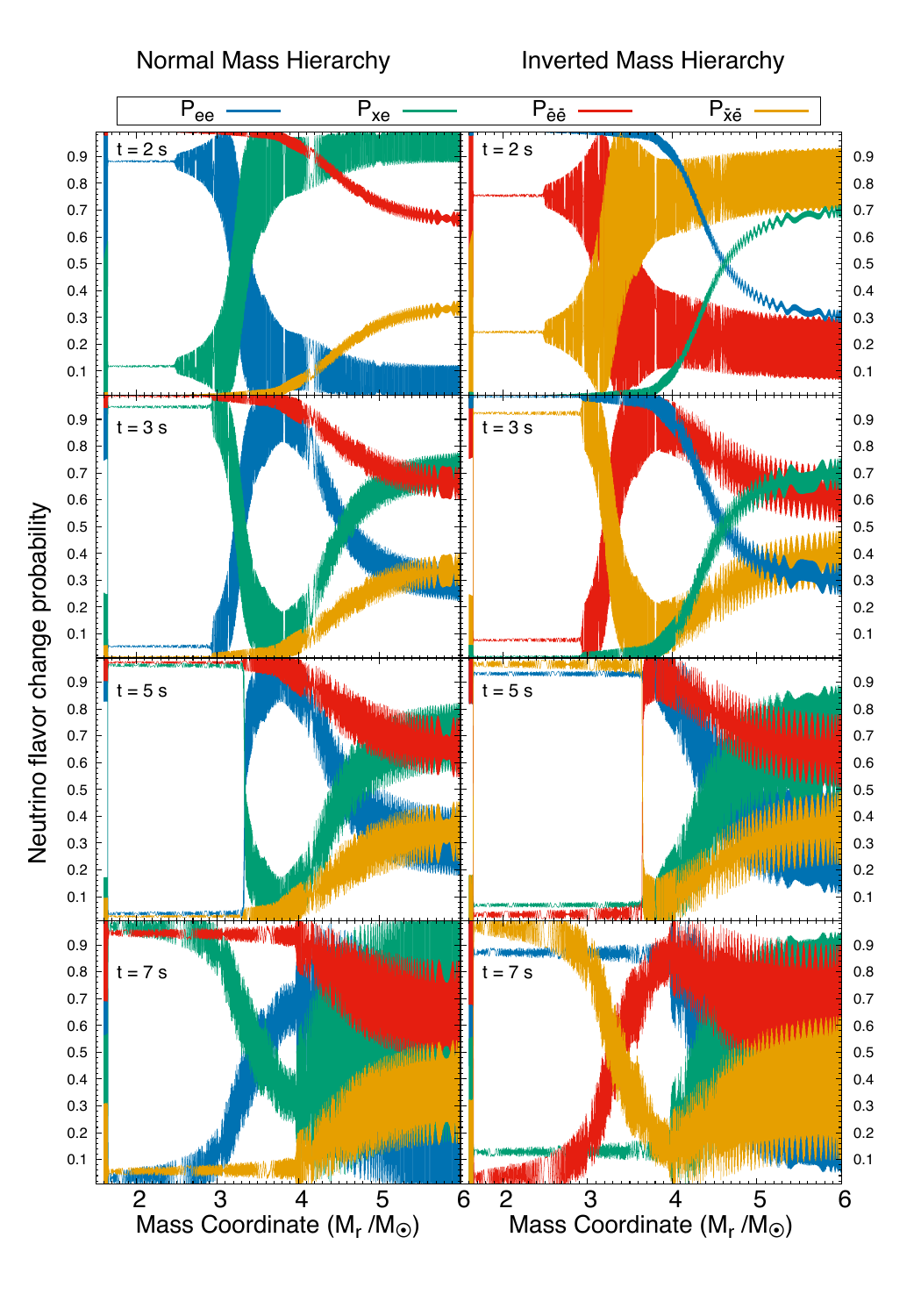}
		\end{center}
		\caption{The flavor change probability including shock propagation effect in the \cite{2019ApJ...872..164K} model at $t=2,3,5$, and $7\,{\rm s}$ for $E_\nu= 15$ MeV. The labels denote the probabilties as in Figure \ref{MSW}, i.e., P$_{\rm ee}=$P$_{\nu_e \rightarrow \nu_e}$ and P$_{\rm xe}=$P$_{\nu_\mu \rightarrow \nu_e}$+P$_{\nu_\tau \rightarrow \nu_e}$. Left and right panels correspond to the NH and IH, respectively.}
		 \label{shock_prob}
	 \end{figure*}

Within the shock propagation effect, the snapshots of the flavor change probabilities are shown in Figure \ref{shock_prob}. Starting off the first resonance at $M_r \simeq$ 1.65 $M_\odot$, the next resonance occurs at the region of $M_r \simeq$ 3.3 $M_\odot$. Due to the second resonance, the $\nu_e$ energy spectra returns to the initial flux. After that, the last resonance appears at $M_{r} \sim 4.7 M_\odot$. The multiple resonances by the shock propagation change the neutrino flux and affect the yields of the $\nu$-process. We discuss the yields of the nucleosynthesis including the shock effect in Section \ref{sec:abun}.

\section{Neutrino Self-Interaction} \label{sec:nu_SI}
\begin{figure}[h!] 
		\includegraphics[width=9.4cm]{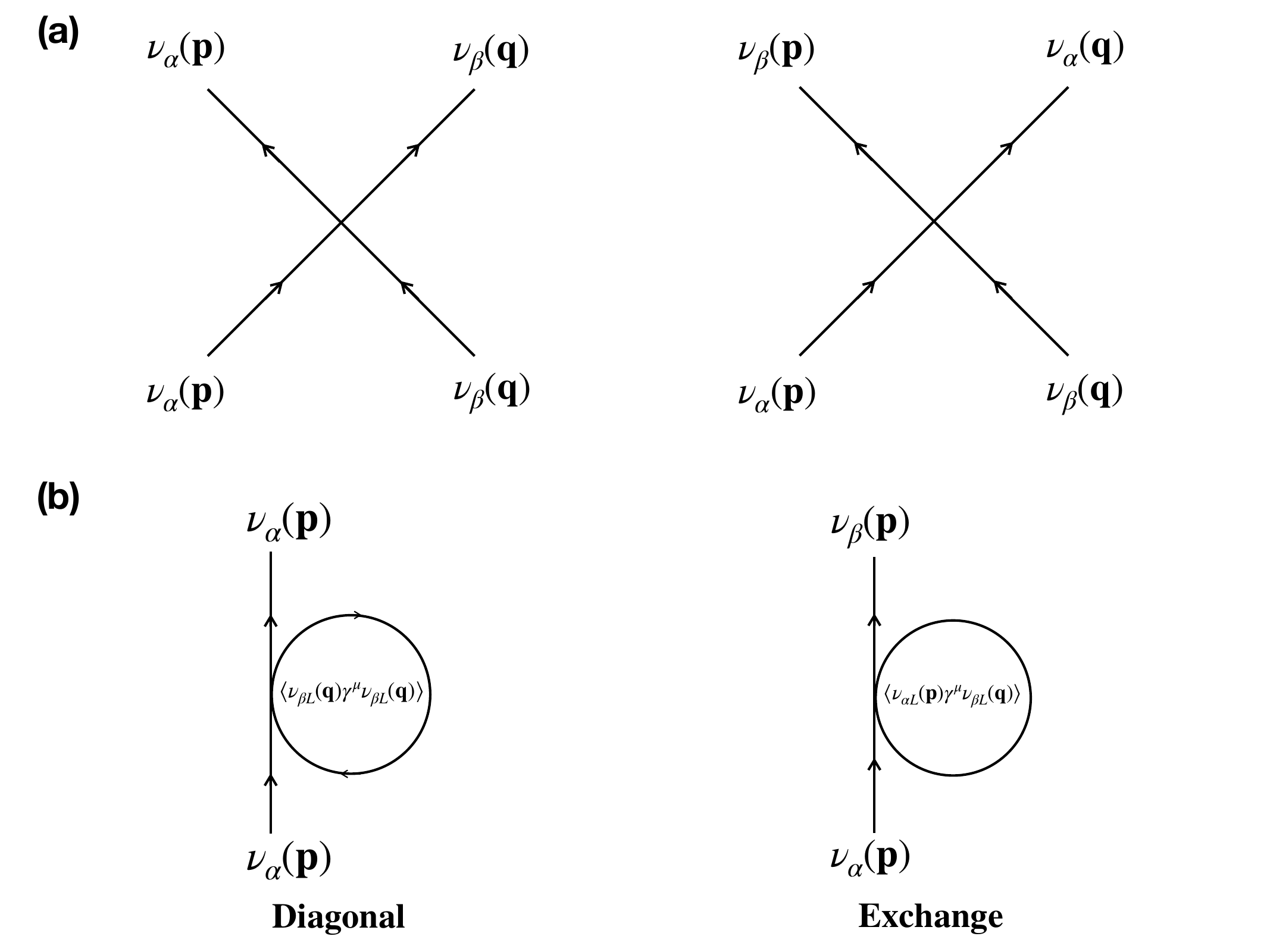}
		\centering
		\caption{(a) The $\nu$-$\nu$ interaction diagrams. The left and right diagrams correspond to the diagonal and exchange terms, respectively. (b) The mean-field approximation for the four-fermion interaction of the diagonal and exchange terms in (a).}
       \label{HF}
	\end{figure}

Near the neutrino sphere, the neutrinos trapped by dense matter formed via the gravitational collapse stream violently, so that the neutrino number densities are high enough to consider their SI during emission \citep{2007PhR...442...38J}. The considerable neutrino background causes the NC interaction to be comparable to the vacuum or electron matter potential for all neutrino flavors \citep{1993NuPhB.406..423S,2006PhRvD..74j5014D,2011JPhG...38c5201D}. The Hamiltonian density for the neutrino-neutrino interaction is written as \citep{1992PhLB..287..128P,1993PhRvD..48.1462S}
	\begin{eqnarray}
	\label{eq:Hamiltonian NC}
        \mathcal{H}^{NC}_{\nu\nu}(x)
  			&=&\frac{G_F }{\sqrt{2}}\sum_{\substack{\alpha,\beta,\gamma,\eta}}
			    ( \delta_{\alpha \gamma}\delta_{\beta \eta}
				 +\delta_{\alpha \eta}\delta_{\beta \gamma}
				  (1-\delta_{\alpha \beta}) ) \nonumber \\
			 	& \times &
				\left [
				      \bar{\nu}_{\gamma L}(x) \gamma^\mu \nu_{\alpha L}(x)
			    \right ]
				\left [
			           \bar{\nu}_{\eta L}(x) \gamma^\mu \nu_{\beta L}(x)
				\right ],
    \end{eqnarray}
where $\delta$ denotes the Kronecker delta describing the diagonal and momentum exchange interactions in Figure \ref{HF} (a) and its subindexes of $\alpha,\beta,\gamma$, and $\eta$ stand for the neutrino flavors. By adopting the mean-field approximation to the background neutrinos, we obtain the following background potential with three momenta ${\bf q}$ (See Appendix \ref{app_nunu})

    \begin{eqnarray}
    	&\hat{\mathcal{V}}&_{\nu\nu}
	=\sqrt{2}G_F \frac{1}{V}\sum_{\mathbf{p}\mathbf{q}}\frac{[a_{\nu_{\beta}}^{\dagger}(\mathbf{p}) a_{\nu_{\alpha}}(\mathbf{p})-b_{\nu_{\alpha}}^{\dagger}(\mathbf{p}) b_{\nu_{\beta}}(\mathbf{p})]}{2E_{p}V} \nonumber \\
	&\times&(1-\mathbf{\hat{p}}\cdot\mathbf{\hat{q}})  \left [
					      f_{{\rm dist}}(t,{\bf q})\rho_{\beta\alpha}({t,\bf q})-\bar{f}_{{\rm dist}}(t,{\bf q})\bar{\rho}_{\beta\alpha}(t,{\bf q})
					      \right],~~~
	\label{pot_nunu}
    \end{eqnarray}
where $\hat{\bf p}$ is the momentum direction of the propagating neutrino. The factor of $(1-\hat{\bf p} \cdot \hat{\bf q})$ stems from the weak current interaction, which prevents the scattering in the same trajectory \citep{2011PhRvD..84f5008P}. $f_{{\rm dist}}$ ($\bar{f}_{{\rm dist}}$) and $\hat{\rho}(\bar{\rho})$ are the normalized distribution and the density matrix for neutrinos (anti-neutrinos), respectively.

	Diagrams of the diagonal and exchange interactions giving rise to the potential are shown in Figure \ref{HF} (b). The diagonal term, as a linear contribution, is related to the number density of neutrinos. As an example, $f_{{\rm dist}}(t,{\bf q})\rho_{ee}(t,{\bf q}) / V$ means $\nu_e$ distribution affected by the flavor change probability $\rho_{ee}$, whose summation over ${\bf q}$ leads to the number density of $\nu_e$. In particular, the momentum exchange interactions, which act as the off-diagonal terms and initially set to be zero, affect the neutrino flavor transformations. The dominant off-diagonal potential known as Background Dominate Solution was studied in \citet{2006PhRvD..73b3004F}.
	
	$\sum_{\mathbf{q}}f_{{\rm dist}}(t,{\bf q}) / V$ in Equation (\ref{pot_nunu}) is the local neutrino number density denoted by $\int d n_{\nu}$. To determine the $d n_{\nu}$, as a neutrino emission model, we adopt the neutrino bulb model \citep{2006PhRvD..74j5014D} assuming the uniform and half-isotropic neutrino emission \citep{2011PhRvD..84b5002C} and obtain $d n_{\nu}$ as follows:
	\begin{eqnarray}
		dn_{\nu_\eta} &=& \frac{1}{\pi R^2_\nu} \frac{L_{\nu_\eta}}{\langle E_{\nu_\eta} \rangle}
					  	  f_{FD}(E_q,T_{\nu_\eta}) E_q^2  dE_q d \Omega_{q}  ,
		\label{diff_nu_num}					  	
	\end{eqnarray}
where $R_\nu$, $L_{\nu_\eta}$ and $\langle E_{\nu_\eta} \rangle$ are the neutrino sphere radius, neutrino luminosity, and averaged neutrino energy for $\eta$ flavor, respectively. We take the normalized distribution function of the neutrinos as the Fermi-Dirac distribution with zero chemical potential, i.e., $f_{FD}(E_q,T_{\nu_\eta}) \equiv (4\pi T_{\nu_\eta}^3 F_2(0))^{-1} (\exp(E_q/T_{\nu_\eta})+1)^{-1} $, where  $F_2(0)$ is 1.80309.  The solid angle part is given by the z-axis symmetry as $\int(1-\hat{\bf p} \cdot \hat{\bf q}) d \Omega_{\bf q} = 2\pi \int^\theta_0 (1-\cos \theta_{p}\cos \theta_{q}) \sin \theta_{q} d \theta_{q} $ (Appendix \ref{app_nunu}). In the single-angle approximation, since $\theta_{p}$ is neglected, the integration of the angular term becomes $(1-\cos \theta)^2/2$. The maximum $\theta$ is the tangential direction at neutrino emission point and satisfies the relation of $\sin \theta_{\rm max} = R_\nu/r$. Although the single-angle approximation may be proper in the early universe satisfying the isotropic and homogeneous conditions \citep{1994PhRvD..49.1740K}, it is not enough to describe the decoupling of neutrinos near the SN core as explained in \citet{2011JPhG...38c5201D}. Hence, in this study, we perform the multi-angle calculation involving the $\theta _{p}$.

	When we consider the broken azimuthal symmetry \citep{2011PhRvD..84e3013B,2013PhRvL.111i1101R}, the azimuthal angle ($\phi_q$) dependent term is not negligible in the factor of $(1-\hat{\bf p} \cdot \hat{\bf q})$. Therefore, in the accretion phase, the violation of the azimuthal symmetry results in the multi-azimuthal angle instability \citep{2014PhRvD..90c3004C} and fast $\nu$-flavor conversion \citep{2019PhRvD.100d3004A} (See also \citet{2008PhRvD..78c3014D} for effects of non-spherical geometry). That is, such a realistic SN model brings about the neutrino emission unpredicted in the symmetric model.

	However, the $\nu$-process is more significantly affected by the neutrino cooling phase---occurring on a long neutrino emission time---rather than the burst and accretion phases as explained in Section \ref{sec:neq}. Although the asymmetric models can also affect the neutrino cooling phase, there have not been appropriate models for applying to the $\nu$-process. Moreover, the asymmetry requires to describe the $\nu$-process by three-dimensional hydrodynamical models, which is beyond the scope of the present study. Hence, we leave the effects of the asymmetric neutrino emission on the $\nu$-process as a future work. As the CCSN model, we adopt the one-dimensional (1D) spherical symmetric model with the multi-angle effect for the collective neutrino oscillation.

	Setting up the model, we solve the following equations of motions for neutrino density matrices with the three-flavor multi-angle calculation \citep{2017PhRvD..96d3013S},
\begin{eqnarray}
	\cos\theta_{p}\frac{d}{dr}\rho(r,E_{p},\theta_{p})
         &=&-i[H_{\rm Vacuum}+\mathcal{V}_{\nu},~\rho(r,E_{p},\theta_{p}) ],~~~~~~\label{eom}\\
	\cos\theta_{p}\frac{d}{dr}\bar{\rho}(r,E_{p},\theta_{p})
	&=&-i[-H_{\rm Vacuum}+\mathcal{V}_{\nu},~\bar{\rho}(r,E_{p},\theta_{p}) ],~~~~~\\
	\mathcal{V}_{\nu}
	&=&\mathcal{V}_{\nu_{e} e} + \mathcal{V}_{\nu\nu},
\end{eqnarray}
where $H_{\rm Vacuum}$ and $\mathcal{V}_{\nu_{e} e}$ are the vacuum term and the MSW matter potential, respectively, in Equation (\ref{Ham_tot}) and $\mathcal{V}_{\nu \nu}$ is the potential for the neutrino SI ($\nu$-SI) in Equation (\ref{eq: V_nunu}).

The study of the collective neutrino oscillation requires the lepton density profiles near the neutrino sphere to outside. For the baryon matter density from $r=$10 km to around 2000 km, we adopt the empirical parametrization of the shock density profile \citep{2003PhRvD..68c3005F}. For $t>1$\,s, the density profile is given as
	\begin{eqnarray}
    	&&\rho_b(r,t)  =  10 \times \left( 10^{14} {\rm g/cm^3} \right) \left(\frac{\rm 1 km}{r} \right )^{2.4} \nonumber \\
     &&\times
          \exp{\left [ \left( 0.28-0.69 \ln \left (\frac{x_s}{\rm 1 km} \right ) \right )
          \left(\sin^{-1}\left(1-\frac{r}{x_s}\right)\right)^{1.1} \right] },~~~~~
		\label{fogli_den}
	\end{eqnarray}
where the shock front position is defined by $x_s(t)=x^0_s+v_st+0.5a_st^2$ with $x_s^0 \simeq -4.6 \times 10^{3}$\,km, $v_s \simeq 11.3 \times 10^3\,{\rm km/s}$, and $a_s \simeq 0.2 \times 10^3\,{\rm km/s^2}$. For $t \le 1$\,s, we use $\rho_b(r,t) =  10^{14} (g/cm^{3}) \left({\rm 1 km}/r \right)^{2.4}$. The electron number density is given in the same way, i.e., $n_e(r,t)= \rho_b(r,t)Y_e  N_A$.

\begin{figure}[h!] 
	\centering
	\includegraphics[width=8.4cm]{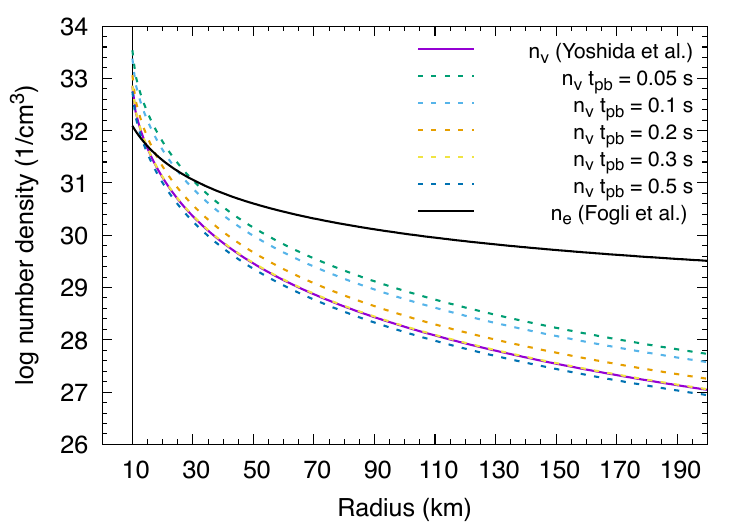}
	\caption{The number density of electron and electro-neutrino. Each dashed line stands for the number density at post bounce time $t_{pb}$=0.05, 0.1, 0.2, 0.3 and 0.5 s as labeled.}
	\label{lep_num_den}
\end{figure}

	Compared to the electron density, the neutrino number density obtained by the integration of Equation (\ref{diff_nu_num}) relies not only on the radius, but also the neutrino luminosity and averaged energy. Within the single-angle approximation without flavor mixing between the neutrinos, the neutrino density is given as follows (See Equation (\ref{sin_ang_geometry})):
	\begin{eqnarray}
		n_\nu (r) &\approx& (2 \pi R^2_\nu)^{-1}  \left (1-\sqrt{1-\left ( R_\nu/r \right) ^2} \right )^2 \\[12pt] \nonumber
		&&\times
		\left [ L_{\nu_e}/\langle E_{\nu_e} \rangle - L_{\bar{\nu}_e} \langle E_{\bar{\nu}_e} \rangle \right].
	\end{eqnarray}	
Figure \ref{lep_num_den} shows the lepton and the neutrino number densities estimated by the above equation, which is similar to the result in \citet{2006PhRvD..74j5014D}. This figure implies that the $n_\nu$ and $n_e$ are sufficient to consider the $\nu$-SI. Furthermore, the valuable change of the neutrino flux by the SI would affect the $\nu$-process. Here we note that
the density hinges on the SN simulation model. For example, \cite{2011PhRvD..84b5002C} exploited the SN simulation of \cite{2010A&A...517A..80F}, where the neutrino sphere blows up to about 100 km during the accretion phase and drops to about 30 km at the cooling phase. Consequently, the neutrino flux is much smaller than the present case where the neutrino sphere is assumed as about 10 km. Moreover, the electron density in \cite{2011PhRvL.107o1101C} is larger than that evaluated by \cite{2003PhRvD..68c3005F} adopted in this calculation. In this perspective, the small neutrino sphere radius of $\sim$ 10 km and a simple fast-decreasing power law density profile proportional to $r^{-2.4}$ in the present work may not be compatible with those typically seen during the accretion phase of SN simulations. Smaller neutrino density and higher electron density during the accretion phase from the SN simulations than the present work may lead to the suppression of the neutrino collective motion. The relevant uncertainty during the accretion phase is discussed quantitatively in the conclusion.

	To determine the neutrino number density with SI effects, we study two kinds of neutrino luminosity and averaged energies. The first one is the flavor independent neutrino luminosity, i.e., each neutrino has the equivalent luminosity. In this model, with the Fermi-Dirac distribution, the averaged energy is determined by $\langle E_\nu \rangle = 3.1514 ~T_\nu$, and we take the temperatures constrained from the $^{11}$B abundance \citep{2005PhRvL..94w1101Y}. As the second model, we take the parameters for the luminosity obtained from the SN simulation results.

\subsection{Equivalent (EQ) luminosity model}
	The first case is the equivalent neutrino luminosity, which is given as \citep{2004ApJ...600..204Y}
	\begin{equation}
		L_{\nu_\alpha}(t)= \frac{1}{6} \frac{E_\nu}{\tau_\nu} \exp \left (-\frac{t-r}{\tau_\nu} \right) \Theta(t-r),
		\label{lum_yoshi}
	\end{equation}
where we take the total explosion energy $E_\nu = 3 \times 10^{53}$ erg and the decay time of luminosity $\tau_\nu = 3$ s. The quantities $t$ and $r$ are time after explosion and radius from the core, respectively. The exponential decay form of the luminosity stands for the cooling phase, in which most of neutrino energy is brought out \citep{1990ApJ...356..272W,2005PhLB..606..258H,2012arXiv1201.1637R,2019ApJ...876..151S}. $\Theta$ is the Heaviside step function, which makes the luminosity zero before the arrival of the neutrinos at the position $r$. In this EQ model, we assume the Fermi-Dirac distribution with $T_{\nu_e}=3.2$ MeV, $T_{\bar{\nu}_e}=5$ MeV, and $T_{\nu_{\mu,\tau},\bar{\nu}_{\mu,\tau}}=6$ MeV.

	By adopting this model, the differential neutrino flux is given as
\begin{eqnarray}
	&& {\phi}^\prime (t,r;E_{\nu},T_{\nu_{\alpha}})
      \equiv  \frac{d}{dE_{\nu}}\phi(t,r;E_{\nu},T_{\nu_{\alpha}}) \nonumber \\
	&&   =     \frac{L_{\nu_\alpha}(t)}{4 \pi r^2} \frac{1}{T_{\nu_\alpha}^4 F_3(0)}
		   \frac{E_{\nu}^2 }{\exp(E_{\nu}/T_{\nu_\alpha})+1}
	       \langle \rho_{\alpha\alpha}(r;E_{\nu}) \rangle, ~~
	\label{diff_flux}
\end{eqnarray}
where $F_3(0)$= 5.6882 in the Fermi  integration and $\langle \rho_{\alpha\alpha} ({r;\epsilon}_{\nu}) \rangle$ is the angular averaged neutrino density matrix of $\nu_\alpha$ from Equation (\ref{eom}).

\begin{figure}[h!]
		\begin{center}
			\includegraphics[width=8.4cm]{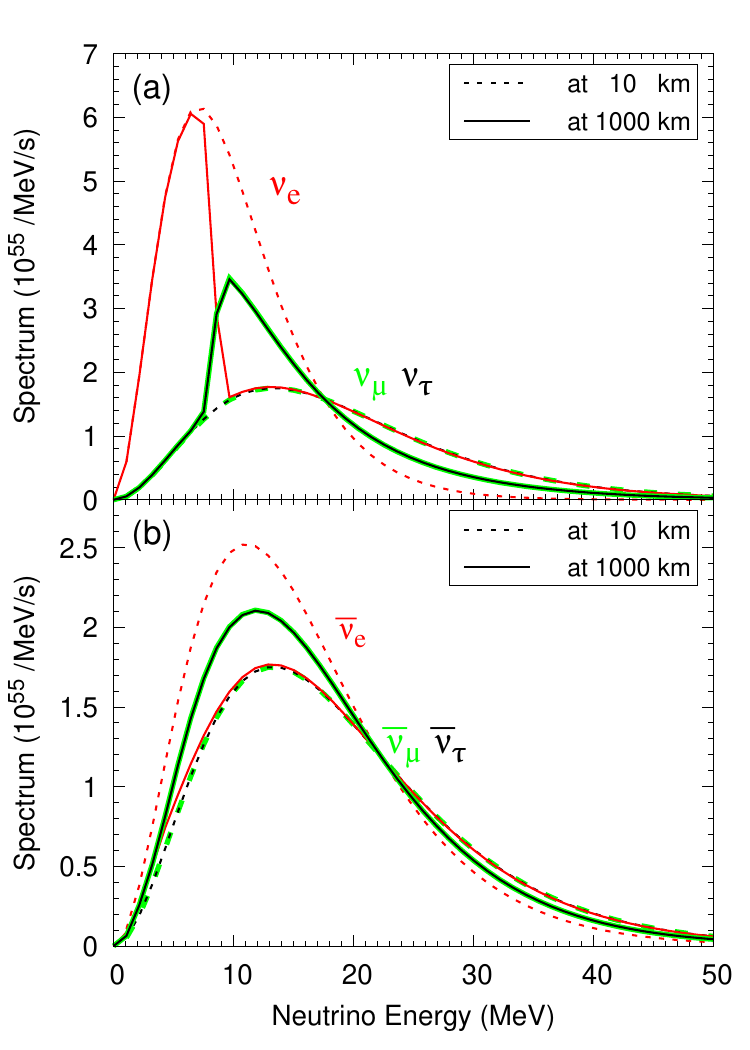}
		\end{center}
		\caption{The neutrino spectra $(\equiv 4 \pi r^2 {\phi}^\prime)$ with SI effects for the IH. Upper and lower panels show the $\nu$ and $\bar{\nu}$ cases, respectively.}
		\label{EQ_SI}
	\end{figure}

	For the IH, Figure \ref{EQ_SI} shows the neutrino spectra obtained by $4\pi r^2 {\phi}^\prime (r,E_{\nu},T_{\nu_{\alpha}})$  at $t = 1$ s in Equation (\ref{diff_flux}). At $r=10$\,km, we take the spectra as the Fermi-Dirac distribution. As shown in Figure \ref{EQ_SI}, over $E_{\nu}=$ 17\,MeV region, the $\nu_e$ distributes more than those before the SI occurs. This enhanced high energy tail is originated from the reduced $\nu_\mu$ and $\nu_\tau$ distributions, so that the total number of the neutrinos is conserved. On the other hand, $\bar{\nu}_e$ changes from $\bar{\nu}_\mu$ and $\bar{\nu}_\tau$ similarly to the MSW resonance effect. For the NH, the SI effect is suppressed, and consequently, there is no difference compared to the case without the SI term.

\subsection{Non-equivalent (NEQ) luminosity model} \label{sec:neq}
The neutrino luminosity of each flavor may have different epochs in CCSN explosions which are classified into three phases by neutrino signals \citep{2010A&A...517A..80F,2012arXiv1201.1637R}: burst, accretion, and cooling phases.
\begin{table*}[ht!]
\begin{center}
	    \begin{tabular}{c | c c c | c c c || c}
	      \hline
	      \hline
	      Time &  ~$L_{\nu_e}$  & $L_{\bar{\nu}_e}$ & $L_{\nu_x}$ &
	      ~~$\langle E_{\nu_e} \rangle $ & $ \langle E_{\bar{\nu}_e} \rangle $ & $ \langle E_{\nu_x} \rangle $
	      & Interval  \\
	      \hline
	      (s)  & ~~~~~~~~~~~~~~~~~~  &   ($ 10^{52}$ erg/s) &   &
	       &  (MeV) & & section(ms)  \\ 
	      \hline
	     0.05  ~&~ ~6.5 (4.1)\footnote{The values in parentheses mean ($10^{58} $ MeV/s) units.}
                             & 6.0 (3.8) & 3.6 (2.3) ~&~  9.3 	& 12.2	& 16.5	&    0-75 \\
	     \hline
	     0.1   ~&~ 7.2 (4.5) & 7.2 (4.5) & 3.6 (2.3) ~&~  10.5	& 13.3	& 16.5	&   75-150 \\
	     \hline
	     0.2   ~&~ 6.5 (4.1) & 6.5 (4.1) & 2.7 (1.7) ~&~  13.3	& 15.5	& 16.5	&  150-250 \\
	     \hline
	     0.3   ~&~ 4.3 (2.7) & 4.3 (2.7) & 1.7 (1.1) ~&~  14.2	& 16.6	& 16.5	&  250-350 \\
	     \hline
	     0.5   ~&~ 4.0 (2.5) & 4.0 (2.5) & 1.3 (0.8) ~&~  16.0	& 18.5	& 16.5	&~~350-500 
	     \footnote{After 500 ms the exponential decay, $\exp(-(t-r)/\tau_\nu)$, is assumed.}  \\
	     \hline
	    \end{tabular}
	    \end{center}
	    \caption{Neutrino luminosity data at given time intervals taken from \citet{2018JPhG...45j4001O}}
		\label{Lumino}
	\end{table*}
In the present study, we ignore the $\nu_e$ burst phase, in which the prompt deleptonization process of electrons occurs at post-bounce time $t_{pb}~\sim$ 0.01 s. The shock propagation from the inner core drives the dissociation of heavy nuclei into protons and neutrons. Consequently the electron captures ($e^-+p \to n + \nu_e$) produces the $\nu_e$ burst.

The sensitivity of the $\nu_e$ burst phase on the $\nu$-process was studied by varying the electron neutrino luminosity in \cite{2019ApJ...876..151S}.
According to the study, however, the yields of synthesized elements in the $\nu$-process increase only up to maximally 20 \% for $^{138}$La and 5 \% for $^{11}$B when an FD distribution is adopted. The other elements of $^{7}$Li, $^{15}$N, $^{19}$F, and $^{180}$Ta increase by less than about 1 \%. Hence, in our study, we only consider the cooling phase presumed by the exponential decay \citep{2004ApJ...600..204Y,2019ApJ...876..151S,2005PhLB..606..258H,1990ApJ...356..272W} and a partial accretion phase, when the shock wave stagnates due to the falling materials and deposits its energies.
	
\begin{figure}[h]	
        	\begin{center}
		\includegraphics[width=8.4cm]{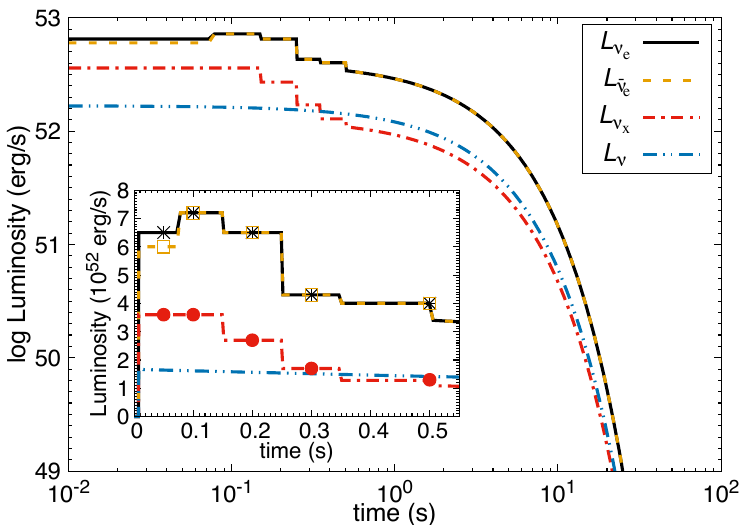}
            \end{center}		
		\caption{The neutrino luminosities for each flavor: $\nu_e,~\bar{\nu}_e$, and $\nu_x (=\nu_\mu,\nu_\tau, \bar{\nu}_\mu$ and $\bar{\nu}_\tau$) in the region of
		$M_r \sim 1.6 M_\odot ($corresponding to $r\approx 2300 $ km). The inset shows an enlarged figure of the x- and y-axes.
		The yellow line denotes the luminosity calculated by Equation (\ref{lum_yoshi}) and others are adopted from Table \ref{Lumino}. }
		\label{fig_lumino}
	\end{figure}

To investigate the accretion phase before $t_{\rm pb} \simeq 1 $ s, we take the SN simulation data in \citet{2018JPhG...45j4001O}.
This study compared the results by six individual SN simulation models performed by different study groups and showed consistency in the CCSN explosion mechanism. We take neutrino luminosities and averaged energies by selecting five post-bounce times, i.e., 0.05, 0.1, 0.2, 0.3, and 0.5 s. These time-dependent data are necessary not only for the neutrino propagation but also for the calculation of the $\nu$-process. For simplicity, we set the time interval between the five selected times as the numbers at the 8th column at Table \ref{Lumino} and use a step function form shown in Figure \ref{fig_lumino}. As mentioned already, we ignore the $\nu_e$ burst in the 0.00 $< t_{pb}< $ 0.05 region in \citet{2018JPhG...45j4001O}. We treat the luminosities after 0.5\,s in the cooling phase as the exponential decay, i.e., $L_\nu(t) = L_\nu(500\,{\rm ms}) \times \exp(-(t-r)/{\tau_\nu}) \Theta(t-r)$.

\begin{figure*}[h!]
		\centering
		\includegraphics[width=15cm]{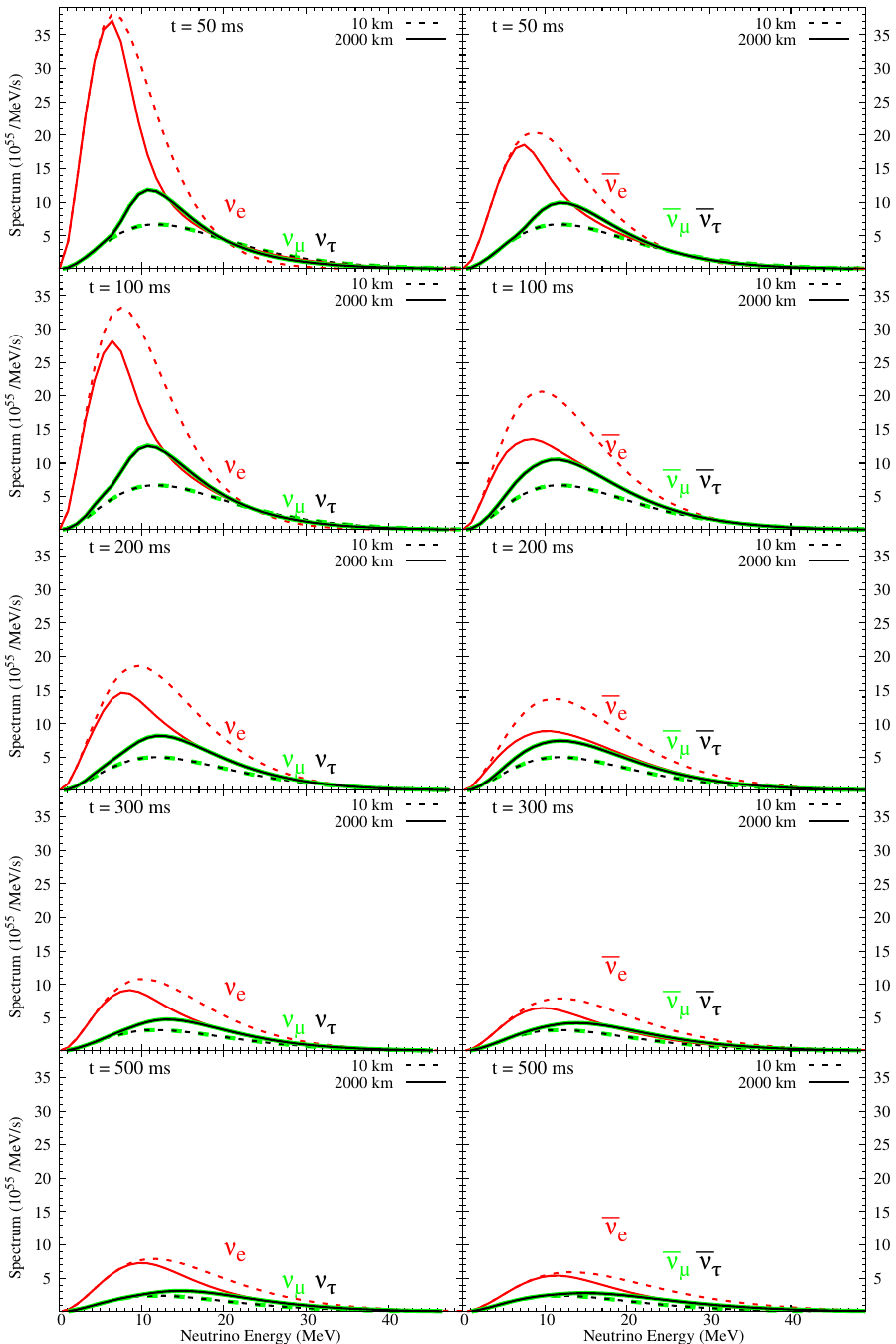}
		\caption{The neutrino spectra including neutrino SI at $t=50, 100, 200, 300\ {\rm and}\ 500\,{\rm ms}$ for NH case.
		The dashed and solid lines denote the spectra at $r=$10 km and 2000 km, respectively. Red, green, and black lines denote $\nu_e$, $\nu_\mu$, and $\nu_\tau$.}
	\label{spectra_NH}
	\end{figure*}

\begin{figure*}[ht!]
		\centering
		\includegraphics[width=15cm]{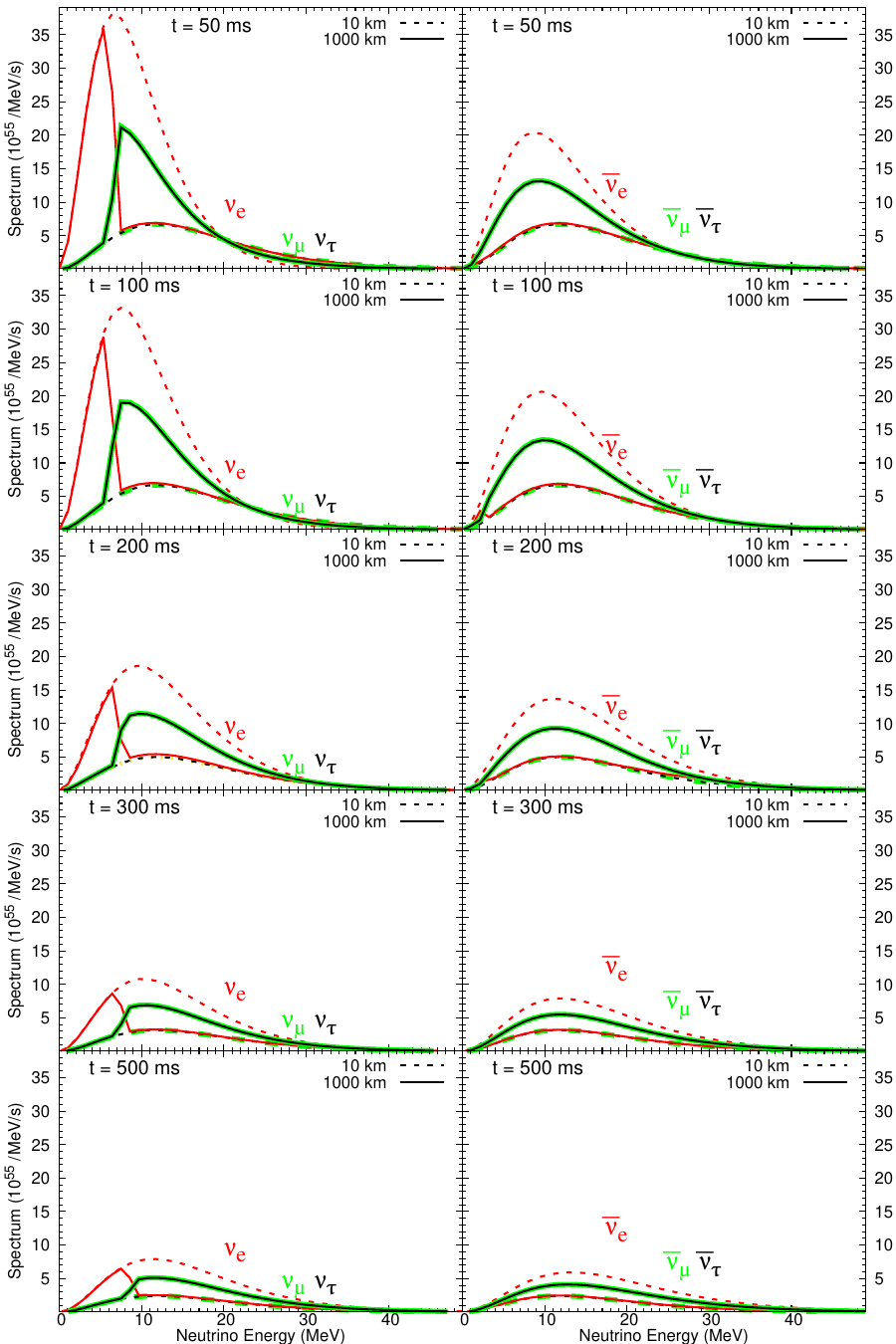}
		\caption{Same as Figure \ref{spectra_NH}, but for the IH case at 10 km and 1000 km. }
		\label{spectra_IH}
	\end{figure*}

	Using the parameters at Table \ref{Lumino}, we solve the equation of motion for neutrinos described in Equation (\ref{eom}). Figures \ref{spectra_NH} and \ref{spectra_IH} show the snapshots of the neutrino spectra at the denoted time. Contrary to the equivalent luminosity model, in the NEQ model, we found that the neutrino SI affects the neutrino spectra for both mass hierarchy cases.

	For the NH case, neutrinos propagating from the neutrino sphere ($r \simeq 10\,{\rm km}$) interact with the background neutrinos. Considering the neutrino SI with Equation (\ref{eom}), we get the neutrino spectra. From $r=10\,\rm{km}$ to $r=2000\,{\rm km}$, the $\nu$-SI affects each neutrino flavor spectrum, and after $r=2000\,{\rm km}$, the $\nu$-SI does not contribute to the neutrino spectra. Hence, we compare the spectra at $r=10\,{\rm km}$ and $r=2000\,{\rm km}$ (Figure \ref{spectra_NH}).

	At $t=50$\,ms and $t=100$\,ms, the spectra of $\nu_e$ and $\nu_x (x =\mu$ and $\tau)$ intersected at $E_\nu \sim 20 \,{\rm MeV}$ and $23\,{\rm MeV}$, respectively. After these cross points, the high energy tail of the $\nu_e$ spectra is enhanced while the spectra in the $E_\nu<$ 5 MeV region are same. As a result, the first and second panels of Figure \ref{spectra_NH} disclose that the comparable luminosity of $\nu_x$ forces the high energy tail of the $\nu_e$ spectra. Also, this trend is seen in the case of the antineutrinos. On the other hand, at the next time steps of $t = 200, 300, {\rm and}\ 500\,{\rm ms}$, although the averaged energy of $\nu_x$ becomes higher than the previous time steps, the spectra of $\nu_e$ are reduced from the initial spectra owing to the lower luminosity of $\nu_x$.

	In the case of IH, the neutrino spectra are shown for $r=10\,{\rm km}$ and $r=1000\,{\rm km}$ in Figure \ref{spectra_IH} . The spectra are not changed after $r=1000$ km. Similar to the equivalent luminosity model, the IH case shows that the spectra at $t = 50$\,ms and 100\,ms are split in the low energy region. Although this behavior is similar to the equivalent luminosity model, the difference occurs in the high energy tail. Analogous to the previous NH case, at $t = 50\,{\rm ms}$ and $t=100\,{\rm ms}$, the spectra of $\nu_e$ increases at $E_\nu=20\,{\rm MeV}$ and $E_\nu=23\,{\rm MeV}$, respectively. Furthermore, at the region of $E_\nu>7\,{\rm MeV}$, the $\nu_e$ spectra changes to $\nu_{x}$. This change is distinct from the previous NH case involving only the partial change in that energy region. Also, although the spectra at $t = 200, 300, {\rm and}\ 500\,{\rm ms}$ indicate the splitting of spectra, the $\nu_e$ spectra are lower than their initial spectra. It is because the luminosity is too low at that time as in the NH case. On the other hand, the $\bar{\nu}_e$ is fully changed to the $\bar{\nu}_x$ in the whole region. Consequently the spectra are the same as the initial $\bar{\nu}_x$ spectra shown in the right panels in Figure \ref{spectra_IH}.
	
\clearpage

\section{Explosive Nucleosynthesis} \label{sec:nusyn}
The stellar temperature ranging from $10^6$\,K to $10^9$\,K is high enough that thermonuclear reactions operate and become the main input of the nucleosynthesis. We perform the network calculation containing about 38,000 nuclear reactions for 3,080 nuclides up to the $^{232}{\rm Fr}$ isotopes by taking the result of the SN1987A progenitor model. After the helium and carbon burning as well as the weak $s$-process of the progenitor model, we make an explosion described by the hydrodynamics, from which the time evolution of abundances is considered. The initially emitted neutrinos react with nuclides generated in the pre-SN during the CCSN explosion. After a few seconds, the shock propagation from the core arrives at each layer and increases the density, radius, and temperature affecting the nucleosynthesis. This section addresses the thermonuclear reaction formulae for the explosive nucleosynthesis.

\subsection{Thermonuclear Reactions}
For the nuclear reactions in the stellar environments, assuming that reacting nuclei stay in local thermal equilibrium, we use thermal averaged reaction rates with the Maxwell-Boltzmann distribution \citep{1999NuPhA.656....3A}. The standard reaction rates are taken from in the JINA REACLIB database \citep{2010ApJS..189..240C}. Based on the averaged reaction rates, the time evolution of the abundance $Y_i$ defined as $Y_i=N_i/(\rho_b N_A)$ with the baryon density $\rho_b$ and the number density $N_i$ for a species $i$ is described in terms of forward (+) and reverse (--) reactions as follows

	\begin{eqnarray}
		\frac{dY_i}{dt} &=&  + Y_{j} \lambda_{j,i} -Y_{i}  \lambda_{i,k}  \\
		&& + Y_m Y_n (\rho_b N_A) \langle \sigma v \rangle_{mn,i} -Y_i Y_o (\rho_b N_A) \langle \sigma v \rangle_{io,p} \nonumber \\
		&& + Y_q Y_r Y_s (\rho_b N_A)^2 \langle qrs \rangle_i \nonumber \\
		&& - Y_i Y_t Y_u (\rho_b N_A)^2 \langle itu \rangle_v \nonumber \\
		&& + Y_w Y_x Y_y Y_z (\rho_b N_A)^3 \langle wxyz \rangle_i \nonumber \\
		&& - Y_i Y_a Y_b Y_c (\rho_b N_A)^3 \langle iabc \rangle_d, \nonumber
		\label{network}
	\end{eqnarray}

where $\left\langle \sigma v \right\rangle_{ij,k}$ is the thermal average of the product of the cross section for $i+j \rightarrow k$ and relative velocity $v$. For the case including identical particles, to avoid the double-counting, the coefficients of 1/2! (for an identical pair among three species) or 1/3! (for all identical particles) is multiplied. $\langle qrs \rangle_i$ (cm$^6$ s$^{-1}$) is the three-body reaction rate, and $\langle wxyz \rangle_i$ (cm$^9$ s$^{-1}$) is the four-body reaction rate.

The yields of the explosive nucleosynthesis are determined by this equation, and therefore it is important what we adopt for the reaction rates. In the present works, in addition to the JINA REACLIB database, we adopt modified neutron capture and neutrino-induced reaction rates (See  Appendix \ref{sec:app_c}).

\subsubsection{Neutron capture $(n,\gamma)$ reaction rate }
Neutron capture reactions are important to produce heavy elements in neutron-rich environments via the $r$- and $s$-processes \citep{1957RvMP...29..547B}. Around $A \sim 100$, we utilize the neutron capture reaction rates calculated by the Monte Carlo method for the particle emission from the compound nucleus based on the Hauser-Feshbach statistical model \citep{2010EPJWC...209001K, 1999NuPhA.656....3A}. The comparison between the adopted reaction rates and those in the JINA database is shown in Figure \ref{fig_ng}. For those nuclei, in the present calculation, the partition functions for the excited states are taken into account. The $(n,\gamma)$ reactions and their reverse reactions are displayed in Figure \ref{fig_ng} and \ref{fig_gn}, respectively, at Appendix \ref{sec:app_c}. In particular, the increased temperature and density by the shock propagation effect are relevant to the photodisintegration reactions in the region inside $M_r \sim 2 M_\odot$.

\subsubsection{Neutrino-induced reaction rates}
The weak interaction of neutrinos is feeble compared to other interactions. However, in the CCSN environment, the neutrino-induced reactions are considerable because of the high neutrino fluxes and crucially affect yields of the nucleosynthesis. The thermal averaged neutrino reaction rate for $\nu_\alpha$ is given as
	\begin{eqnarray}
		&&\lambda_{\nu_\alpha} (t,r)
		= \frac{1}{4 \pi r^2} \sum_{\beta=e,\mu,\tau}
		\left[
			\frac{L_{\nu_\beta}(t)}{F_3(0) T_{\nu_\beta}^4} \right. \nonumber \\
 		&&	\times \left.
			\int \frac{E_\nu^2 \sigma_{\nu_\alpha} (E_\nu) \langle \rho_{\alpha \alpha}(r_c;E_{\nu}) \rangle P_{\nu_\beta \nu_\alpha}(r;E_\nu)}
			{\exp(E_\nu/T_{\nu_\beta})+1} dE_\nu
		\right],~~~
		\label{eq_lambda_nu}
	\end{eqnarray}
where $L_{\nu_\beta}$ is the luminosity and $T_{\nu_\beta}$ is the temperature of $\nu_\beta$. The cutoff radius $r_c$ is set as 1000 km or 2000 km, where the $\nu$-SI is no longer effective. The cross section $\sigma_{\nu_\alpha}$ between a nucleus and a neutrino depends on the nuclear structure model and is an important input to determine the neutrino-induced reaction rate.

In order to obtain the cross sections between nucleus and neutrinos, there are two kinds of approaches-- the nuclear shell model (SM) and quasi-particle random phase approximation (QRPA). Compared to the SM, the QRPA is advantageous because the Bardeen-Cooper-Schrieffer theory can be used for the nuclear ground state and makes it possible to perform efficient calculations of the nuclear excitations of the medium and heavy nuclei. Despite the efficiency, the results between SM and QRPA do not show any significant difference within the error bar \citep{2008ApJ...686..448Y, 2013JPhG...40h3101S, 2010PhRvC..81b8501C}.
For $^4$He and $^{12}$C, we take the cross sections by the SM from \citet{2008ApJ...686..448Y}; for nuclei of $^{13}$C to $^{80}$Kr from \url{http://dbserv.pnpi.spb.ru/elbib/tablisot/toi98/www/astro/hw92_1.htm}; for the heavy elements of $^{92}$Nb, $^{98}$Tc, $^{138}$La, and $^{180}$Ta by the QRPA from \citet{2010PhRvC..82c5504C, 2012PhRvC..85f5807C}.

\clearpage

\section{Nuclear Abundances by $\nu$-Process} \label{sec:abun}
\begin{figure}[h!]
		\centering
		\includegraphics[width=8.4cm]{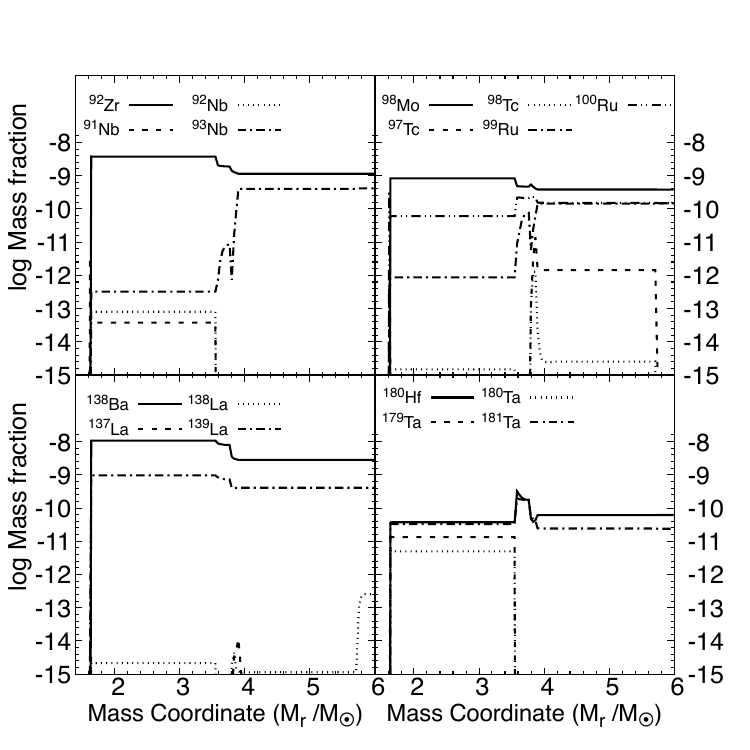}
		\caption{The logarithmic nuclear mass fractions in the pre-SN model used in the present work \citet{2015PTEP.2015f3E01K}.}
		\label{preSN}
	\end{figure}

In this section, with the nuclear reaction rates described in the previous section, we analyze the elements produced prominently by neutrino-induced reactions during CCSN explosion. The main factors of the CCSN $\nu$-process are pre-SN seed abundances, the number densities of electrons and neutrinos, and shock propagation during the explosion. As seed elements for the CCSN $\nu$-process, we adopt the pre-SN model including modified neutron capture rates for $A \sim 100$ nulcei \citep{2010EPJWC...209001K}, which are shown in  Figure \ref{preSN}. Based on the pre-SN model, we discuss the production mechanism for the SN $\nu$-process elements in three different aspects:  effects of hydrodynamics, shock propagation, and neutrino collective oscillation by the $\nu$-SI. Since the neutrino oscillation behavior depends on the neutrino mass hierarchy, we also investigate each model with the two kinds of neutrino mass hierarchy. Consequently, twelve models are studied for the yields of the $\nu$-process elements.
	
\subsection{Hydrodynamics dependence}
\begin{figure}[h]
	\centering
	\includegraphics[width=8.4cm]{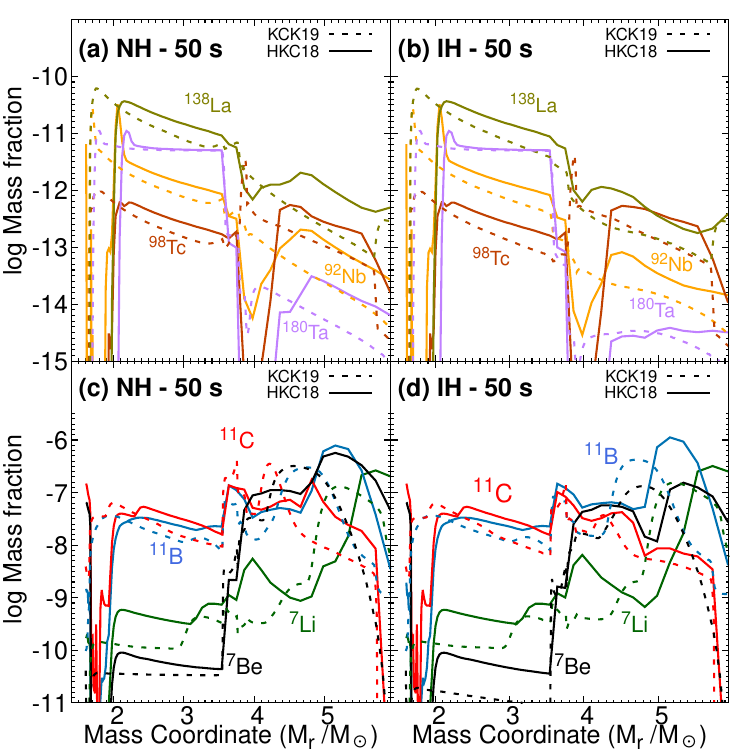}
		\caption{Logarithmic mass fractions of heavy nuclei: $^{92}$Nb, $^{98}$Tc, $^{138}$La and $^{180}$Ta versus the mass coordinate. Panels (a) and (b) correspond to the yields for the NH and IH, respectively. Panels (c) and (d) denote light nuclei : $^{7}$Li, $^{7}$Be, $^{11}$B and $^{11}$C. Dashed and Solid lines, respectively, stand for the results by the hydrodynamics of {\it HKC18} and {\it KCK19} models. All results are the nuclear abundances synthesized after about 50 s from the CCSN explosion.}
		\label{abun_hydro}
	\end{figure}

First, we discuss the effects of different hydrodynamics models on the CCSN $\nu$-process yields. For the hydrodynamics models, we take the two different models described in Section \ref{sec:nu_osc}, i.e., {\it HKC18} and {\it KCK19}. As explained in the previous section, the different temperature and baryon density profiles affect the thermonuclear reaction rates and neutrino oscillation behavior during SN explosion. Figure \ref{abun_hydro} shows the synthesized abundances of heavy ($^{92}$Nb, $^{98}$Tc, $^{138}$La, and $^{180}$Ta) and light nuclei ($^{7}$Li, $^{7}$Be, $^{11}$B, and $^{11}$C) by the two models.  One may confirm that the heavy (apart from $^{98}$Tc) and light nuclei are, respectively, mainly produced in O-Ne-Mg layer ($1.6 M_\odot \lesssim M_r \lesssim 3.7 M_\odot $) and He-layer ($3.9 M_\odot \lesssim M_r \lesssim 6 M_\odot $), irrespective of the adopted hydrodynamics.

\subsubsection{Heavy elements synthesis}
The main reactions producing the heavy elements are neutrino CC reactions. Neutrinos propagating from the SN core react with seed nuclei in the pre-SN model (See Figure \ref{preSN}). The main reactions producing $^{92}$Nb and $^{138}$La are $^{92}{\rm Zr}(\nu_e , e^-)^{92}{\rm Nb}$ and $^{138}{\rm Ba} (\nu_e, e^-) ^{138}{\rm La}$, respectively (See Figures 24 and 26 at Appendix D). Since the neutrino flux decreases with an increasing radius from the core, the rates of the main reactions related to those abundances decrease. As a result, the yields of $^{92}$Nb and $^{138}$La decrease from the flat abundance pattern of the adopted pre-SN model by the neutrino reactions as they go outward. The main production reactions of $^{180}$Ta are $^{179}$Ta$(n,\gamma)$ $^{180}$Ta and $^{180}$Hf$(\nu_e,e^-)$ $^{180}$Ta shown in Figure \ref{fig_rea_180ta}. However, the preexisting abundance of $^{180}$Ta in Pre-SN is larger than any productions from nearby seed nuclei during the SN explosion. So $\nu$-process has only a minor effect on $^{180}$Ta production in the present models.

On the other hand, the $^{98}$Tc nuclei are evenly produced in the whole region by reactions of $^{98}{\rm Mo}(\nu_e, e^-)^{98}$Tc (inner region; $M_r \lesssim  4M_\odot$) and $^{97}{\rm Tc}(n,\gamma)^{98}$Tc (outer region; $M_r \gtrsim 4M_\odot$) (See Figure 25). Also, unlike other heavy elements, the CC reactions induced by $\bar{\nu}_e$ contribute maximally about 12\% and 8\% in {\it HKC18} and {\it KCK19} models, respectively. The present results by {\it HKC18} model differs a bit from our previous study in \cite{2018PhRvL.121j2701H}. This is because the abundance of seed nucleus, $^{97}$Tc, is a bit larger in the current pre-SN (Figure \ref{preSN}).

In Figure \ref{abun_hydro}, a valley appears around the region of ${M}_r \sim 4 M_{\odot}$ in the {\it HKC18} model, but it disappears in the {\it KCK19} model. In this region, the neutron capture rates are important not only for production but also for destruction. The major production of $^{98}$Tc is from $^{97}$Tc$(n,\gamma)^{98}$Tc and the destruction comes from $^{98}$Tc$(n,\gamma)^{99}$Tc. The valley arises from the density discrepancy between the {\it HKC18} and {\it KCK19} models (Figure \ref{hydro_tem}). The density in the {\it HKC18} is one order of magnitude higher than the {\it KCK19} case until the shock passes. As a result, the $^{98}$Tc is more  destroyed by the reaction $^{98}$Tc$(n,\gamma)^{99}$Tc.
    	
Another discrepancy between the hydrodynamics appears at $M_r \sim 2 M_{\odot}$, where threshold positions of producing elements are different. This fact arises from the coincidence between hydrodynamics and pre-SN models. The pre-SN calculation gives the initial elemental abundances in each region. According to Figures \ref{hydro_den} and \ref{hydro_tem}, however, the initial density and temperature of {\it HKC18} are higher than those of the pre-SN model adopted in the {\it KCK19} model. Consequently, the higher density and temperature profiles in {\it HKC18} increase most of the reaction rates especially the photodisintegration rate, which forms the valley in this region.

The trend of $\nu$-process yields is similar in the both hydrodynamics models aside from the valley and photodisintegration region, and the total abundances of $^{92}$Nb, $^{98}$Tc, and $^{138}$La differs by less than about 13\% as shown at Table \ref{inte_mass}. In our previous study of \citet{ 2018PhRvL.121j2701H, 2020ApJ...891L..24K}, the hydrodynamics of {\it HKC18} was used. However, it was found that there was the discrepancy of the density profile between {\it HKC18} model and the adopted pre-SN, which spawns the discontinuities of physical quantities as a function of time. In order to be consistent with the pre-SN model \citep{2015PTEP.2015f3E01K}, in this paper, we exploit {\it KCK19} model for more systematic investigations because the two models are consistent with each other. At Table 4, we compare the results with the previous results by {\it HKC18} at the last two lines.

\subsubsection{Light elements synthesis}
Light elements, $^{7}$Li and $^7$Be, are synthesized mostly in the outer region, while $^{11}$B and $^{11}$C are produced in the whole region. In the inner region ($M_r < 4 M_\odot$), the light elements are produced by the neutrino reaction, and outside the region alpha capture reactions at He-layer mainly contribute to the production of light elements. The hydrodynamics models also affect the synthesis of light elements by changing the $\nu$-process environment.

First, the dependence on the hydrodynamics stems from the radius profiles (Figure \ref{hydro_radi}). As seen in Equation (\ref{eq_lambda_nu}), the neutrino reaction rate is inversely proportional to the squared radius. The discrepancy of the radius profiles, ${\rm R_{\it KCK19}} > {\rm R_{\it HKC18}}$, makes the neutrino reaction rates different. As a consequence, the results of {\it KCK19} show lower abundances than those of the {\it HKC18} model in the O-Ne-Mg layer, which is shown in Figure \ref{abun_hydro} (c) and (d).

Second, the different density profiles in He-layer also affect the nucleosynthesis. Within the same neutrino oscillation parameters, the discrepancy of density profiles causes the different MSW regions, which results in the change of $\nu$-induced reaction rates. The change of MSW region by the hydrodynamics models is closely related to the synthesis of the light elements {---}especially for $^7$Be and $^{11}$C \citep{2019ApJ...872..164K}.

Third, temperature profiles are also important to determine the alpha capture reactions as well as their destruction rates. Higher temperature implies higher reaction rates for both production and destruction. For example, for $^{11}$B, in the region of $4.5 < M_r/M_\odot < 5$, the temperature profile of the {\it HKC18} model is higher than the {\it KCK19} model, but the abundance of $^{11}$B does not follow the temperature pattern. This is because the higher temperature stimulates more the destruction of $^{11}$B. For a detailed analysis of light elements, we should investigate the main reactions for the elements on given hydrodynamics condition, which is discussed in the subsection \ref{sec:5.2.2}.

\subsection{Shock propagation effect}
As delineated in Section \ref{sec:nu_osc}, the matter potential related to the MSW resonance is determined by the electron density. During the explosion, the matter density is changed by the shock propagation, and subsequently the neutrino oscillation probability varies (Figure \ref{shock_prob}) and affects the neutrino reaction rates. We refer to this effect here as `shock effect'.
	
The $\nu$-process elements are sensitive to the neutrino flavor distribution and luminosity. For example, for $E_\nu=15$ MeV, the multiple resonances occur at $t\sim 3 s$  in Figure \ref{shock_prob} \citep{2019AcPPB..50..385K}. At $t =3 s$, the neutrino luminosity is about 0.37 times smaller than the initial luminosity (Equation (\ref{lum_yoshi})). Then there may happen a competition between the luminosity and flavor change probabilities in the neutrino reaction rate. The neutrino luminosity decreases over time, resulting in a decrease in the reaction rates. However, a decrease in a CC reaction rate can be partially compensated by the effect of the flavor change to $\nu_e$ caused by the shock propagation. Hereafter we adopt the KCK19 hydrodynamics for the forthcoming results.

\subsubsection{Heavy elements synthesis}
Since the heavy elements are produced mainly by the CC reactions, it is important to estimate the quantity of $\nu_e$ as a function of the mass coordinate. The multiple  resonances caused by the shock promote the flavor change from $\nu_\mu$ and $\nu_\tau$ to $\nu_e$. This change of neutrino flavor transition region is more significant for the NH case rather than the IH case.
	
Figure \ref{abun_shock} (a) shows the production of heavy elements for the NH case. The resonance in the inner region increases the CC reaction rates. Therefore, in the region $M_r \sim 1.6$ - $3.9 M_\odot$, the abundances of the heavy elements---except for $^{180}$Ta---increase by the shock propagation. However, by the second resonances at $M_r \sim 3.7 M_\odot$, the distribution of $\nu_e$ returns to its initial distribution. Consequently, the shock effect enhances the production of heavy elements only within $M_r \sim 1.6$ - $3.9 M_\odot$. Precisely, the shock propagation enhances the abundances of $^{92}$Nb, $^{98}$Tc, and $^{138}$La by about 9\%, 8\%, and 11\%, respectively. For the IH case, the shock effect has relatively less impact on the yields, that is, abundances are increased by maximally about 1 \% as shown in Figure \ref{abun_shock} (b).

\subsubsection{Light elements synthesis} \label{sec:5.2.2}
\begin{figure}[h!]
		\centering
		\includegraphics[width=8.4cm]{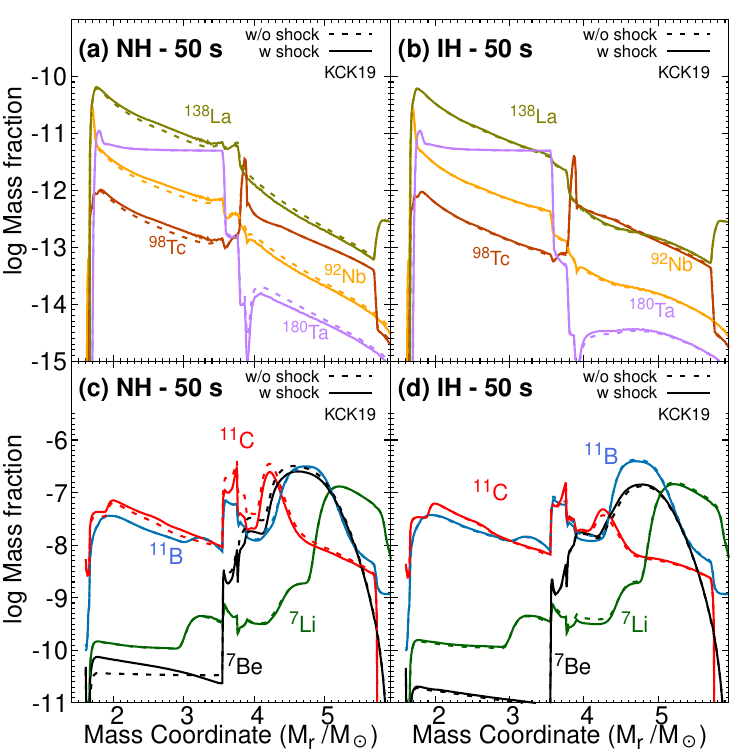}
		\caption{Logarithmic mass fractions of heavy and light elements at 50 s after core bounce over the mass coordinate for the shock effect. Panels (a) and (b) show the heavy elements, and panels (c) and (d) are for the light elements. In each panel, solid and dashed lines denote the case with and without shock effect, respectively.}
		\label{abun_shock}
	\end{figure}

The production mechanism of the light elements has been studied by \citet{2019ApJ...872..164K}. We briefly review the main production reactions of light elements and investigate the shock effect. Table \ref{tab_main_reactions} and Figures 20 - 23 at Appendix \ref{sec:app_d} show the relevant nuclear reactions. In Figure \ref{abun_shock} (c) and (d), around the region of $M_r \sim 1.6$ - $3.9 M_\odot$, the NC reaction in the $^{12}$C $+\nu$ reactions mainly produces $^{11}$B and $^{11}$C. Since all flavors contribute to the NC reaction, the neutrino oscillation probability does not affect the reaction rate. Therefore, in this region, the production of $^{11}$B and $^{11}$C is independent of the shock propagation and mass hierarchy. The various channels in the $^{12}$C $+\nu$ reactions are explained in detail at Table \ref{reaction_eq_light}.

On the other hand, in the region of $M_r \sim 3.9$ - $6 M_\odot$, most of $^{11}$B and $^{11}$C are produced by alpha capture reactions of $^{7}$Li and $^{7}$Be, respectively. The dominant production processes for $^{7}$Li are $^4$He$(\nu,\nu'p)^3$H$(\alpha,\gamma)^{7}$Li and $^4$He$(\bar{\nu}_e,e^+n)^3$H$(\alpha,\gamma)^{7}$Li, and $^{7}$Be is mainly produced by $^4$He$(\nu,\nu'n)^3$He$(\alpha,\gamma)^{7}$Be and $^4$He$(\nu_e,e^-p)^3$He$(\alpha,\gamma)^{7}$Be. The contribution of each reaction is succinctly presented at Figures \ref{fig_rea_7li} and \ref{fig_rea_7be} at Appendix \ref{sec:app_d}.

In the NH case, as shown in Figure \ref{shock_prob}, the $\bar{\nu}_e$ distribution with the shock propagation is similar to that without considering the shock. Therefore, $^{7}$Li and $^{11}$B abundances are also not affected by the shock effect. The increments of $^{7}$Li and $^{11}$B are 0.04\% and 2.7\%, respectively. On the other hand, $^{7}$Be and $^{11}$C abundances are more sensitive to the shock effect, and those abundances decrease by 22\% and 21\%, respectively. In the case of IH, because the distribution of $\nu_e$ is rarely varied, the shock effect is relatively small and the change tendency is opposite. The decreases of $^{7}$Li and $^{11}$B are 6.7\% and 5.7\%, and the increases $^{7}$Be and $^{11}$C are 9.8\% and 6.1\%, respectively.

\subsection{Collective neutrino oscillation effect}
As described in Section \ref{sec:nu_SI}, the $\nu-$SI affects the neutrino reaction rates through the change of neutrino spectra. For the $\nu-$SI, we investigate the $\nu$-process with two different neutrino luminosity models---equivalent and non-equivalent luminosity models---explained in Section \ref{sec:nu_SI}. Note that we do not consider the shock effect simultaneously in this section to understand only the $\nu-$SI effect. The shock effects are less than 11{\%} for the heavy elements and 22{\%} for the light elements as discussed in the last section.

\subsubsection{Equivalent (EQ) Luminosity}
\begin{figure}[h!]
	\centering
		\includegraphics[width=8.4cm]{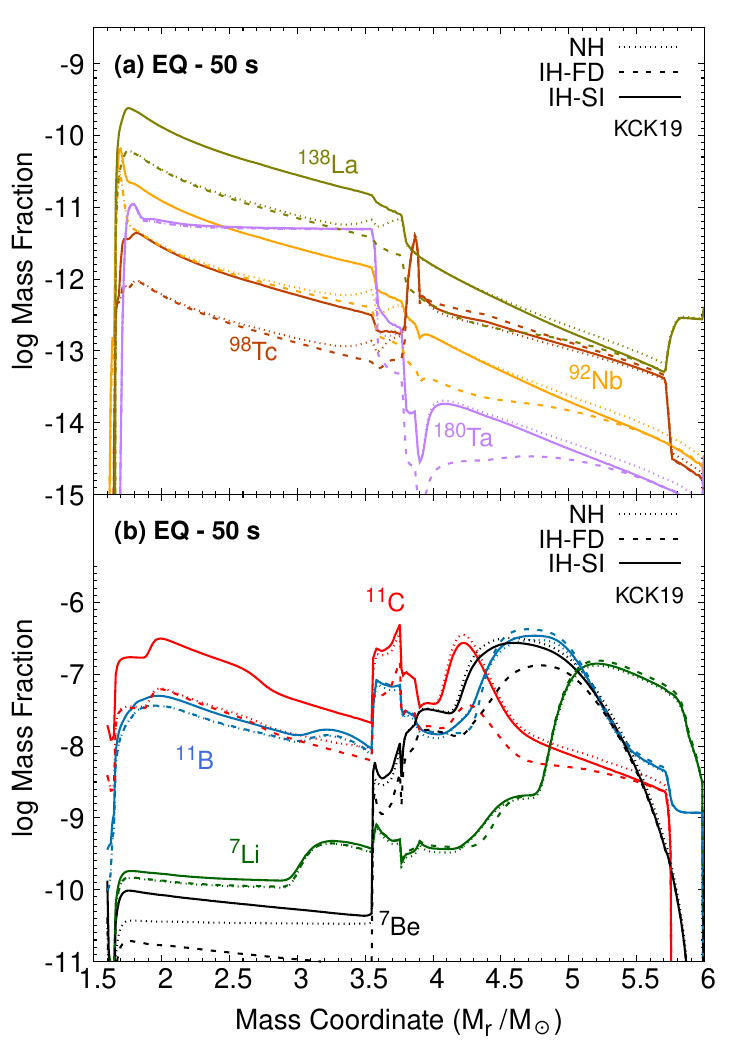}
		\caption{Logarithm of mass fractions of heavy (a) and light (b) elements over the mass coordinate.
		Solid and dashed lines denote the case with and without $\nu-$SI, respectively.
		Dotted line stands for the NH case. All results are shown after CCSN explosion 50 s.
		The results for the NH are given without the SI effect.}
		\label{abun_EQ}
	\end{figure}

In the equivalent luminosity model, each neutrino luminosity exponentially decreases with time owing to a flavor-dependent temperature. As shown in Figure \ref{EQ_SI}, $\nu_e$ and $\bar{\nu}_e$ have high energy tails in their distributions which can increase CC reaction rates. As a result, Figure \ref{abun_EQ} (a) shows that the $\nu-$SI effects increase the synthesis of the heavy elements for both NH and IH cases. Since the $\nu-$SI effect in neutrino spectra is not so significant for the NH case, we only show the results without the $\nu-$SI. The SI increases the abundances of $^{92}$Nb, $^{98}$Tc, and $^{138}$La by factors of 3.6, 2.7, and 4, respectively. The $^{180}$Ta abundance is rarely changed due to its high initial abundance (See Figure \ref{preSN}). We also find that there is only a small difference between the results for the NH and IH scheme by the MSW effect on the outer region.

Figure \ref{abun_EQ} (b) shows three cases of light element abundances. For the cases of the NH and the IH with FD distribution (IH-FD), the key NC reactions in the inner region are the same as those explained in Section \ref{sec:5.2.2}. In the case of IH including the $\nu-$SI, the main reactions to produce the light elements are changed to CC reaction.

The $\nu-$SI effects on light elements are explained as follows. Before undergoing the MSW resonance, $2 M_\odot \lesssim M_r \lesssim 4 M_\odot$, production of all light elements is increased by the $\nu-$SI, whose main reaction is changed from NC to CC reactions for $^{12}$C as tabulated at Table \ref{reaction_eq_light}. As the spectra of ${\nu_e}$ increase at higher energy tails, the CC reactions at Table \ref{reaction_eq_light} become significant. In particular, the $^7$Be and $^{11}$C abundances are increased 2.1 and 4.3 times, respectively, in this region for the IH case. The abundances of $^7$Li and $^{11}$B are increased by 7\% and 15\%, which are small compared to $^7$Be and $^{11}$C.
	
	\begin{table}[h!]
		\centering	
		\caption{Main production channels for $^{11}$C and ${^7}$Be in $ 2 M_\odot \lesssim M_r \lesssim  4 M_\odot$.}
	    \begin{tabular}{c |c}
    	\hline
     	Elements & Main reactions with $\nu-$SI~ \\
        \hline 	
        $^{11}$C & $^{12}$C$(\nu_e,e^-)^{12}$N$(\gamma,p)^{11}$C\\
                 & $^{12}$C$(\nu_e,e^-)^{12}$N$^*$, N$^* \rightarrow ^{11}$C + p\\
        $^{7}$Be & $^{12}$C$(\nu_e, \nu_e a)^{7}$Be ; $a = (\alpha + n, 3p + 2n, ..)$  \\
                 & $^{12}$C$(\nu_e, e^-)^{12}$N,$^{12}$N $\rightarrow ^8$B + $a$ ; $a = (\alpha, 2p + 2n, ...)$ \\
    	\hline
        \end{tabular}
		\label{reaction_eq_light}
	\end{table}

\subsubsection{Non-equivalent (NEQ) Luminosity} \label{sec:5.3.2}
\begin{figure}[h!]
		\centering
        \includegraphics[width=8.4cm]{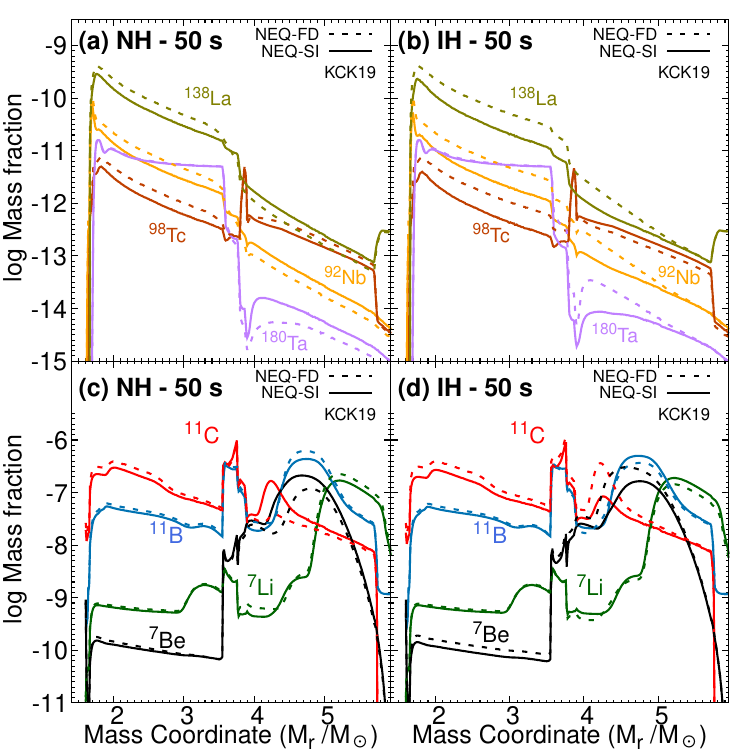}
		\caption{Panels (a) and (b) show the mass fractions of heavy elements as a function of the mass coordinate at 50 s for the NH and IH, respectively.
		         Panels (c) and (d) denote the mass fraction of light elements.  }
		\label{abun_NEQ}
    \end{figure}

Hereafter, we investigate the $\nu-$SI effects with the non-equivalent luminosity model described at Table \ref{Lumino} and Figure \ref{fig_lumino} with {\it KCK19} hydrodynamics. In this model, the luminosities of $\nu_e$ and $\bar{\nu}_e$ are higher than those by the equivalent model. The higher luminosity activates related neutrino reactions. Hence, the abundances of both heavy and light elements, which are shown in Figure \ref{abun_NEQ}, are larger than those in the equivalent model. Note that the index of 'NEQ-FD' means the 'non-equivalent luminosity with Fermi-Dirac distribution'.
 	
On the other hand, the abundances of heavy elements are reduced by the $\nu-$SI regardless of mass hierarchy, which is opposite to the trend in the equivalent luminosity model. The heavy element synthesis is mostly affected by the cooling phase (exponential decay time) due to its longer duration of neutrino emission than other explosion phases. The spectral change by the $\nu-$SI at $t \gtrsim$ 0.5 s from $\nu_e$ to $\nu_\mu$ and $\nu_\tau$ (and vice versa) decreases the $\nu_e$ number distribution, which is shown at the lowermost panels in Figures \ref{spectra_NH} and \ref{spectra_IH}, and reduce the production of the heavy elements.

Here we note the difference in abundance according to MH. Unlike the equivalent model, the NH case undergoes the spectral change (Figure \ref{spectra_NH}), by which the number of $\nu_e$ is more than those of other flavors in the beginning at $r = 10$ km. However, after $\nu-$SI interaction, $\nu_e$ turns into $\nu_\mu$ and $\nu_\tau$, while the inverse flavor change does not sufficiently compensate the initial $\nu_e$ number. Consequently, the production of heavy elements decrease by the reduction of CC reactions rates.

In the IH case, the tendency of the spectra is similar to that of the equivalent luminosity case in Figure \ref{spectra_IH}. Despite these similarities, the flavor change cannot increase the neutrino CC reaction rate because $\nu_\mu$ and $\nu_\tau$ have lower initial luminosities than $\nu_e$ at $t \gtrsim 200$ ms. Considering the spectra above $E_\nu \sim $10 MeV, the NH case has larger $\nu_e$ number distribution than the IH case. As a result, the heavy element abundance in the NH case is larger than in the IH case by the $\nu-$SI.

Another interesting aspect of heavy element synthesis is the competition of the $\nu-$SI and the outer region MSW effect. The elements decrease in both NH and IH cases inside the MSW region. On the other hand, outside the MSW region, the elements increase by $\nu-$SI in the NH cases because the  $\nu_e$ number distribution is higher in the SI cases than the FD case due to the exchange of the $\nu_e$ and $\nu_x$ spectra. However, in the IH case, the spectra are not fully exchanged. As shown in Figure \ref{MSW} (d) for $E_\nu$=15 MeV, the electron neutrino spectrum is a mixture of about 30\% $\nu_e$ and about 70\% $\nu_x$ after the MSW resonance. Consequently, the $\nu_e$ spectrum in the FD case has a larger number distribution than that in the SI case.

For the light elements, the production of $^{11}$B and $^7$Li in the inner region, where NC reactions predominantly contribute, is less affected by the change of spectra. But for $^{11}$C and $^{7}$Be synthesis, both CC and NC reactions are important. As shown in Figures \ref{abun_NEQ} (c) and (d), when the non-equivalent luminosity model is adopted, the outermost peak of $^{11}$C is affected although it is subdominant in the both NH and IH case. Since $^{11}$C decays into $^{11}$B, $^{11}$B can be produced over the whole region.
	
In the region $M_r \gtrsim 4 M_\odot$ (Figure \ref{abun_NEQ}), there is a novel feature in which the trend of solid and dotted lines is opposite to each other. Namely, the MH dependence clearly appears. For the NH case in Figure \ref{abun_NEQ} (c), we find the increase of $^{11}$C and $^{7}$Be abundances and the decrease of $^{11}$B and $^{7}$Li abundances by $\nu-$SI. In this region, the neutrinos pass the MSW resonance region and the $\nu_e$ spectra for NEQ-FD and NEQ-SI cases follow the dashed and solid lines of $\nu_\mu$ and $\nu_\tau$ spectra in Figure \ref{spectra_NH}, respectively. As a result, $\nu_e$ CC reaction rate is greater for NEQ-SI than that for NEQ-FD. The flavor change of ${\bar \nu}_e$ partially occurs by the MSW effect.

On the other hand, for the IH case, the $\bar{\nu}_e$ is fully converted to the $\bar{\nu}_x$ by considering $\nu-$SI, and the subsequent MSW resonance converts the $\bar{\nu}_x$ back to $\bar{\nu}_e$. Therefore, the initial $\bar{\nu}_e$ spectrum recovers in the outer region. Due to the abundant $\bar{\nu}_e$, the abundances of $^{7}$Li and $^{11}$B increase compared to NEQ-FD shown in Figure \ref{abun_NEQ} (d). We note that the closer to the initial $\bar{\nu}_e$ spectrum, the more abundances are produced.

Finally, we summarize the whole yields as the masses integrated from $M_r \sim$ 1.6$M_\odot$ to $\sim$ 6.0$M_\odot$ at 50 s at Table \ref{inte_mass}. First, we discuss the dependence on the mass hierarchy. For the light elements by {\it HKC18} and {\it KCK19}, the MSW effects appear explicitly, regardless of the hydrodynamics model, if we note the increase (decrease) of $^{7}$Li and $^{11}$B ($^{7}$Be and $^{11}$C) abundances in the IH case compared to those in the NH. But the heavy elements are less dependent on the MH. As already commented, the results by {\it HKC18} are larger than those by {\it KCK19}.

Second, the shock effect increases $^7$Be and $^{11}$C in the NH case. It may indicate that $\nu_e$ spectra are much more affected  by the shock propagation rather than ${\bar \nu}_e$ spectra.

Third, the SI effects show interesting features depending on the MH and the neutrino luminosity. For example, the results in the NEQ-FD case in the last row of Table \ref{inte_mass} indicate the decrease (increase) of $^{7}$Li and $^{11}$B ($^{7}$Be and $^{11}$C) in the IH case compared to those for the NH case, which is contrary to those in the {\it KCK19} case with the EQ luminosity. In brief, the neutrino luminosity plays a vital role in the $\nu$-process. However, if we include the $\nu-$SI, the change of the light elements in the NEQ-SI case compensates the differences, and final abundances resemble the trend by {\it KCK19}. Therefore, nuclear abundances depend on not only the MSW effect, but also on the $\nu-$SI effect sensitively.

\subsection{Yield ratio of $^{7}$Li to $^{11}$B in each model}
Here we report abundance ratios of some nuclei produced in the $\nu$-process tabulated at the rightmost two columns at Table \ref{inte_mass}. The ratio might cancel the aforementioned model dependences in the respective nucleus yield.

Moreover, meteorite analyses data for the ratios provide important information about the $\nu$-process abundances. In the following, we compare our results for the ratio with the observational data from the Bayesian analysis of Silicon Carbide X (SiC X) grains. Here we presumed that elements in SiC X grains have been uniformly mixed before condensing in the SN ejecta long before solar system formation. The analysis of SiC X grains constrains the ratio of $^{7}$Li/$^{11}$B produced by the $\nu$-process as $ - 0.31 \pm 0.42$ and the upper limit is given by 0.53  under $2\sigma$ error bar \citep{2012PhRvD..85j5023M}. Based on the above assumption, we integrate the yields of $^{7}$Li and $^{11}$B over the whole mass region after the decay of unstable nuclei and present the yield ratio of $^{7}$Li to $^{11}$B, which is calculated as $N({\rm Li})/N({\rm B}) = (M_{^7{\rm Li}}/7)/(M_{^{11}{\rm B}}/11)$.

A previous study suggested that the abundance ratio $^{7}$Li/$^{11}$B is sensitive to the MH \citep{2006PhRvL..96i1101Y}. In our previous paper \citep{2020ApJ...891L..24K}, the $^{7}$Li/$^{11}$B ratio was changed by the $\nu-$SI effect from 0.671 to 0.413 in the IH case, and from 0.343 to 0.507 in the NH case. The $^{7}$Li/$^{11}$B ratio in the NH case was larger than that in IH by about 23\%, which are shown in the last two rows at Table \ref{inte_mass}. However, when the {\it KCK19} hydrodynamic model is adopted, the ratios are different from those of the {\it HKC18} model. That is, the ratio is increased (decreased) in the NH (IH) case (See the results by SI NEQ and FD NEQ with {\it KCK19} at Table \ref{inte_mass}.). We note that the results by the {\it HKC18} hydrodynamics in the last two rows \citep{2020ApJ...891L..24K} used an old pre-SN model \citep{2000ApJ...532.1132B} while the others in the present work exploit the updated pre-SN model \citep{2015PTEP.2015f3E01K} with {\it KCK19} explained in Figure \ref{preSN}.

On the other hand, when we adopt the EQ luminosity, the SI effect decreases the ratio from 0.771 to 0.695 in the IH case, while there is no change in NH case. Therefore, the $\nu-$SI turns out to depend on the luminosity and plays a vital role in understanding the light nuclear abundances. We note that the results by using EQ luminosity are not favorable for the meteorite data within the 2$\sigma$ range \citep{2012PhRvD..85j5023M}.

\subsection{Production Factor ratio of $^{138}$La to $^{11}$B for each model} \label{sec:5.5}
Here we discuss the production factor (PF) ratio of $^{138}$La to $^{11}$B \citep{Heger02}, i.e., PF($^{138}$La) to PF($^{11}$B), where PF[A] = $X_A/X_{A\odot}$ with $X_A$ and $X_{A\odot}$ defined as the mass fractions of A in the SN ejecta and in the Solar system, respectively. The previous study on the $\nu$-process without considering both the $\nu-$SI and the MSW effects \citep{2005PhLB..606..258H} concluded that enough $^{138}$La is produced by the $\nu$-process, while the $^{11}$B is overproduced. Our previous study showed that the $^{138}$La abundance was decreased by a factor of about 2, whereas the $^{11}$B abundance was nearly unaffected by the $\nu-$SI \citep{2020ApJ...891L..24K}. By using {\it KCK19} hydrodynamics, this trend is still apparent for $^{11}$B and $^{138}$La. We present the PF ratio for each case at the last column at Table \ref{inte_mass}. These results used the normalization to $^{16}$O. (PF ratios normalized to other nuclei are also presented at Tables \ref{PF_16O}, \ref{PF_24Μg}, and \ref{PF_28Si} at Appendix \ref{sec:app_e}.)

In particular, for the two different hydrodynamics models with NEQ-SI, the ratio of PF is approximately 0.2671 (NH) and 0.1899 (IH) for {\it HCK18}; 0.4776 (NH) and 0.3672 (IH) for {\it KCK19}; The results indicate that the ratio for the NH case is larger than for the IH case by a factor of about 1.4 and 1.3 for respective models. This difference comes from the fact that $^{138}$La is predominantly produced by $\nu_e$ but $^{11}$B production is less sensitive to the $\nu-$SI as discussed above.

The ratio can be compared with the expected value R$_{ex} = f_{metal} f_{La}^{SN} / f_{B}^{SN} $. Here $f_{metal}$ is the metallicity used in this work which roughly scales as the abundance of $^{138}$Ba, the seed of $^{138}$La in the $\nu$-process. The quantity $f_{La}^{SN} \sim 1$ is the fraction of the solar system abundance of $^{138}$La originating from the SNe, while $f_{B}^{SN}$ is the fraction of $^{11}$B originating from SNe which is deduced to be $0.41_{-0.42}^{+0.21}$ by the observed isotopic ratio of $^{11}$B/$^{10}$B =$(0.7 \pm 0.1)/(0.3 \pm 0.1)$ for cosmic ray yields \citep{1990PhR...191..351S}; $^{11}$B/$^{10}$B =3.98 for the solar abundance ratio \citep{liu2010}. From these values, the ratio is R$_{ex}$ = 0.41 - ${\infty}$.
Our theoretical values are 0.2671$-$0.4776 for the NH and 0.1899$-$0.3672 for the IH case, where the former values (0.2671 and 0.1899) come from {\it HKC18} and the latter from {\it KCK19}. Consequently, the $^{138}$La/$^{11}$B ratio is more consistent with the NH case within 1 $\sigma$.

This trend originates from the fact that the abundance change by the $\nu-$SI in the IH case is larger than that in the NH case. After the $\nu-$SI, the $\nu_e$-flux in the NH case is higher than that in the IH case by a factor of 2$\--$3 (see Figure \ref{spectra_IH} and \ref{spectra_NH}) in the energy range appropriate to $^{138}$La production (10$\--$20 MeV). As discussed previously, $^{138}$La production depends strongly on the $\nu_e$-flux. Thus, even if the initial neutrino energy spectra are changed from that assumed here, the trend is expected to be preserved, so that the PF($^{138}$La)/PF($^{11}$B) ratio including the $\nu-$SI effect in the NH case is higher than that in the IH case. It implies that the NH scheme is favorable to explain the empirical ratio. For comparison, we note that the results of the EQ luminosity case are 0.2354 (NH) and 0.5838 (IH).
	\begin{table*}[h]
		\centering
        \caption{Integrated masses of the nuclei after 50 s in the mass range, $M_r =1.6$--6 ($M_\odot$). We used two hydrodynamics models (HKC18 and KCK19), two luminosity models (EQ and NEQ) and two cases without the $\nu-$SI (FD) and with the $\nu-$SI (SI) for the NH and IH case, by which the results for twelve different cases are tabulated. The last two results are quoted from our previous results. See texts for the details.}        
        \begin{tabular}{r|ccccc|cccc||cc}
        \hline
        \hline
        \cline{2-12}
        \multicolumn{1}{c|}{} &
        \multicolumn{1}{c|}{Mass} & $^{7}$Li &   $^{7}$Be  &  $^{11}$B   &   $^{11}$C
                                                  &  $^{92}$Nb  &  $^{98}$Tc  & $^{138}$La  & $^{180}$Ta
        						                     & {Yield ratio} & {PF ratio} \\
        \cline{3-12}
        \multicolumn{1}{c|}{} &
        \multicolumn{1}{c|}{Hierarchy} &
        \multicolumn{4}{c|}{($10^{-7}~M_\odot$)}   &
        \multicolumn{2}{c|}{($10^{-12}~M_\odot$)}  &
        \multicolumn{2}{c||}{($10^{-11}~M_\odot$)} &
        \multicolumn{1}{c} {N($^{7}$Li)/N($^{11}$B)} & {$^{138}$La/$^{11}$B} \\
        \hline
        \multicolumn{1}{c|}{FD EQ} &
        \multicolumn{1}{c|}{NH} &
			{1.256} & {4.953} & {5.576} & {2.048} & {4.903} & {1.048} & {3.395} & {0.845} & {1.280} & {0.1288}\\
        \multicolumn{1}{c|}{(HKC18)} &
        \multicolumn{1}{c|}{IH} &
			{1.496} & {1.461} & {7.141} & {1.218} & {4.760} & {1.112} & {3.267} & {0.843} & {0.556} & {0.1130}\\
        \hline
        \multicolumn{1}{c|}{FD EQ} &
        \multicolumn{1}{c|}{NH} &
			{0.861} & {2.428} & {2.480} & {2.139} & {4.551} & {1.180} & {3.760} & {1.016} & {1.119} & {0.2354}\\
        \multicolumn{1}{c|}{(KCK19)} &
        \multicolumn{1}{c|}{IH} &
			{1.017} & {0.936} & {3.099} & {0.883} & {4.226} & {1.218} & {3.436} & {1.012} & {0.771} & {0.2495}\\
        \hline
        \multicolumn{1}{c|}{FD EQ Shock} &
        \multicolumn{1}{c|}{NH} &
			{0.861} & {1.904} & {2.546} & {1.701} & {4.973} & {1.271} & {4.164} & {1.017} & {1.023} & {0.2835}\\
        \multicolumn{1}{c|}{(KCK19)} &
        \multicolumn{1}{c|}{IH} &
			{0.949} & {1.027} & {2.922} & {0.937} & {4.271} & {1.215} & {3.485} & {1.012} & {0.805} & {0.2611}\\
        \hline
        \multicolumn{1}{c|}{SI EQ\footnote{Same as FD EQ (KCK19) NH result}}  &
        \multicolumn{1}{c|}{NH} &
			{0.861} & {2.428} & {2.480} & {2.139} & {4.551} & {1.180} & {3.760} & {1.016} & {1.119} & {0.2354}\\
        \multicolumn{1}{c|}{(KCK19)} &
        \multicolumn{1}{c|}{IH} &
			{0.920} & {2.057} & {2.852} & {3.874} & {15.07} & {3.259} & {13.58} & {1.052} & {0.695} & {0.5838}\\
        \hline
        \hline
        \multicolumn{1}{c|}{SI NEQ} &
        \multicolumn{1}{c|}{NH} &
			{1.132} & {1.601} & {4.276} & {4.920} & {16.44} & {3.559} & {15.19} & {1.295} & {0.467} & {0.4776}\\
        \multicolumn{1}{c|}{(KCK19)} &
        \multicolumn{1}{c|}{IH} &
			{1.261} & {1.206} & {4.623} & {4.283} & {12.29} & {2.854} & {11.31} & {1.281} & {0.435} & {0.3672}\\
        \hline
        \multicolumn{1}{c|}{FD NEQ} &
        \multicolumn{1}{c|}{NH} &
			{1.483} & {0.841} & {5.407} & {5.258} & {25.44} & {5.367} & {23.14} & {1.323} & {0.342} & {0.6274}\\
        \multicolumn{1}{c|}{(KCK19)} &
        \multicolumn{1}{c|}{IH} &
			{0.959} & {2.303} & {3.946} & {6.566} & {26.15} & {5.302} & {23.94} & {1.331} & {0.488} & {0.6585}\\
        \hline
        \hline
        \multicolumn{1}{c|}{SI NEQ \cite{2020ApJ...891L..24K}} &
        \multicolumn{1}{c|}{NH} &
			{1.643} & {3.347} & {9.332} & {6.138} & {17.92} & {3.511} & {14.29} & {1.363} & {0.507} & {0.2671} \\
        \multicolumn{1}{c|}{(HKC18)} &
        \multicolumn{1}{c|}{IH} &
			{1.792} & {2.372} & {10.33} & {5.524} & {13.59} & {2.720} & {10.41} & {1.358} & {0.413} & {0.1899} \\
        \hline
        \multicolumn{1}{c|}{FD NEQ \cite{2020ApJ...891L..24K}} &
        \multicolumn{1}{c|}{NH} &
			{2.400} & {1.860} & {12.46} & {7.080} & {27.56} & {5.361} & {22.62} & {1.349} & {0.343} & {0.335} \\
        \multicolumn{1}{c|}{(HKC18)} &
        \multicolumn{1}{c|}{IH} &
			{1.640} & {5.270} & {8.382} & {7.804} & {27.83} & {5.318} & {22.94} & {1.353} & {0.671} & {0.410} \\
        \hline
        \hline
        \end{tabular} \label{inte_mass}
    \end{table*}

\begin{table*}[h]
		\centering
        \caption{{Integrated masses for FDSI NEQ (KCK19) case, which is the case of the complete matter suppression of the collective neutrino oscillation during the accretion phase ($t_{pb} < 0.5 s$) represented by FD distribution and the collective oscillation during the cooling phase ($t_{pb} > 0.5 s$), whose neutrino distribution is changed by the SI. Difference is the change of the element abundances by the FDSI NEQ case compared to those by the SI NEQ (KCK19) case in Table 4. The others are the same as Table 4. See texts for the details.}}
        \begin{tabular}{r|ccccc|ccc}
        \hline
        \hline
        \cline{2-9}
        \multicolumn{1}{c|}{} &
        \multicolumn{1}{c|}{Mass} & $^{7}$Li &   $^{7}$Be  &  $^{11}$B   &   $^{11}$C
                                                  &  $^{92}$Nb  &  $^{98}$Tc  & $^{138}$La  \\
        \cline{3-9}
        \multicolumn{1}{c|}{} &
        \multicolumn{1}{c|}{Hierarchy} &
        \multicolumn{4}{c|}{($10^{-7}~M_\odot$)}   &
        \multicolumn{2}{c|}{($10^{-12}~M_\odot$)}  &
        \multicolumn{1}{c}{($10^{-11}~M_\odot$)} \\ 
        \hline
        \multicolumn{1}{c|}{FDSI NEQ} &
        \multicolumn{1}{c|}{NH} &
			{1.131}&{1.601}&{4.275}&{4.887}&{16.73}&{3.600}&{15.47} \\
        \multicolumn{1}{c|}{(KCK19)} &
        \multicolumn{1}{c|}{IH} &
			{1.261}&{1.206}&{4.622}&{4.242}&{12.80}&{2.927}&{11.89}\\
		\hline
        \multicolumn{1}{c|}{Difference (\%)} &
        \multicolumn{1}{c|}{NH} &
			{0.09} & {0.00} & {0.02} & {0.67} & {1.76} & {1.15} & {1.84} \\
        \multicolumn{1}{c|}{} &
        \multicolumn{1}{c|}{IH} &
			{0.00} & {0.00} & {0.02} & {0.96} & {4.15} & {2.56} & {5.13} \\
        \hline
        \hline
        \end{tabular} \label{inte_mass_2}
    \end{table*}

\clearpage

\section{Summary and Conclusion} \label{sec:sum_con}

\subsection{Summary}
In this work, we investigated the multifaceted features in the $\nu$-process of the CCSN from due to the choice of the various physics models. First, we update the hydrodynamics model from {\it HKC18} to {\it KCK19}. The density, temperature, and radius in the former case are a bit larger than the latter. Due to the differences, the MSW resonance region of {\it HKC18} occurs around $M_r \sim 4 M_\odot$, whereas that of {\it KCK19} appears around $M_r \sim$ 3.7 $M_\odot$ for $E_\nu = 15$ MeV. In addition, the nuclear abundances in the {\it HKC18} model were generally larger than those of {\it KCK19}. Howeber, both models have similar production patterns to each other.

For the neutrino reactions, we adopted the results from the shell model and QRPA calculation tabulated at Appendix \ref{c2}, which have been shown to properly account for available data related to the neutrino-induced reactions on the relevant nuclei. Other nuclear reaction rates are taken from the JINA REACLIB database \citep{2010ApJS..189..240C}. Since the $(n,\gamma)$ reactions turn out to be important for the $\nu$-process, we also utilized recent $(n,\gamma)$ reaction calculations developed by the Monte Carlo method (see Appendix \ref{c1}).

Second, we investigated the MSW effect in the outer region. The different hydrodynamics models show the shift of the MSW resonance region. As a result, light element abundances are increased outside the MSW resonance region, while heavy element abundances are less affected by the MSW resonance.

Third, we examined the shock propagation effect peculiar to the CCSN. We found a neutrino flavor change resonance around $M_r \sim 1.6 M_{\odot}$ by the shock effect. Most of the neutrino spectra go back to the initial neutrino flux at the resonance. But neutrino luminosities are exponentially decreasing as a function of time. Thus, the nuclear abundances are less affected by the shock effect than that by the SI effect. Heavy elements are increased maximally by about 17\% (1.1\%) for the NH (IH) case, respectively. Light elements are more changed, maximally 32 \%(11 \%) for the NH (IH) case, respectively.

Fourth, we analyzed the effects of the neutrino luminosity on the element abundances. The neutrino luminosity was deduced by recent simulations of the neutrino transport model  for five post-bounce time intervals. We adopted the results in \cite{2018JPhG...45j4001O}, which compared six CCSN simulations and provided luminosities and averaged energies of neutrinos emitted from the neutrino sphere. In particular, the neutrino self-interaction (SI) strongly hinges on the neutrino luminosity and neutrino sphere radius depending on the adopted SN simulation model. We termed it as non-equivalent (NEQ) luminosity and studied in detail the SI effect with the comparison to the results of the equivalent (EQ) neutrino luminosity model.

We used the multi-angle approach to derive the SI effect in the $\nu$-oscillation. The spectra of neutrinos emitted from the neutrinosphere are modified by the $\nu-$SI. In this study, we investigated the two different neutrino luminosity, EQ and NEQ. When we adopt EQ luminosity the $\nu-$SI effect shows only in the IH case, where a neutrino splitting phenomenon occurs around $E_\nu \sim 10$ MeV. Below this energy, the neutrino spectra remain as initial $\nu_e$, and the $\nu_e$ spectra above the energy are following the initial $\nu_x$ spectra (Figure \ref{EQ_SI}). As a result, inside the MSW resonance region, nuclear abundances are increased by the $\nu$-process. In particular, heavy nuclear abundances show the trend.

On the other hand, the $\nu-$SI effect with NEQ luminosity appears for both mass hierarchies. At the initial propagation, the $\nu-$SI effect in the IH case is analogous to that in the EQ luminosity case. The difference is that the $\nu_x$ luminosity decreases faster than that of $\nu_e$ with the time evolution. Therefore, the changed $\nu_e$ spectra have lower values above the splitting energy $E_\nu \sim$ 5-7 MeV (Figure \ref{spectra_IH}). These lead to the decrease of the neutrino CC reaction by the MSW effect, which affects the light element synthesis. In the NH case, the splitting phenomenon is less clear than in the IH case. But for the same reason as before, the neutrino CC reactions are decreased by SI effect.

Finally, we discussed the ratio of $N(^{7}{\rm Li})/N(^{11}{\rm B})$ and PF$( ^{138}{\rm La})/{\rm PF}( ^{11}{\rm B})$ by using the final abundance results tabulated at Table \ref{inte_mass}. Present results for the both ratios imply that the NH case is favored by more advanced models, that is, NEQ SI luminosity model using {\it KCK19} hydrodynamics.

\subsection{Conclusion}
In conclusion, 1) elemental abundances produced by the $\nu$-process strongly depend on the hydrodynamics model and the pre-SN model. 2) Shock propagation effect are not as large as other effects, but they give maximally 22\% difference for a specific nucleus abundance. 3) MSW effects are still important for understanding the yield differences between the light and heavy element abundances. 4) Neutrino luminosity is more important than other factors in the $\nu$-process, which are critically sensitive to the neutrino transport model and their simulations. 5) Ratio of some specific nuclei like $N(^7$Li)/N($^{11}$B) and PF($^{138}$La)/PF($^{11}$B) could be valuable indicators of the $\nu$-process because they are less sensitive to the models exploited in the $\nu$-process calculations. 6) Our systematic calculations support the nucleosynthesis results for the NH neutrino mass hierarchy. 7) We remind that the neutrino sphere radius $\sim$ 10 km and the power law density profile could differ from the SN simulations. The increase of the neutrino sphere during the accretion phase may lead to the complete suppression of the collective neutrino oscillation. Therefore, we tested the case of the complete matter suppression for the collective neutrino oscillation during the accretion phase and the collective oscillation during the cooling phase, termed as FDSI NEQ. The results are presented in Table 5. The difference turns out to be less than 5 \% maximally. It means that the uncertainty from the neutrino sphere radius and the power law density profile during the accretion phase can be retained within 5 \% level.

Finally, we note that recent three-dimensional hydrodynamical SN simulations predicted asymmetric radiations of $\nu_e$ and $\bar{\nu}_e$ \citep{2021MNRAS.506.1462N}. The subsequent studies taking the neutrino angular distribution into account suggest that the different angular distributions of $\nu_e$ and $\bar{ \nu}_e$ cause a fast neutrino flavor transformation by the crossing of $\nu_e$ and $\bar{\nu_e}$. Other symmetry violations due to the asymmetric flux and the convection layer can cause the fast flavor conversion compared to the flavor change due to the mater effect \citep{2019PhRvD.100d3004A,Glas19-2}. It may happen in the CCSN, and affect the neutrino observation \citep{Dasgupta17, Tamborra17} and the diffuse SN neutrino background \citep{Mirizzi16}. In this case, the energy exchange may occur earlier and bring about the larger difference in luminosities between $\nu_e$ and  $\nu_x$. A hypothetical sterile neutrino may also cause such a fast neutrino flavor change \citep{2020ApJ...894...99K}. This can enhance the MH dependence of the $\nu$-process abundances. In other words, the constraints from the analysis of elemental abundances for the $\nu$-process can be a good test-bed to evaluate the many interesting facets in the present neutrino physics model. However, for more definite conclusion, detailed $\nu$-process calculations should involve the realistic neutrino emission models in CCSN with a more precise evaluation of the fast neutrino-conversion effects as well as the advanced models beyond the standard model. We leave them as future works.

\acknowledgments
	This work was supported by the National Research Foundation of Korea (Grant No. NRF-2021R1A6A1A03043957, NRF-2020K1A3A7A09080134 and NRF-2020R1A2C3006177).
	This work was supported by Grants-in-Aid for Scientific Research of JSPS (17K05459, 20K03958).
 	The work of DJ is supported by Institute for Basic Science under IBS-R012-D1.
	MK was supported by NSFC Research Fund (grant No. 11850410441).
	Work of GJM was supported in part by DOE nuclear theory grant DE-FG02-95-ER40934 and in part by the visitor program at NAOJ.

\clearpage
\appendix
\section{Neutrino field and Total Hamiltonian for CC Interaction}\label{app_Basic formula}
	We introduce neutrino field, which can be expressed in a finite volume to describe one particle states with appropriate normalization. In a box having width L and momentum $\vec{p}=({2\pi}/ {L}) \vec{n}$  (n is integer), continuum states can be discretized as $({Vd^3 \vec{p}}) / {(2\pi)^3} \rightarrow \sum_{\bf p}$ and $(2\pi)^3 \delta^3({\bf p}-{\bf p'}) \rightarrow V \delta_{{\bf p p'}}$ with the volume $V=L^3$ \citep{2007fnpa.book.....G, 2017aptp.book.....S}. 
    Then, the field operator for the left-handed neutrino $\nu_{\alpha}(\alpha=e,\mu,\tau)$ is quantized as follows:
	\begin{eqnarray} \label{dirac_field}
    	\nu_{\alpha L}(x) = \sum_{\mathbf{p}}\frac{1}{2E_{p}V}
    					(
    					   \hat{a}_{\nu_\alpha} (\mathbf p) u^{(-)}(p) e^{-i p \cdot x}
    					 + \hat{b}^\dagger_{\nu_\alpha}(\mathbf p) v^{(+)}(p) e^{+i p \cdot x}
    					),
        \end{eqnarray}
where $u^{(-)}(p)$ and $v^{(+)}(p)$ are spinors of the particle with negative helicity and the antiparticle with positive helicity. The operators $\hat{a}_{\nu_\alpha}({\bf p})$ and $\hat{a}^\dagger_{\nu_\alpha}({\bf p})$ are annihilation and creation operators, respectively, for $\nu_\alpha$, and $\hat{b}_{\nu_\alpha}({\bf p})$ and $\hat{b}^\dagger_{\nu_\alpha}({\bf p})$ for $\bar{\nu}_{\alpha}$. The dispersion relation for neutrinos in free space is given as $E_p = |{\bf p}|$. Here, we normalize the anti-commutation relation of the operators as $\{a_{\nu_\alpha}({\bf p}),a^\dagger_{\nu_\alpha}({\bf p'})\}=\{b_{\nu_\alpha}({\bf p}),b^\dagger_{\nu_\alpha}({\bf p'})\}=2E_p V \delta_{{\bf pp'}} $. As a result, the neutrino and anti-neutrino states are expressed as $|\nu_\alpha({\bf p}) \rangle = 1/\sqrt{2E_p V} a_{\nu_\alpha}^\dagger({\bf p}) |0 \rangle$ and $|\bar{\nu}_\alpha ({\bf p}) \rangle = 1/\sqrt{2E_p V} b^\dagger_{\nu_\alpha}({\bf p})$, where  $|0 \rangle$ is the vacuum for the neutrino field.

For the CC interaction, the Hamiltonian density $\mathcal{H}^{CC}$ between neutrinos and background leptons is written as follows:
	\begin{eqnarray} 	\label{Ham_CC}
		\mathcal{H}^{CC}_{\nu_l l} (x)
		=2\sqrt{2}G_F
 	            \left[ \bar{\nu}_{lL}(x)\gamma^\lambda \nu_{lL}(x) \right]
 	            \left[ \bar{l}_{L}(x) \gamma_\lambda l_{L}(x) \right],~~
	\end{eqnarray}
where $G_F$ is the Fermi constant describing the effective interaction strength and $l$ stands for the leptons such as an electron, muon, and tau, respectively. The subscript $L$ means the left-handed projection of neutrino field defined by $\nu_{lL}(x) \equiv  (1-\gamma^5)/2\  \nu_{l}(x)$ in which we follow the convention of $\bar{\nu}_{lL}(x) \equiv \nu^\dagger_{lL}(x) \gamma^0$ and $\gamma_5 \equiv \gamma^5 \equiv i  \gamma^0 \gamma^1 \gamma^2 \gamma^3$.

By taking the thermal average to the electron background, the effective Hamiltonian density is reduced to $\mathcal{H}^{CC}_{\rm eff}(x)=\sqrt{2}G_F n_e \nu_{eL}^\dagger(x) \nu_{eL}(x)$ \citep{2007fnpa.book.....G}, and the effective potential for the CC interaction is given by
\begin{eqnarray} \label{Ham_CC eff}
	 \hat{\mathcal{V}}_{\nu_e e}&= &\int\mathrm{d}^{3}x\ \mathcal{H}^{CC}_{\rm eff}(x)
	 =\sqrt{2}G_F n_e \sum_{\bf{p}}
	 \frac{[a_{\nu_{e}}^{\dagger}(\mathbf{p}) a_{\nu_{e}}(\mathbf{p})-b_{\nu_{e}}^{\dagger}(\mathbf{p}) b_{\nu_{e}}(\mathbf{p})]}{2E_{p}V},
\end{eqnarray}
where $a_{\nu_{e}} (a^{\dagger}_{\nu_{e}})$ and $b_{\nu_{e}}(b^{\dagger}_{\nu_{e}})$ are annihilation (creation) operators of  $\nu_{e}$ and $\bar{\nu}_{e}$. $E_{p}$ and  $V$ denote the energy of neutrinos and the volume with the factor of $(2 \pi)^3$. These quantities come from the second quantization of  $\nu_{eL}(x)$ in Equation (\ref{dirac_field}). Because of the charge neutrality condition, the net electron density is given by $n_e= \rho_b N_A Y_e$, where $\rho_b$, $N_A$, and $Y_e$ are baryon density, Avogadro's number, and electron fraction, respectively. The matrix components of the MSW matter potential are derived from $\langle\nu_\beta({\bf p})|\hat{\mathcal{V}}_{\nu_e e}|\nu_\alpha({\bf p})\rangle (\alpha,\beta=e,\mu,\tau)$ in Equation (\ref{Ham_CC eff}).

\section{Potential for the neutrino SI}\label{app_nunu}

     The Hamiltonian density for the neutrino SI  is given as Equation~(\ref{eq:Hamiltonian NC}). Similar to the derivation of the effective Hamiltonian in Equation (\ref{Ham_CC eff}), the one-body effective Hamiltonian for the neutrino SI is given by the average of the neutrino background. We introduce the average of the neutrino operators \citep{1993NuPhB.406..423S,2013PhRvD..87k3010V},
\begin{eqnarray}
\langle a^\dagger_{\nu_\alpha}({\bf p'}) a_{\nu_\beta}({\bf p})\rangle &=& 2E_p V \delta_{{\bf pp'}}f_{{\rm dist}}(t,{\bf p})\rho_{\beta\alpha}(t,{\bf p}), \\
\langle b^\dagger_{\nu_\alpha}({\bf p'}) b_{\nu_\beta}({\bf p})\rangle &=& 2E_p V \delta_{{\bf pp'}}\bar{f}_{{\rm dist}}(t,{\bf p})\bar{\rho}_{\alpha\beta}(t,{\bf p}),
\end{eqnarray}
 where $f_{{\rm dist}}$ ($\bar{f}_{{\rm dist}}$) and $\rho$($\bar{\rho}$) are the normalized distribution and the density matrix of neutrinos (anti-neutrinos), respectively. Here, we normalize the traces of the matrices, $\sum_{\alpha}\rho_{\alpha\alpha}(t,{\bf p})=\sum_{\alpha}\bar{\rho}_{\alpha\alpha}(t,{\bf p})=1$, thus the average of the neutrino background is given by
  \begin{eqnarray}\label{eq:average neutrino background}
 \langle\bar{\nu}_{\alpha L}(x)\gamma^\mu\nu_{\beta L}(x)\rangle &=& \sum_{\mathbf{q}\mathbf{q'}}\frac{
    					 					   \bar{u}^{(-)}(q')\gamma^{\mu}u^{(-)}(q)\langle a^\dagger_{\nu_\alpha}({\bf q'}) a_{\nu_\beta}({\bf q})\rangle e^{i(q'-q)\cdot x}-\bar{v}^{(+)}(q')\gamma^{\mu}v^{(+)}(p)e^{-i(q'-q)\cdot x}\langle b^\dagger_{\nu_\beta}({\bf q}) b_{\nu_\alpha}({\bf q'})\rangle
					     }{(2E_{q}V)(2E_{q'}V)} \nonumber \\
					      &=&\frac{1}{V}\sum_{\mathbf{q}}\frac{q^{\mu}}{E_{q}}  \left [
					      f_{{\rm dist}}(t,{\bf q})\rho_{\beta\alpha}(t,{\bf q})-\bar{f}_{{\rm dist}}(t,{\bf q})\bar{\rho}_{\beta\alpha}(t,{\bf q})
					      \right],	
 \end{eqnarray}
where the averages of expectation values of $ \langle a^{\dagger}b^{\dagger} \rangle$ and $\langle ba \rangle$ are ignored. Without flavor mixing between neutrinos and anti-neutrinos, the diagonal term in Figure~\ref{HF} does not contribute to neutrino oscillations. Therefore, the effective Hamiltonian for the neutrino SI is written as
    \begin{eqnarray} \label{eq:pot_nunu2}
         \hat{\mathcal{V}}_{\nu\nu}
                      &=&\frac{G_F}{ \sqrt{2} }\int d^{3}x\
				    \left [ \bar{\nu}_{\beta L} (x) \gamma^\mu\nu_{\alpha L}(x) \right ]
    	            2\langle \bar{\nu}_{\alpha L}(x) \gamma^\mu\nu_{\beta L}(x) \rangle.
	              \end{eqnarray}
Finally, Equation~(\ref{pot_nunu}) is derived from Equations~(\ref{eq:pot_nunu2}), (\ref{eq:average neutrino background}), and (\ref{dirac_field}). The number density of neutrino background term, $ \frac{d^3\vec{q}}{(2\pi)^3}f_{{\rm dist}}(t,{\bf q}) = \sum_{\eta}dn_{\nu_\eta}$, depends on the angle between neutrinos. We follow the uniform and isotropic neutrino emission model, which is called the bulb model described in \citep{2006PhRvD..74j5014D}.
    The differential neutrino number density can be written as
	\begin{eqnarray}
		dn_{\nu_\eta} = \frac{L_{\nu_\eta}}{\pi R^2_\nu} \frac{1}{\langle E_{\nu_\eta} \rangle}
					  	\frac{1}{(T_{\nu_\eta})^3 F_2(0)} \frac{E_q^2 dE_q}{\exp(E_q/T_{\nu_\eta})+1} \left( \frac{d\Omega_q}{\int d\Omega_q} \right),
	\end{eqnarray}
	where $\left( \frac{d\Omega_q}{\int d\Omega_q} \right) = \frac{d(\cos \theta_q) d\phi_q}{4\pi} $.	
	Here we assume a cylindrical symmetry along the z-axis, so that the integration of only $\phi_q$ direction is done as
	$\int^{2\pi}_0 (1-\hat{\bf p} \cdot \hat{\bf q}) d\Omega_q = 2 \pi (1-\cos \theta_p\cos \theta_q) d(\cos \theta_q) $.
    Finally, the potential for the neutrino SI in the Schr\"odinger-like equation is described by
	\begin{eqnarray}\label{eq: V_nunu}
		\mathcal{V}_{\nu\nu}
				&= & \sqrt{2} G_F  \int
			\left [\sum_{\eta=e,\mu,\tau}
			    \frac{L_{\nu_\eta}}{2\pi R^2_\nu} \frac{1}{\langle E_{\nu_\eta} \rangle}
				\frac{1}{(T_{\nu_\eta})^3 F_2(0)} \frac{E^2_q}{\exp(E_q/T_{\nu_\eta})+1} \rho (r,E_q,\theta_q)
			\right . \nonumber \\
		 && ~~~~~~~~~~~~~~~~~ -
			\left .
			    \sum_{\eta=e,\mu,\tau}\frac{L_{\bar{\nu}_\eta}}{2\pi R^2_\nu} \frac{1}{\langle E_{\bar{\nu}_\eta} \rangle}
				\frac{1}{(T_{\bar{\nu}_\eta})^3 F_2(0)} \frac{E^2_q}{\exp(E_q/T_{\bar{\nu}_\eta})+1} \bar{\rho} (r,E_q,\theta_q)
            \right ]
           (1-\cos \theta_{p}\cos \theta_{q}) dE_q d(\cos \theta_{q}).~~~~~~
    \end{eqnarray}	
    	In the case of single angle approximation, the propagation angle $\theta_{p}$ is not considered, by which the integration is given as
	\begin{eqnarray}
		\int^{\theta_{\rm max}}_0 (1-\cos \theta_q) \sin \theta_q d\theta_q
		= \frac{1}{2} (1- \cos \theta_{\rm max})^2 =\frac{1}{2} \left (1-\sqrt{1-\left ( \frac{R_\nu}{r} \right) ^2} \right )^2 ~,
		\label{sin_ang_geometry}
	\end{eqnarray}
where the possible maximum angle of emitted background neutrino is given as $\sin \theta_{\rm max} = {{R_\nu} \over {r}}$;
    $r$ is the radius from the center of core and the emission follows the tangential direction of the neutrino sphere.
    This single-angle approximation is appropriate when $r$ is large enough.

\section{Reaction data} \label{sec:app_c}
	In the present calculation, we exploited updated nuclear reaction rates related to the $\nu$-process from the JINA database \citep{2010ApJS..189..240C}. But parts of them, such as, neutron capture, photonuclear reaction, and neutrino-induced reactions, are different from the JINA data base. In the following we present the numerical results of the updated nuclear reactions in detail.
\subsection{ $(n,\gamma)$ and $(\gamma, n)$ reaction rates in the A $\sim$ 100 region} \label{c1}
\begin{figure*}[h!]
	\centering
		\includegraphics[width=16.4cm]{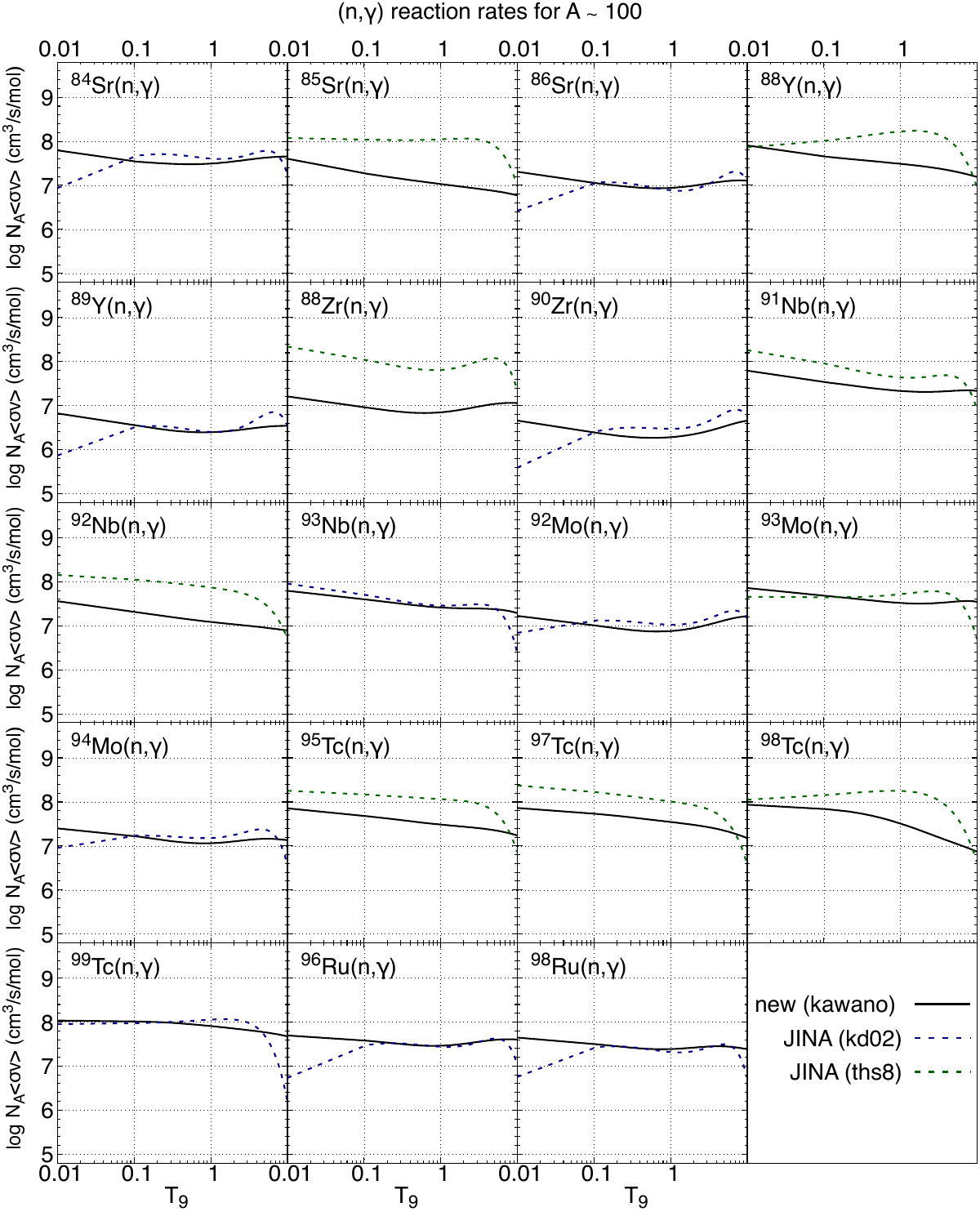}
		\caption{ Forward $(n,\gamma)$ reaction rates.}
		\label{fig_ng}
    \end{figure*}

\begin{figure*}[h!]
	\centering
		\includegraphics[width=16.4cm]{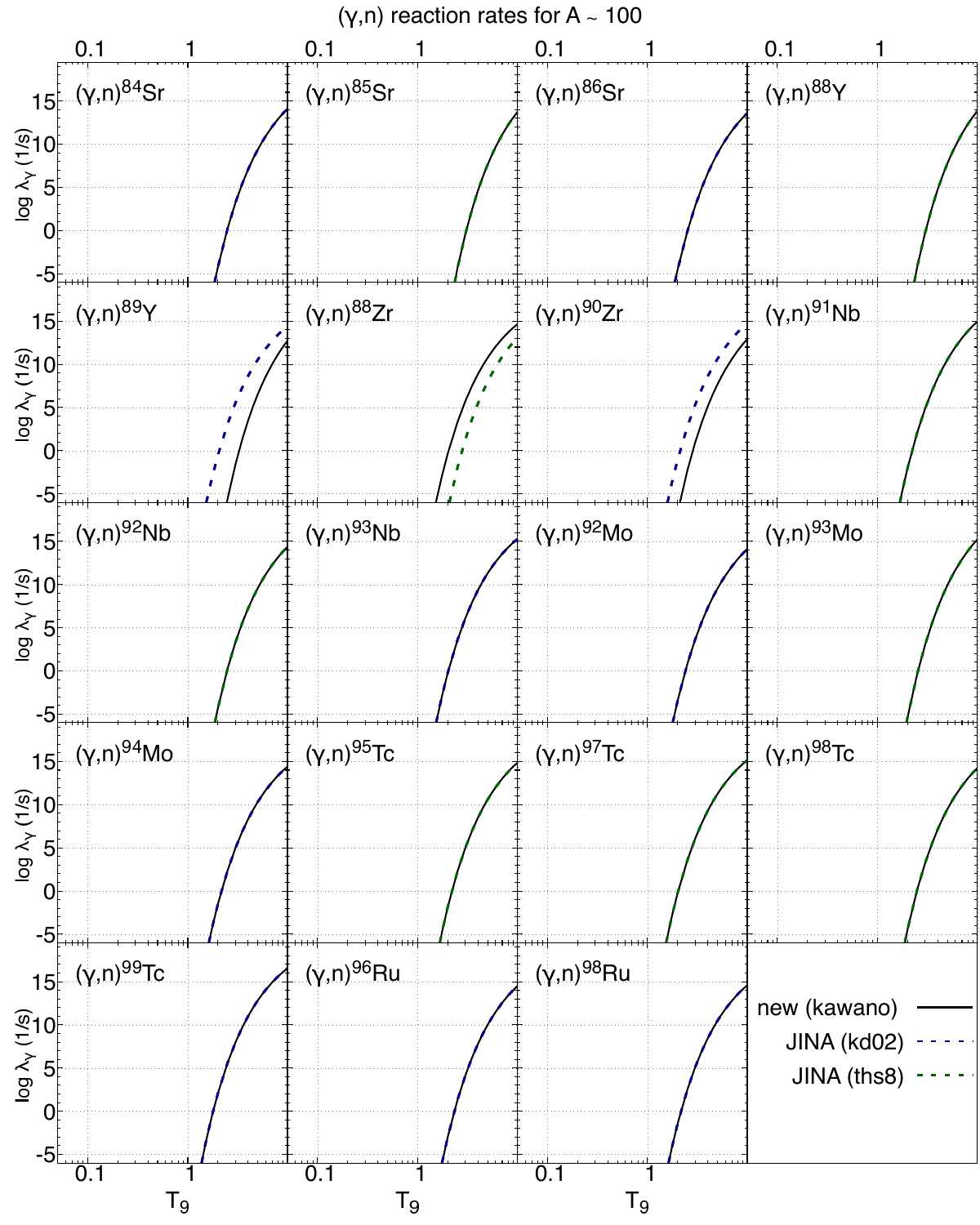}
		\caption{ Inverse $(\gamma,n)$ reaction rates.}
		\label{fig_gn}
    \end{figure*}

	First, the neutron capture reactions turned out to play important roles around the MSW region. For example, the valley in $^{98}$Tc abundance is sensitive to the $(n,\gamma)$ reactions. Therefore we used newly updated calculations by Kawano \citep{2010EPJWC...209001K}, in which local systematics of the Hauser-Feshbach model parameters were carefully investigated to infer the theoretical prediction for nuclear reactions of the relevant unstable nuclei to obtain realistic abundances after the weak s-process. The updated $(n,\gamma)$ and $(\gamma, n)$ reaction data and those by JINA database are presented in Figure \ref{fig_ng} and \ref{fig_gn}, respectively. The new parameters for the temperature dependence {utilized} in the JINA database \citep{2010ApJS..189..240C} are obtained from the new reaction rate functions and tabulated at Tables \ref{tab_ng} and \ref{tab_gn} for $(n,\gamma)$ and $(\gamma,n)$ reactions, respectively.
		
	\begin{longrotatetable}
	\begin{deluxetable*}{ccccccccc}
	\tablecaption{Forward $(n,\gamma)$ reactions and temperature parameters \label{tab_ng}} 	
	\tablehead{
		\multicolumn{1}{c}{reaction} & \multicolumn{7}{c}{JINA database parameter} \\
			& $a_0$ & $a_1$ & $a_2$ & $a_3$ & $a_4$ & $a_5$ & $a_6$ }
	\startdata
    $^{84}$Sr$(n,\gamma)^{85} $Sr & 1.248898E+01 &  2.370158E-02 & -2.452547E+00 &  7.664219E+00 & -4.995054E-01 &  2.502527E-02 & -2.778081E+00 \\
    $^{85}$Sr$(n,\gamma)^{86} $Sr & 1.608127E+01 &  5.866922E-04 &  1.163815E-01 & -4.176012E-02 &  1.914001E-02 & -5.292960E-03 & -1.787551E-01 \\
    $^{86}$Sr$(n,\gamma)^{87} $Sr & 1.005771E+01 &  2.481909E-02 & -2.949004E+00 &  9.353030E+00 & -5.609058E-01 &  2.291346E-02 & -3.463590E+00 \\
    $^{88} $Y$(n,\gamma)^{89} $Y  & 1.673954E+01 &  3.646441E-03 & -1.878931E-01 &  8.603386E-01 & -1.382795E-01 &  8.356414E-03 & -3.971826E-01 \\
    $^{89} $Y$(n,\gamma)^{90} $Y  & 9.051425E+00 &  1.395434E-02 & -2.159787E+00 &  8.338274E+00 & -5.464776E-01 &  2.700101E-02 & -2.933181E+00 \\
    $^{88}$Zr$(n,\gamma)^{89} $Zr & 7.946674E+00 &  2.684651E-02 & -3.545765E+00 &  1.208552E+01 & -7.999320E-01 &  3.881745E-02 & -4.307062E+00 \\
    $^{90}$Zr$(n,\gamma)^{91} $Zr & 1.260702E+01 & -8.700704E-03 &  4.592438E-01 &  1.303912E+00 &  1.371825E-01 & -2.291362E-02 & -2.587433E-01 \\
    $^{91}$Nb$(n,\gamma)^{92} $Nb & 1.590488E+01 &  1.147514E-02 & -1.074272E+00 &  1.982728E+00 &  8.122271E-02 & -1.900906E-02 & -1.176654E+00 \\
    $^{92}$Nb$(n,\gamma)^{93} $Nb & 1.452824E+01 &  9.406627E-03 & -1.076378E+00 &  3.107980E+00 & -2.343066E-01 &  1.445819E-02 & -1.371316E+00 \\
    $^{93}$Nb$(n,\gamma)^{94} $Nb & 1.345852E+01 &  2.241971E-02 & -2.521347E+00 &  6.419594E+00 & -3.573849E-01 &  9.647026E-03 & -2.719755E+00 \\
    $^{92}$Mo$(n,\gamma)^{93} $Mo & 1.057350E+01 &  1.980617E-02 & -2.623842E+00 &  8.157561E+00 & -3.182821E-01 &  1.101292E-03 & -3.138314E+00 \\
    $^{93}$Mo$(n,\gamma)^{94} $Mo & 1.694178E+01 &  1.375932E-02 & -1.233149E+00 &  1.468402E+00 &  1.885581E-01 & -3.051284E-02 & -1.134859E+00 \\
    $^{94}$Mo$(n,\gamma)^{95} $Mo & 7.382442E+00 &  3.701772E-02 & -4.788892E+00 &  1.462631E+01 & -1.019866E+00 &  5.446624E-02 & -5.436827E+00 \\
    $^{95}$Tc$(n,\gamma)^{96} $Tc & 1.472420E+01 &  1.699289E-02 & -1.911592E+00 &  4.739958E+00 & -3.251417E-01 &  1.411674E-02 & -2.073607E+00 \\
    $^{97}$Tc$(n,\gamma)^{98} $Tc & 1.521961E+01 &  1.513890E-02 & -1.715202E+00 &  4.216543E+00 & -3.616674E-01 &  1.916291E-02 & -1.844236E+00 \\
    $^{98}$Tc$(n,\gamma)^{99} $Tc & 2.171829E+01 &  5.139517E-03 &  8.927720E-02 & -4.932672E+00 &  4.558193E-01 & -3.258772E-02 &  7.150464E-01 \\
    $^{99}$Tc$(n,\gamma)^{100}$Tc & 1.831816E+01 &  1.330489E-02 & -1.174862E+00 &  1.101243E+00 & -3.004754E-02 & -2.700345E-03 & -8.823261E-01 \\
    $^{96}$Ru$(n,\gamma)^{97} $Ru & 9.533413E+00 &  4.251006E-02 & -5.004759E+00 &  1.342078E+01 & -8.266547E-01 &  3.943924E-02 & -5.272868E+00 \\
    $^{98}$Ru$(n,\gamma)^{99} $Ru & 8.495533E+00 &  3.960303E-02 & -4.832171E+00 &  1.427723E+01 & -1.056089E+00 &  5.999115E-02 & -5.321814E+00
	\enddata
    \end{deluxetable*}
    \end{longrotatetable}

    \begin{longrotatetable}
	\begin{deluxetable*}{ccccccccc}
	\tablecaption{Inverse $(\gamma,n)$ reactions and temperature parameters \label{tab_gn}} 	
	\tablehead{
		\multicolumn{1}{c}{reaction} & \multicolumn{7}{c}{JINA database parameter} \\
		 & $a_0$ & $a_1$ & $a_2$ & $a_3$ & $a_4$ & $a_5$ & $a_6$ }
	\startdata           	
    $^{85} $Sr$(\gamma,n)^{84}$Sr & 2.941888E+01 & -1.001321E+02 &  2.493498E+01 & -1.485769E+01 &  2.750941E-01 & -2.923625E-03 &  1.345984E+01 \\
    $^{86} $Sr$(\gamma,n)^{85}$Sr & 4.323333E+01 & -1.325321E+02 & -2.101852E+01 &  1.989032E+01 & -8.112554E-01 &  3.261639E-02 & -1.076570E+01 \\
    $^{87} $Sr$(\gamma,n)^{86}$Sr & 2.689858E+01 & -9.802114E+01 & -1.165699E+00 &  1.267696E+01 & -9.167931E-01 &  4.833428E-02 & -2.413698E+00 \\
    $^{89}  $Y$(\gamma,n)^{88} $Y & 4.151676E+01 & -1.333624E+02 &  5.217288E+00 & -4.102714E+00 &  5.340280E-02 &  6.013768E-04 &  4.170938E+00 \\
    $^{90}  $Y$(\gamma,n)^{89} $Y & 3.120161E+01 & -1.339622E+02 &  1.720268E+01 & -8.732321E+00 &  1.194352E-01 & -1.559901E-03 &  9.271344E+00 \\
    $^{89} $Zr$(\gamma,n)^{88}$Zr & 2.642602E+01 & -8.007789E+01 &  7.131969E+00 &  5.421843E+00 & -6.738864E-01 &  3.952511E-02 &  2.308722E+00 \\
    $^{91} $Zr$(\gamma,n)^{90}$Zr & 4.201612E+01 & -1.058214E+02 & -5.260266E+01 &  4.610917E+01 & -1.525340E+00 &  4.487688E-02 & -2.750847E+01 \\
    $^{92} $Nb$(\gamma,n)^{91}$Nb & 3.599003E+01 & -9.091540E+01 & -2.039827E+01 &  2.501031E+01 & -1.094619E+00 &  4.416519E-02 & -1.212790E+01 \\
    $^{93} $Nb$(\gamma,n)^{92}$Nb & 4.245298E+01 & -1.027860E+02 &  1.145714E+01 & -1.381590E+01 &  6.881836E-01 & -3.694381E-02 &  8.814094E+00 \\
    $^{94} $Nb$(\gamma,n)^{93}$Nb & 3.064462E+01 & -8.423374E+01 &  2.103344E+00 &  8.803124E+00 & -7.872117E-01 &  4.469467E-02 & -4.897477E-01 \\
    $^{93} $Mo$(\gamma,n)^{92}$Mo & 3.213648E+01 & -9.263918E+01 & -2.984683E+01 &  3.589655E+01 & -1.549057E+00 &  6.000635E-02 & -1.784346E+01 \\
    $^{94} $Mo$(\gamma,n)^{93}$Mo & 3.932353E+01 & -1.110093E+02 & -3.845938E+01 &  4.237102E+01 & -1.777939E+00 &  7.074719E-02 & -2.260765E+01 \\
    $^{95} $Mo$(\gamma,n)^{94}$Mo & 2.555338E+01 & -8.712851E+01 &  3.671438E+01 & -2.290833E+01 &  4.379981E-01 & -6.649467E-03 &  1.940762E+01 \\
    $^{96} $Tc$(\gamma,n)^{95}$Tc & 3.595634E+01 & -9.160394E+01 &  3.915321E+00 &  1.194696E+00 & -2.927797E-01 &  1.865394E-02 &  2.242460E+00 \\
    $^{98} $Tc$(\gamma,n)^{97}$Tc & 3.673426E+01 & -8.480618E+01 &  6.191082E+00 & -1.572013E+00 & -2.188507E-01 &  1.784643E-02 &  3.732824E+00 \\
    $^{99} $Tc$(\gamma,n)^{98}$Tc & 4.255113E+01 & -1.040263E+02 & -3.006118E+00 &  1.711734E+00 & -3.671844E-02 &  1.554448E-03 & -5.663074E-01 \\
    $^{100}$Tc$(\gamma,n)^{99}$Tc & 3.943086E+01 & -7.846703E+01 & -4.985715E+00 &  9.194351E+00 & -5.917306E-01 &  3.360030E-02 & -2.833361E+00 \\
    $^{97} $Ru$(\gamma,n)^{96}$Ru & 2.504405E+01 & -9.585048E+01 &  3.673857E+01 & -2.135974E+01 &  3.905580E-01 & -5.503320E-03 &  1.882684E+01 \\
    $^{99} $Ru$(\gamma,n)^{98}$Ru & 2.843891E+01 & -8.875171E+01 &  5.316734E+01 & -4.184288E+01 &  1.290534E+00 & -4.610888E-02 &  2.988585E+01
	\enddata
    \end{deluxetable*}
    \end{longrotatetable}

\subsection{Neutrino-induced reaction cross sections data} \label{c2}
	We tabulate neutrino-induced reactions for $A \sim$ 92, 98, 138 and 180 nuclei via NC and CC currents at Tables \ref{tab_main_NC reactions} and \ref{tab_main_CC reactions}. These are calculated from the QRPA which includes all pairing interactions through the Br\"uckner $G$-matrix evaluated from the CD Bonn potential \citep{2010JPhG...37e5101C}.
	\begin{table*}[h]
 	\centering
    \caption{Cross sections for the main neutrino-induced reaction via neutral current for $^{93}$Nb, $^{99}$Ru, $^{139}$La and $^{181}$Ta}
 	\begin{tabular}{r|ccccc}
	\hline
    \multicolumn{1}{c|}{$E_\nu$} &
    \multicolumn{4}{c}{Neutral current cross section,~$\sigma$ (cm$^2$)} \\
          (MeV)   & $^{93}$Nb$(\nu,\nu'n)^{92}$Nb               & $^{93}$Nb$(\nu,\nu'p)^{92}$Zr
 	              & $^{93}$Nb$(\bar{\nu},\bar{\nu}'n)^{92}$Nb   & $^{93}$Nb$(\bar{\nu},\bar{\nu}'p)^{92}$Zr \\
 	\hline
      0  &   0.00000000E-00   &   0.00000000E-00   &   0.00000000E-00   &   0.00000000E-00 \\
      8  &   2.50713618E-50   &   4.36590897E-50   &   7.75514966E-50   &   1.35047621E-49 \\
     16  &   6.41702589E-42   &   1.42274917E-43   &   5.31736472E-42   &   1.18398366E-43 \\
     24  &   1.19460781E-41   &   2.61691966E-42   &   9.26003709E-42   &   2.02888823E-42 \\
     32  &   1.11618046E-41   &   4.05216387E-42   &   8.10204807E-42   &   2.94138929E-42 \\
     40  &   1.15439623E-41   &   4.77290399E-42   &   7.85088848E-42   &   3.24604476E-42 \\
     48  &   1.41992964E-41   &   6.22778850E-42   &   9.12788654E-42   &   4.00355303E-42 \\
     56  &   1.44336157E-41   &   6.65086093E-42   &   8.87325346E-42   &   4.08876745E-42 \\
     64  &   1.40070286E-41   &   6.61187328E-42   &   8.33028829E-42   &   3.93230587E-42 \\
     72  &   1.48701381E-41   &   7.07143664E-42   &   8.65991005E-42   &   4.11826341E-42 \\
     80  &   1.45054088E-41   &   7.02562041E-42   &   8.34253207E-42   &   4.04075106E-42  \\
    \toprule
       \nodata    & $^{99}$Ru$(\nu,\nu'n)^{98}$Ru               & $^{99}$Ru$(\nu,\nu'p)^{98}$Tc
                  & $^{99}$Ru$(\bar{\nu},\bar{\nu}'n)^{98}$Ru   & $^{99}$Ru$(\bar{\nu},\bar{\nu}'p)^{98}$Tc \\
    \hline
      0  &   0.00000000E-00   &   0.00000000E-00   &   0.00000000E-00   &   0.00000000E-00 \\
      8  &   8.31938546E-48   &   3.44319491E-43   &   7.85612776E-48   &   0.00000000E+00  \\
     16  &   3.21807831E-41   &   3.62001976E-42   &   2.88619734E-41   &   5.68020612E-43  \\
     24  &   2.34930998E-41   &   6.96046308E-42   &   2.00095035E-41   &   2.93190765E-42  \\
     32  &   2.56998078E-41   &   5.52408294E-42   &   2.07665520E-41   &   5.41904121E-42  \\
     40  &   2.14218934E-41   &   2.84717287E-42   &   1.64510701E-41   &   5.27353169E-42  \\
     48  &   2.48543962E-41   &   1.94343549E-42   &   1.82246833E-41   &   5.98999512E-42  \\
     56  &   2.30132053E-41   &   1.20297515E-42   &   1.61746490E-41   &   5.75926394E-42  \\
     64  &   2.12816045E-41   &   7.36653543E-43   &   1.43831447E-41   &   5.24028336E-42  \\
     72  &   2.31177290E-41   &   5.51800694E-43   &   1.50636782E-41   &   5.64197366E-42  \\
     80  &   2.37025986E-41   &   3.87751776E-43   &   1.49177096E-41   &   5.54453176E-42  \\
    \toprule
       \nodata    & $^{139}$La$(\nu,\nu'n)^{138}$La             & $^{139}$La$(\nu,\nu'p)^{138}$Ba
                  & $^{139}$La$(\bar{\nu},\bar{\nu}'n)^{138}$La & $^{139}$La$(\bar{\nu},\bar{\nu}'p)^{138}$Ba \\
    \hline
      0  &   0.00000000E-00   &   0.00000000E-00   &   0.00000000E-00   &   0.00000000E-00 \\
      8  &   1.14328551E-58   &   0.00000000E+00   &   1.35047621E-49   &   7.75514966E-50 \\
     16  &   5.26640495E-44   &   2.40232318E-41   &   1.18398366E-43   &   5.31736472E-42 \\
     24  &   5.29198208E-42   &   9.23905206E-42   &   2.02888823E-42   &   9.26003709E-42 \\
     32  &   1.46484515E-41   &   1.22319044E-41   &   2.94138929E-42   &   8.10204807E-42 \\
     40  &   2.24460135E-41   &   1.55437490E-41   &   3.24604476E-42   &   7.85088848E-42 \\
     48  &   2.18293273E-41   &   1.26069097E-41   &   4.00355303E-42   &   9.12788654E-42 \\
     56  &   1.95624622E-41   &   1.06066696E-41   &   4.08876745E-42   &   8.87325346E-42 \\
     64  &   2.42966931E-41   &   1.31735886E-41   &   3.93230587E-42   &   8.33028829E-42 \\
     72  &   2.16489860E-41   &   1.11070692E-41   &   4.11826341E-42   &   8.65991005E-42 \\
     80  &   2.05087962E-41   &   1.03754018E-41   &   4.04075106E-42   &   8.34253207E-42 \\
    \toprule
       \nodata    & $^{181}$Ta$(\nu,\nu'n)^{180}$Ta             & $^{181}$Ta$(\nu,\nu'p)^{180}$Hf
                  & $^{181}$Ta$(\bar{\nu},\bar{\nu}'n)^{180}$Ta & $^{181}$Ta$(\bar{\nu},\bar{\nu}'p)^{180}$Hf \\
    \hline
      0  &   0.00000000E-00   &   0.00000000E-00   &   0.00000000E-00   &   0.00000000E-00 \\
      8  &   7.12415025E-50   &   8.74129047E-44   &   8.10389589E-50   &   8.17122116E-44 \\
     16  &   2.93958839E-45   &   2.68586946E-42   &   2.90647665E-45   &   2.65112708E-42 \\
     24  &   3.67968942E-43   &   4.51747170E-42   &   3.71060172E-43   &   4.56012162E-42 \\
     32  &   1.24984216E-42   &   6.89812693E-42   &   1.27083830E-42   &   7.01667485E-42 \\
     40  &   2.56869094E-42   &   1.11448222E-41   &   2.60570403E-42   &   1.13044001E-41 \\
     48  &   3.59301912E-42   &   1.31187144E-41   &   3.56647425E-42   &   1.30218781E-41 \\
     56  &   4.82760384E-42   &   1.67583423E-41   &   4.66951437E-42   &   1.62094060E-41 \\
     64  &   6.25367586E-42   &   2.06427678E-41   &   5.83161284E-42   &   1.92500015E-41 \\
     72  &   7.63247112E-42   &   2.47097518E-41   &   6.80911481E-42   &   2.20436351E-41 \\
     80  &   8.76210673E-42   &   2.85776514E-41   &   7.44981200E-42   &   2.42971224E-41 \\
    \hline
    \end{tabular}
    \label{tab_main_NC reactions}
	\end{table*}
	\begin{table*}[h]
 	\centering 		
    \caption{Cross sections for the main neutrino-induced reaction via charged current for $^{92}$Nb, $^{98}$Mo, $^{99}$Ru, $^{100}$Ru, $^{138}$Ba and $^{180}$Hf}
 	\begin{tabular}{r|cccccc}
	\hline
    \multicolumn{1}{c|}{$E_\nu$} &
    \multicolumn{5}{c}{Charged current cross section,~$\sigma$ (cm$^2$)} \\
            (MeV) & $^{92}$Zr$(\nu_e,e^-)^{92}$Nb  & $^{92}$Zr$(\nu_e,e^-p)^{92}$Zr & $^{92}$Zr$(\nu_e,e^-n)^{91}$Nb \\ 	
 	\hline
 	   0   &  0.00000000E+00   &   0.00000000E+00   &   0.00000000E+00 &&\\
       8   &  1.09959219E-42   &   5.47587107E-43   &   5.20403126E-45 &&\\
      16   &  2.25014331E-41   &   1.22824926E-41   &   1.25064181E-42 &&\\
      24   &  1.08407931E-40   &   6.67298484E-42   &   5.58604016E-42 &&\\
      32   &  2.78249619E-40   &   4.84525629E-42   &   5.62397084E-42 &&\\
      40   &  5.38772348E-40   &   6.14584557E-42   &   7.85259869E-42 &&\\
      48   &  8.21131516E-40   &   5.50500721E-42   &   7.89996859E-42 &&\\
      56   &  9.73133433E-40   &   4.18051021E-42   &   6.48922060E-42 &&\\
      64   &  1.18909971E-39   &   5.58049470E-42   &   8.66302950E-42 &&\\
      72   &  1.51267048E-39   &   5.48536212E-42   &   8.56701524E-42 &&\\
      80   &  1.85697485E-39   &   5.75426202E-42   &   9.00574682E-42 &&\\
    \toprule
          \nodata & $^{98}$Mo$(\nu_e,e^-)^{98}$Tc  & $^{98}$Mo$(\nu_e,e^-p)^{97}$Mo & $^{98}$Mo$(\nu_e,e^-n)^{97}$Tc
                  & $^{99}$Ru$(\bar{\nu}_e,e^+n)^{98}$Tc & $^{100}$Ru$(\bar{\nu}_e,e^+2n)^{98}$Tc \\
 	\hline
       0   &  0.00000000E+00   &   0.00000000E+00   &   0.00000000E+00   &   0.00000000E+00   &   0.00000000E+00  \\
       8   &  7.70989088E-43   &   7.70989088E-43   &   0.00000000E+00   &   2.22690899E-41   &   7.43796701E-52  \\
      16   &  1.73568589E-41   &   1.73568589E-41   &   1.27629938E-43   &   6.48833284E-40   &   2.09369501E-43  \\
      24   &  1.02710369E-40   &   1.02710369E-40   &   1.71459340E-42   &   1.21296066E-41   &   8.24406827E-41  \\
      32   &  2.85652085E-40   &   2.85652085E-40   &   2.37034880E-42   &   1.00332160E-42   &   1.13478523E-41  \\
      40   &  5.75412528E-40   &   5.75412528E-40   &   2.72131091E-42   &   3.49011336E-44   &   6.81952771E-43  \\
      48   &  9.10315416E-40   &   9.10315416E-40   &   4.04495970E-42   &   1.44088642E-45   &   4.85493147E-44  \\
      56   &  1.15041308E-39   &   1.15041308E-39   &   3.96748000E-42   &   6.64842557E-47   &   2.51775430E-45  \\
      64   &  1.30393579E-39   &   1.30393579E-39   &   3.67176526E-42   &   2.93322999E-48   &   1.42313551E-46  \\
      72   &  1.66758950E-39   &   1.66758950E-39   &   4.63981404E-42   &   1.77322096E-49   &   9.88014554E-48  \\
      80   &  2.05845775E-39   &   2.05845775E-39   &   5.32473753E-42   &   1.06916477E-49   &   7.77608897E-49  \\
    \toprule
          \nodata & $^{138}$Ba$(\nu_e,e^-)^{138}$La  & $^{138}$Ba$(\nu_e,e^-p)^{137}$Ba & $^{138}$Ba$(\nu_e,e^-n)^{137}$La \\
    \hline
       0   &   0.00000000E+00   &   0.00000000E+00   &   0.00000000E+00 \\
       8   &   7.35496305E-42   &   1.94262799E-42   &   6.62576330E-51 \\
      16   &   9.87540687E-41   &   9.53163886E-41   &   3.97731716E-43 \\
      24   &   3.98835084E-40   &   2.27041114E-41   &   1.53258451E-41 \\
      32   &   1.01378882E-39   &   1.66143076E-41   &   1.92037893E-41 \\
      40   &   1.89565025E-39   &   2.11978177E-41   &   2.56088381E-41 \\
      48   &   2.99592402E-39   &   1.83284468E-41   &   2.56488809E-41 \\
      56   &   4.27490906E-39   &   1.73415784E-41   &   2.45188862E-41 \\
      64   &   5.70163925E-39   &   2.50252082E-41   &   3.53790143E-41 \\
      72   &   7.25025285E-39   &   2.42380089E-41   &   3.52754018E-41 \\
      80   &   8.89389789E-39   &   2.57475634E-41   &   3.75243103E-41 \\
      \toprule
          \nodata & $^{180}$Hf$(\nu_e,e^-)^{180}$Ta  & $^{180}$Hf$(\nu_e,e^-p)^{179}$Hf & $^{180}$Hf$(\nu_e,e^-n)^{179}$Ta \\ 	
      \hline
       0   &   0.00000000E+00   &   0.00000000E+00   &   0.00000000E+00 \\
       8   &   7.75463639E-42   &   7.44290806E-42   &   1.84915780E-51 \\
      16   &   8.82760000E-41   &   5.14914969E-41   &   3.63902374E-44 \\
      24   &   4.13377695E-40   &   4.43942334E-41   &   2.95886195E-42 \\
      32   &   1.31540943E-39   &   5.11056133E-41   &   8.45900752E-42 \\
      40   &   2.63588344E-39   &   6.00408430E-41   &   1.18186695E-41 \\
      48   &   4.22215839E-39   &   6.19622075E-41   &   1.41993471E-41 \\
      56   &   5.96889593E-39   &   8.04424856E-41   &   1.88733180E-41 \\
      64   &   7.81247822E-39   &   9.66390861E-41   &   2.26243506E-41 \\
      72   &   9.71554309E-39   &   1.09833483E-40   &   2.61080109E-41 \\
      80   &   1.16463985E-38   &   1.29155821E-40   &   3.11110419E-41 \\
    \hline
    \end{tabular}
    \label{tab_main_CC reactions}
	\end{table*}
\clearpage

\section{Main reactions of nuclei in the $\nu$-process} \label{sec:app_d}
We tabulate the main reactions for the production of the nuclide $^{7}$Be, $^{7}$Li, $^{11}$B, $^{11}$C, $^{92}$Nb, $^{98}$Tc, $^{138}$La, and $^{180}$Ta at Table \ref{tab_main_reactions}, which results are obtained for NH using {\it KCK19} hydrodynamics at $M_r$ = 2.63, 3.25, 3.98, and 4.93 $M_{\odot}$, respectively. The reactions shown in this table contribute to the final yields by more than about 3\% of their main production reaction.

\begin{figure*}[h!]
        \centering
        {
        \includegraphics[width=8.5cm]{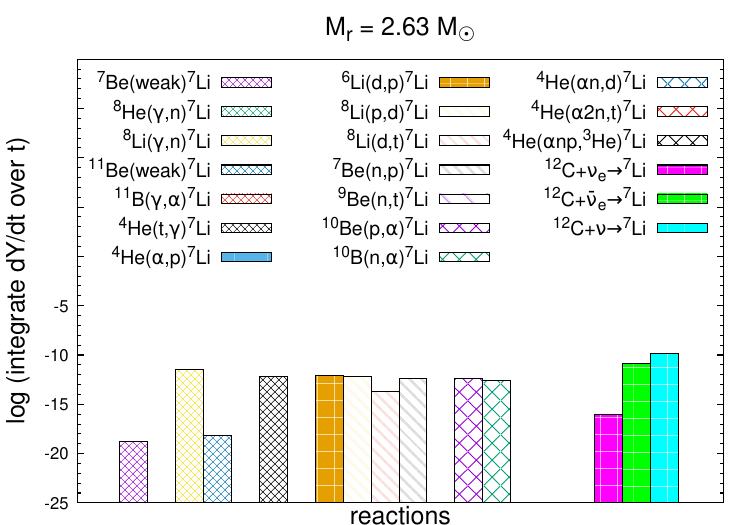}
        \includegraphics[width=8.5cm]{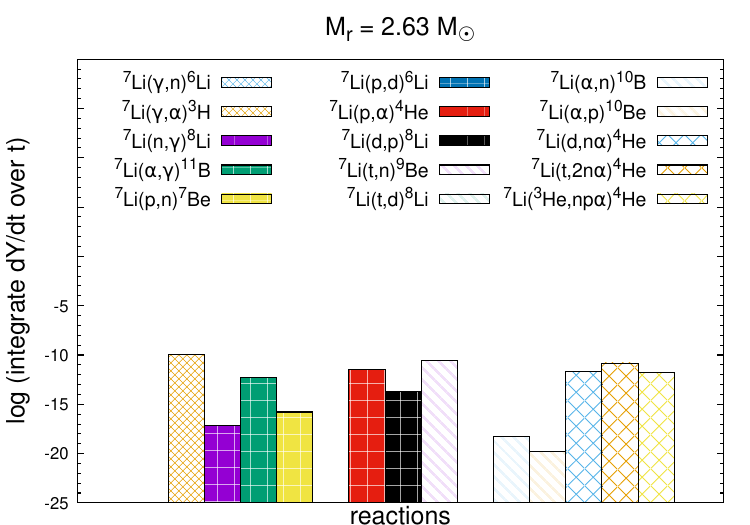}
        \includegraphics[width=8.5cm]{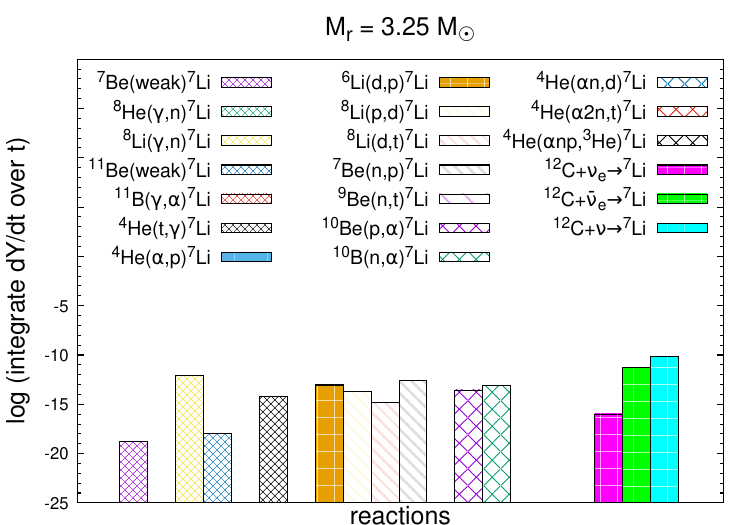}
        \includegraphics[width=8.5cm]{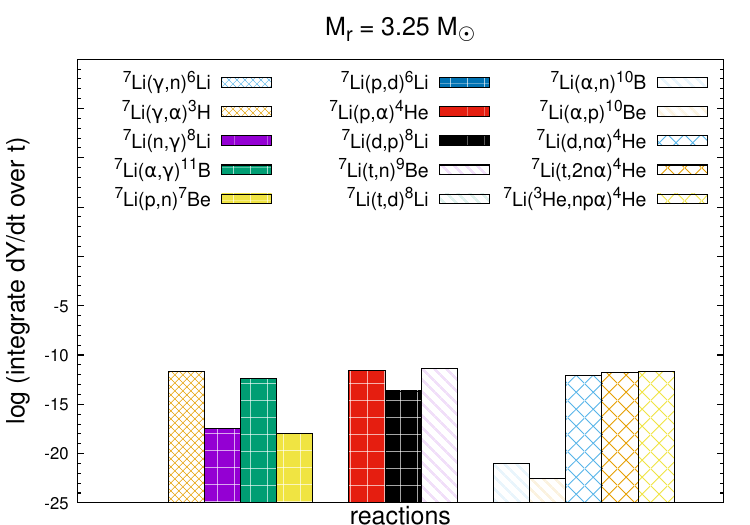}
        \includegraphics[width=8.5cm]{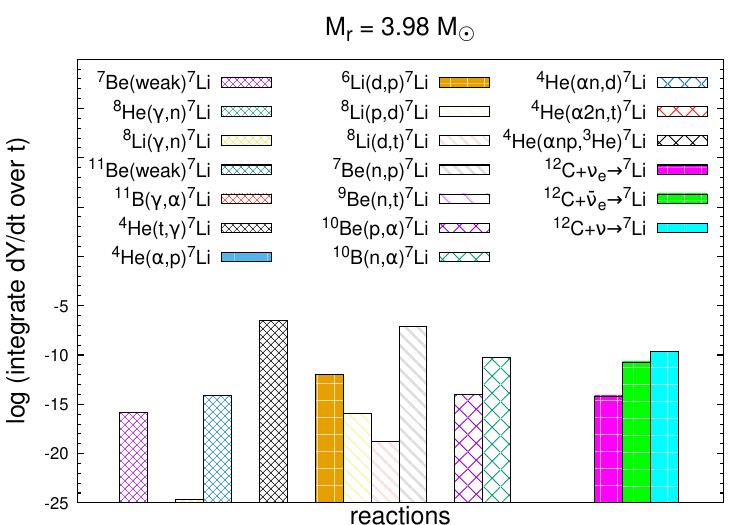}
        \includegraphics[width=8.5cm]{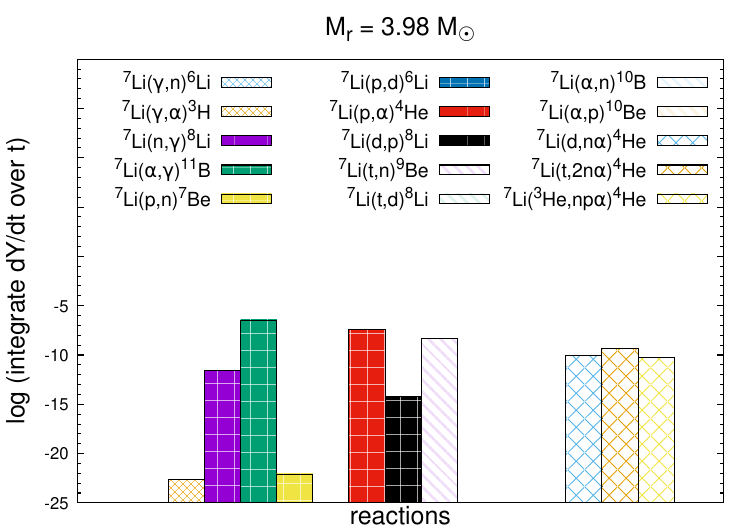}
        \includegraphics[width=8.5cm]{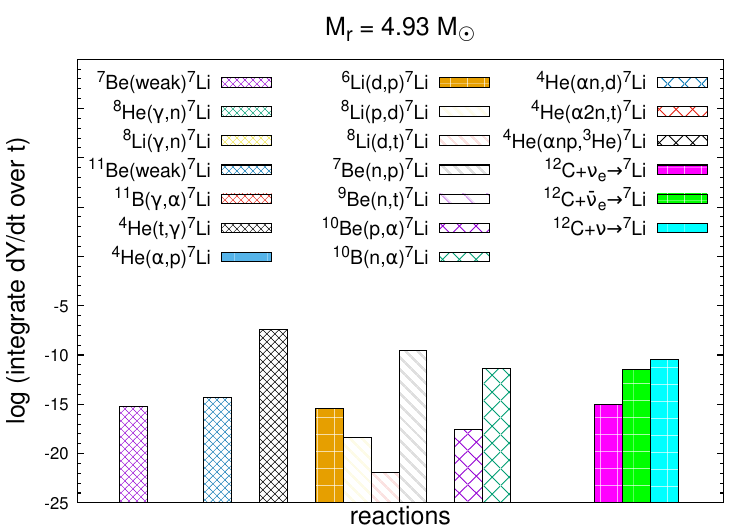}
        \includegraphics[width=8.5cm]{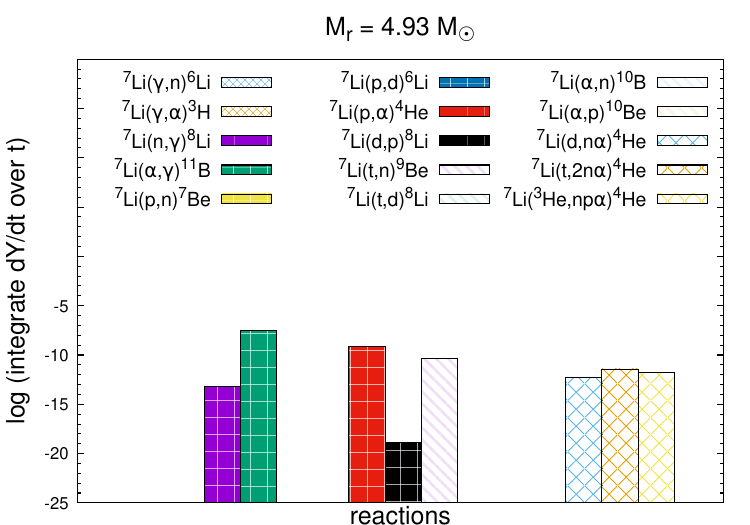}
        }

        \caption{Main nuclear reactions for $^7$Li. The 'weak' means the channels by the weak interaction such as beta-decays and electron-captures. Histograms of some nuclear reactions below a limit are not shown.}
        \label{fig_rea_7li}
    \end{figure*}

\begin{figure*}[h!]
        \centering
        {
        \includegraphics[width=8.5cm]{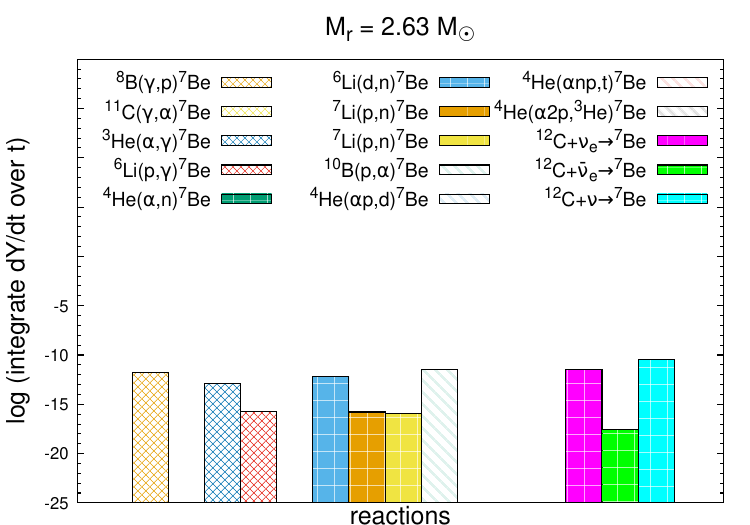}
        \includegraphics[width=8.5cm]{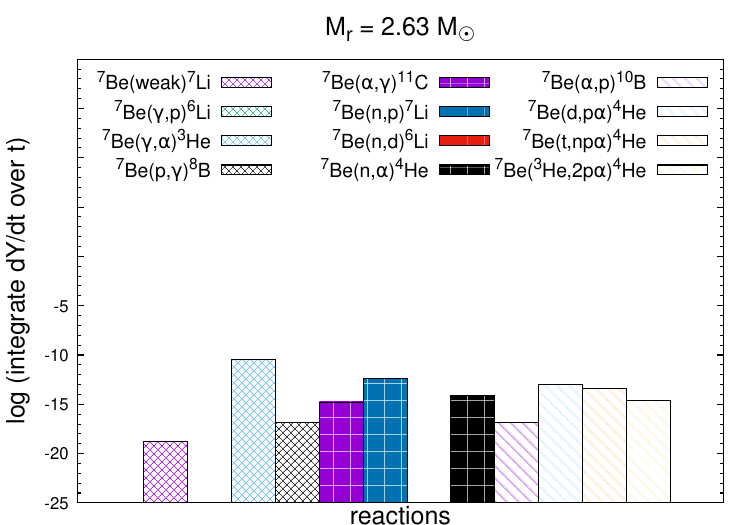}
        \includegraphics[width=8.5cm]{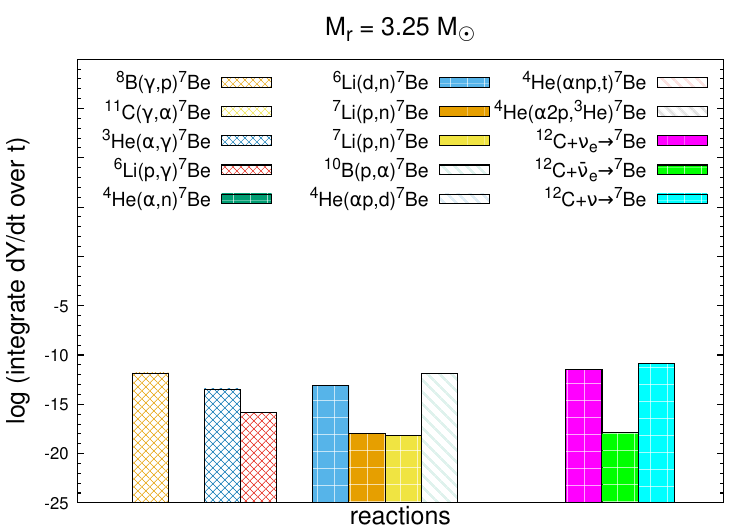}
        \includegraphics[width=8.5cm]{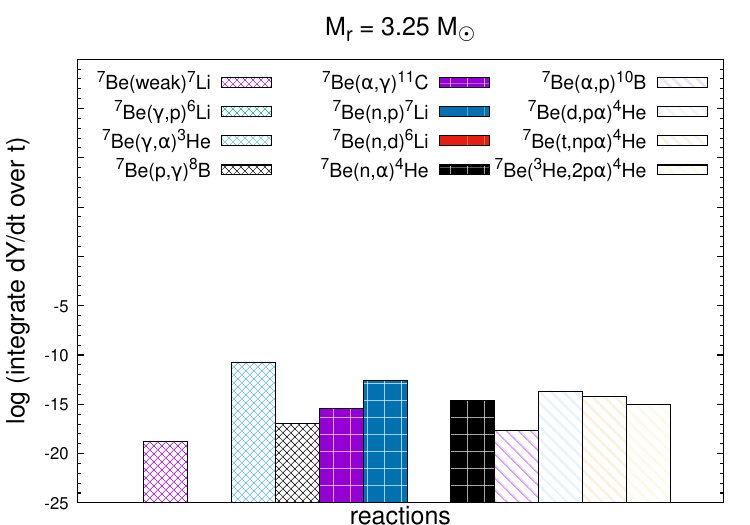}
        \includegraphics[width=8.5cm]{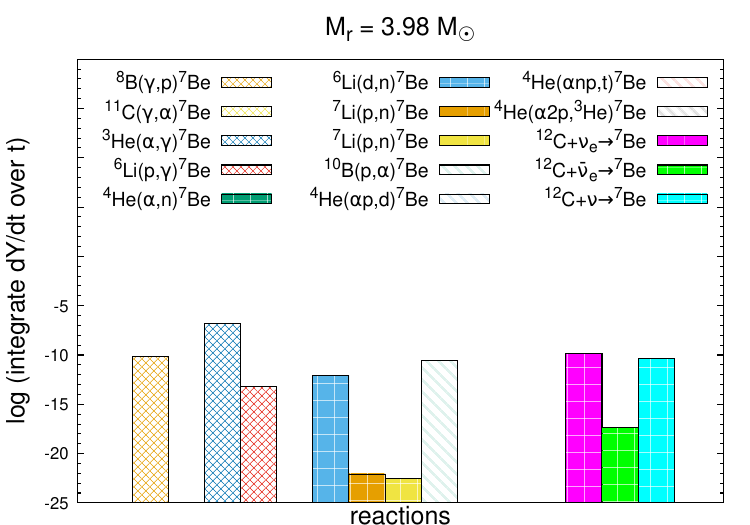}
        \includegraphics[width=8.5cm]{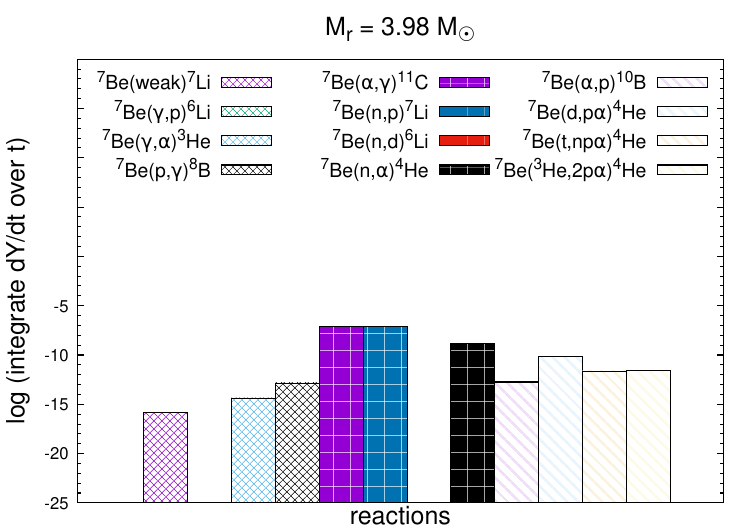}
        \includegraphics[width=8.5cm]{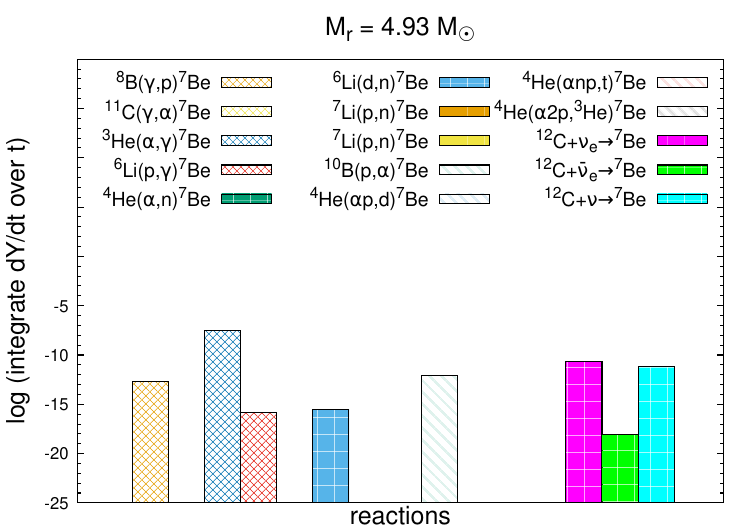}
        \includegraphics[width=8.5cm]{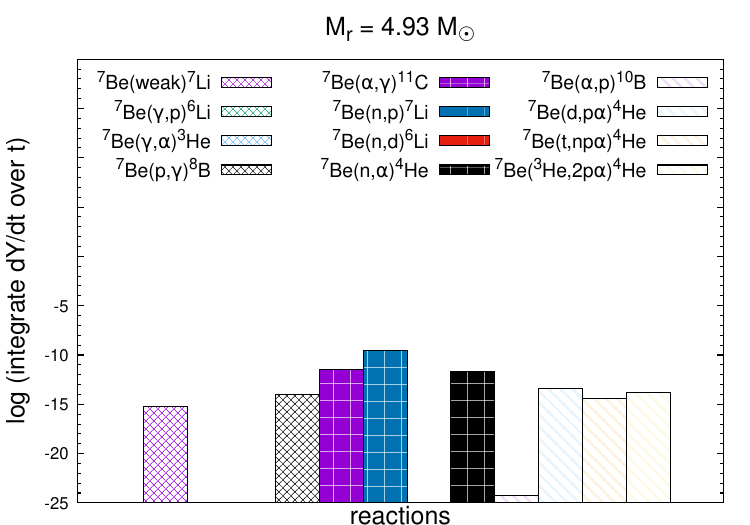}
        }
        \caption{Same as Figure \ref{fig_rea_7li}, but for $^7$Be.}
        \label{fig_rea_7be}
    \end{figure*}

\begin{figure*}[h!]
        \centering
        {
        \includegraphics[width=8.5cm]{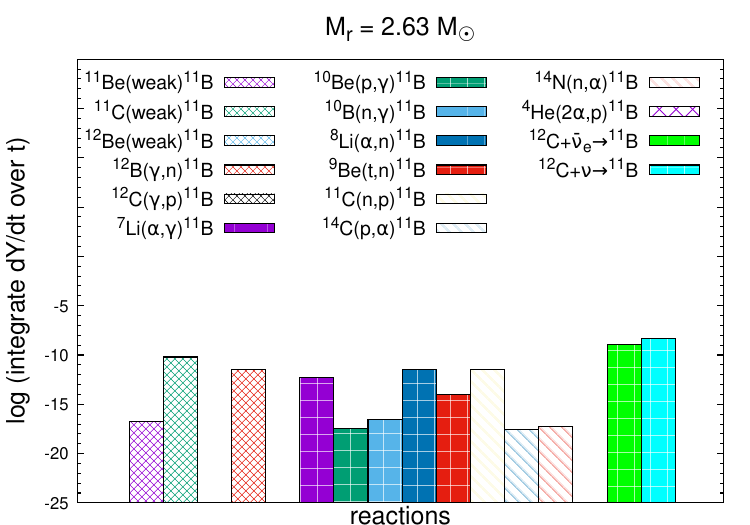}
        \includegraphics[width=8.5cm]{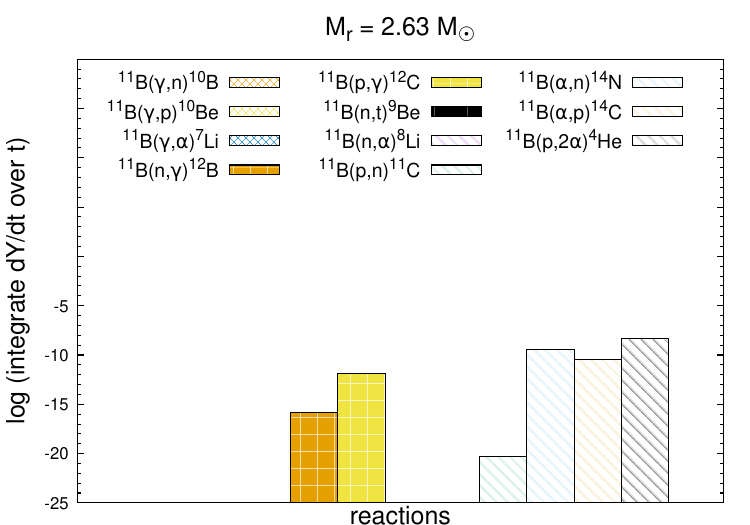}
        \includegraphics[width=8.5cm]{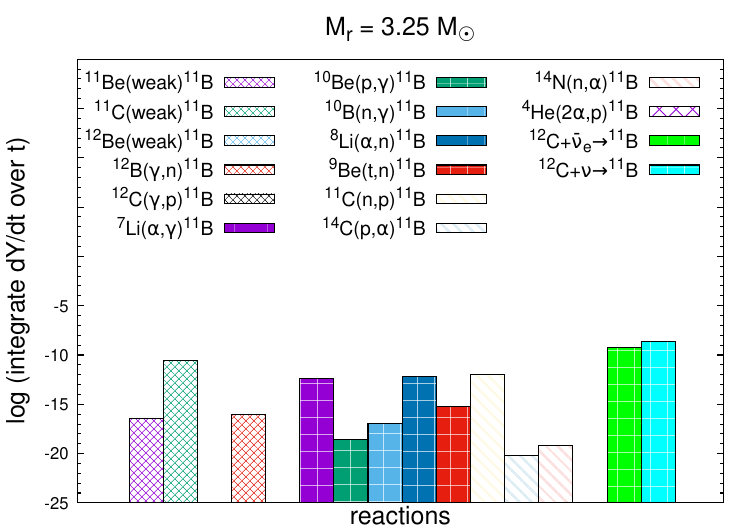}
        \includegraphics[width=8.5cm]{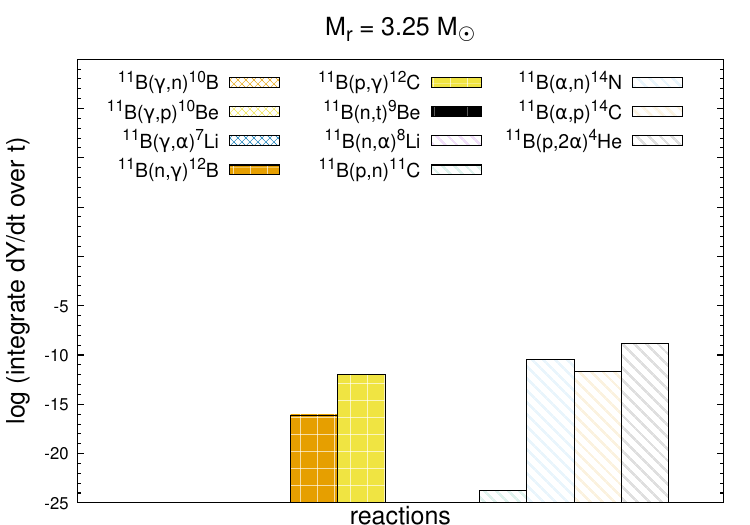}
        \includegraphics[width=8.5cm]{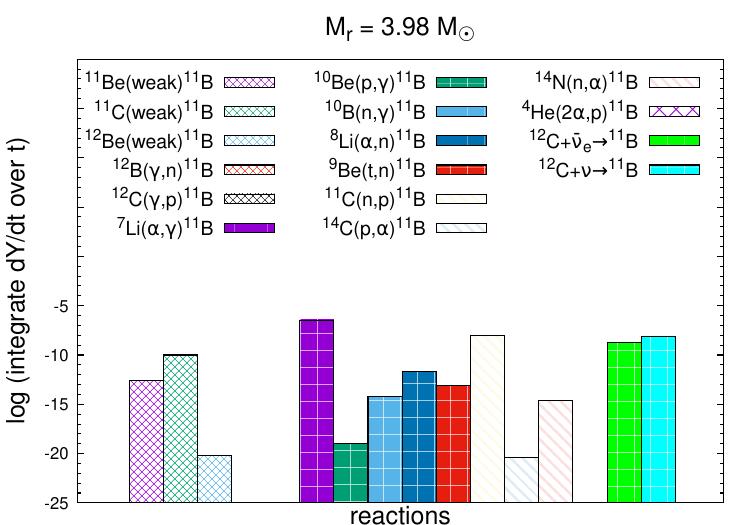}
        \includegraphics[width=8.5cm]{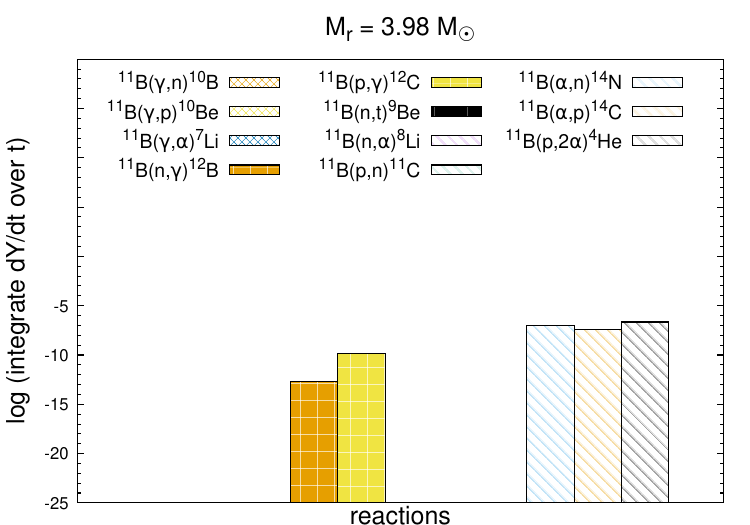}
        \includegraphics[width=8.5cm]{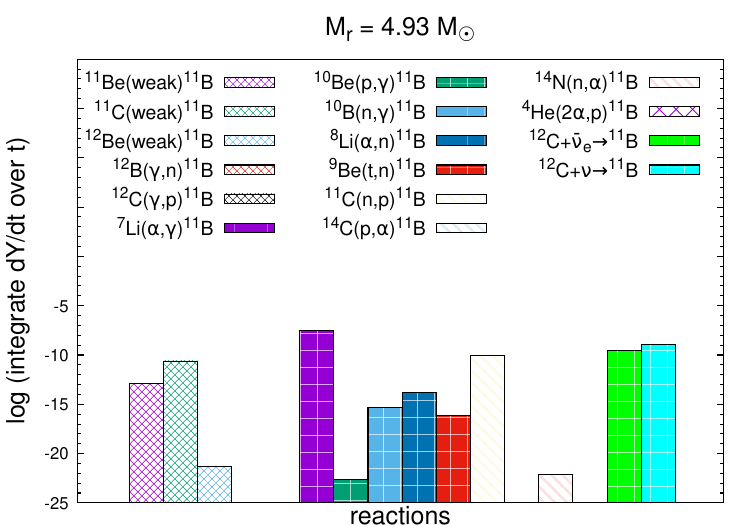}
        \includegraphics[width=8.5cm]{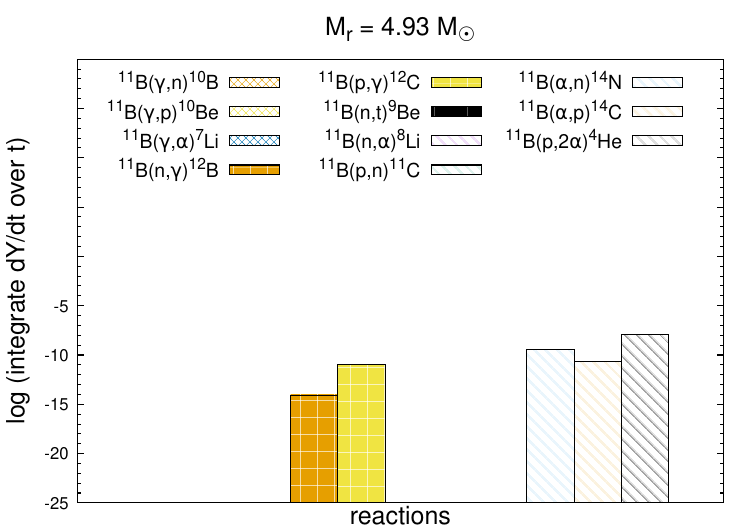}
        }
        \caption{Same as Figure \ref{fig_rea_7li}, but for $^{11}$B.}
        \label{fig_rea_11b}
    \end{figure*}

\begin{figure*}[h!]
        \centering
        {
        \includegraphics[width=8.5cm]{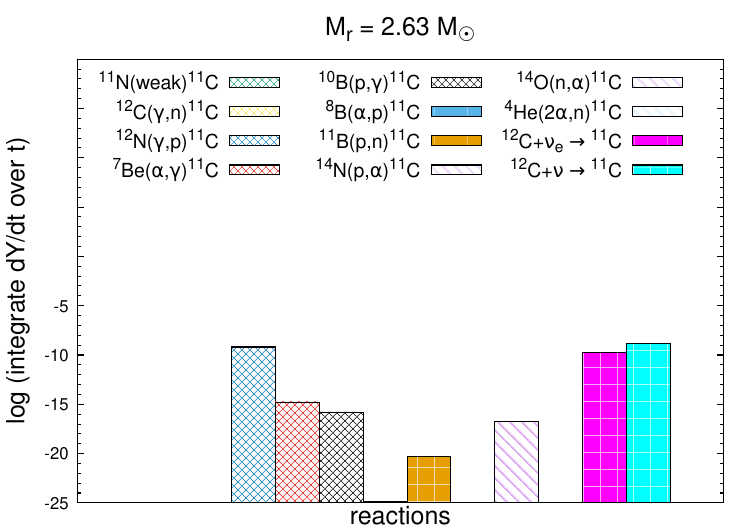}
        \includegraphics[width=8.5cm]{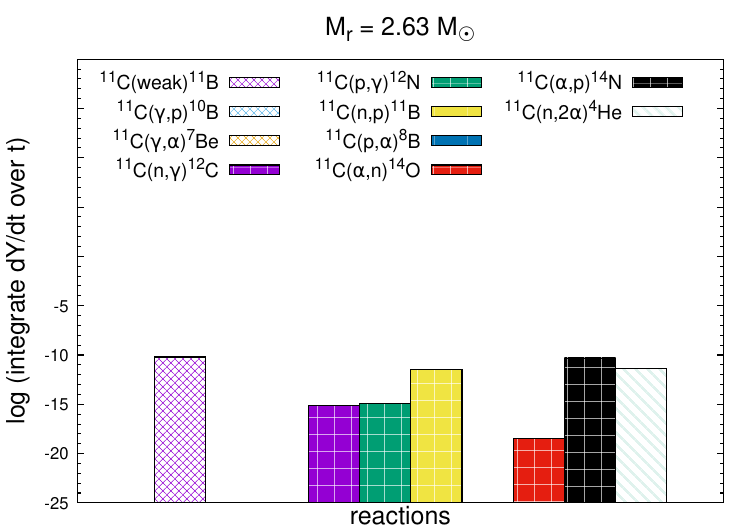}
        \includegraphics[width=8.5cm]{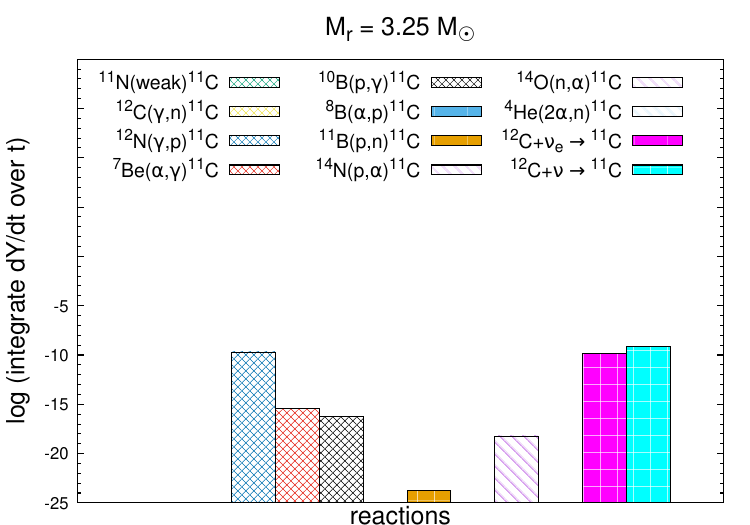}
        \includegraphics[width=8.5cm]{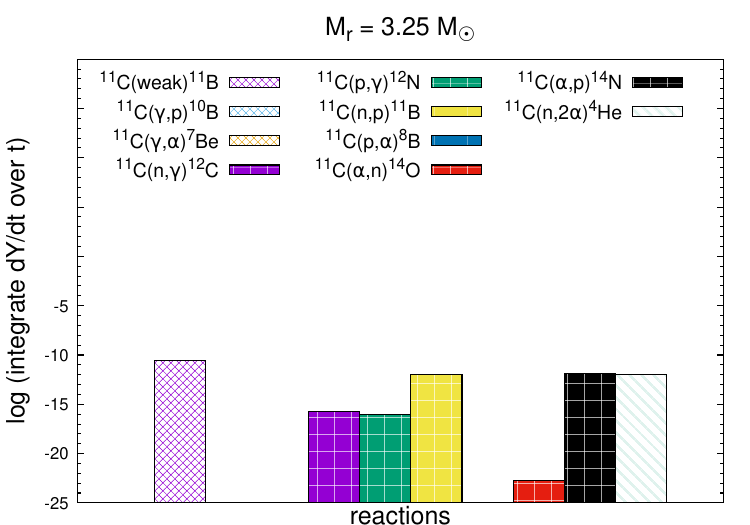}
        \includegraphics[width=8.5cm]{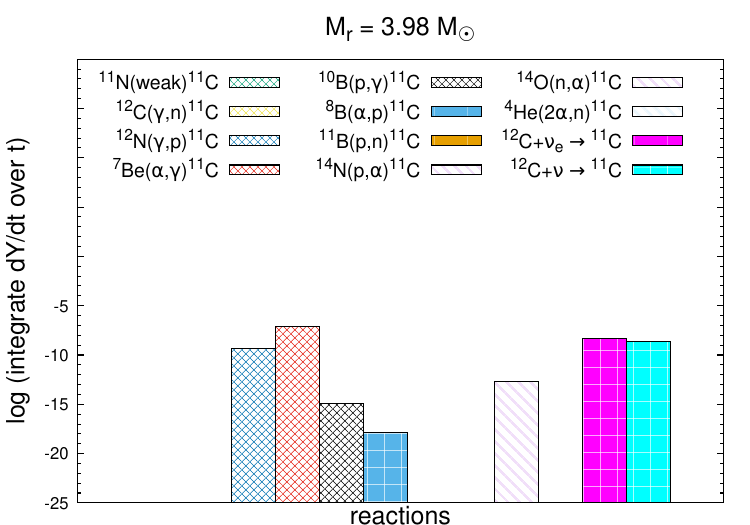}
        \includegraphics[width=8.5cm]{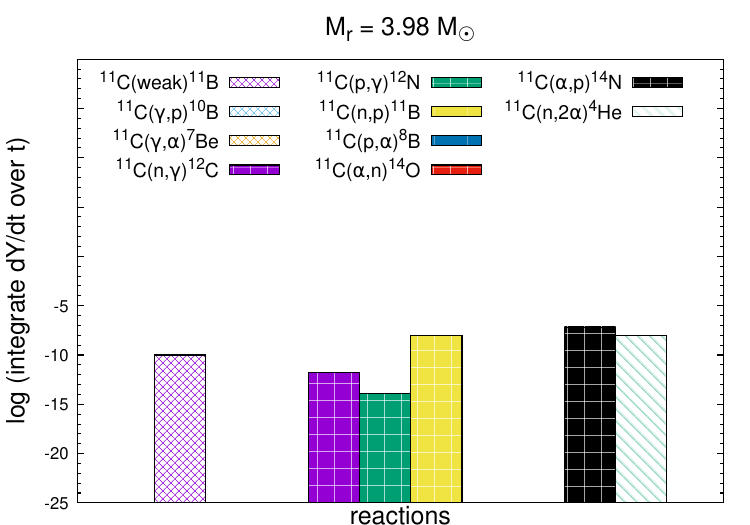}
        \includegraphics[width=8.5cm]{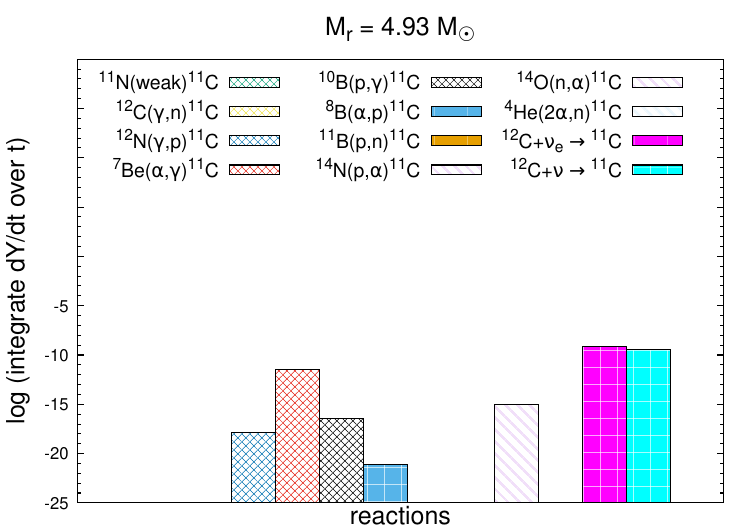}
        \includegraphics[width=8.5cm]{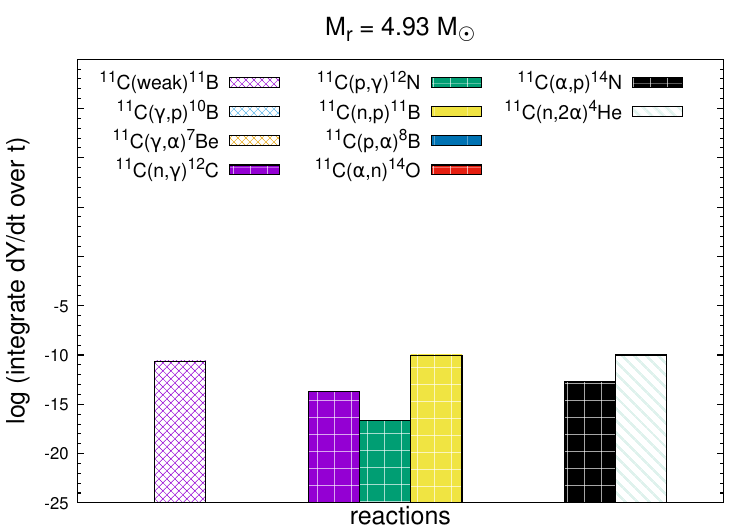}
        }
        \caption{Same as Figure \ref{fig_rea_7li}, but for $^{11}$C.}
        \label{fig_rea_11c}
    \end{figure*}

\begin{figure*}[h!]
        \centering
        {
        \includegraphics[width=8.5cm]{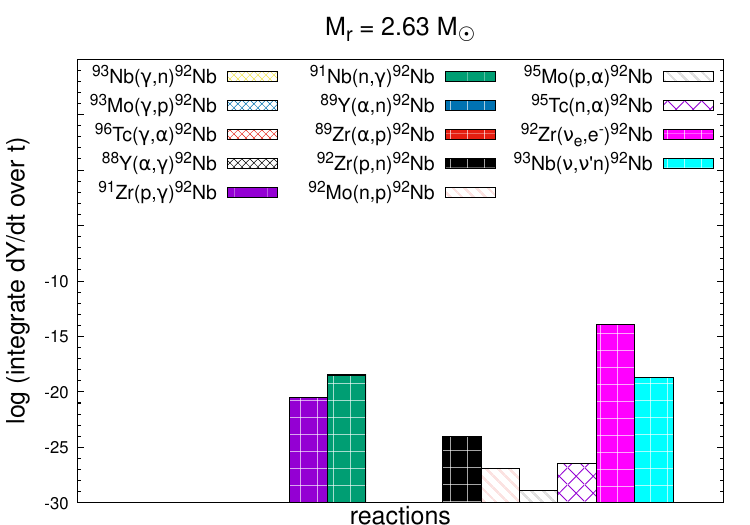}
        \includegraphics[width=8.5cm]{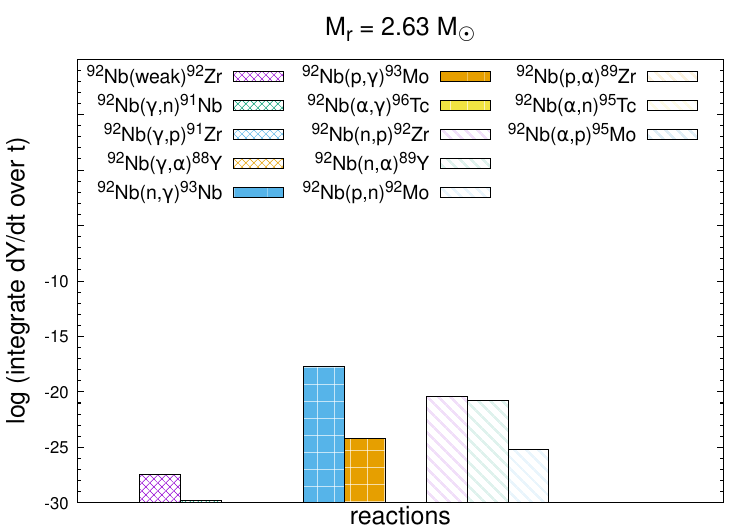}
        \includegraphics[width=8.5cm]{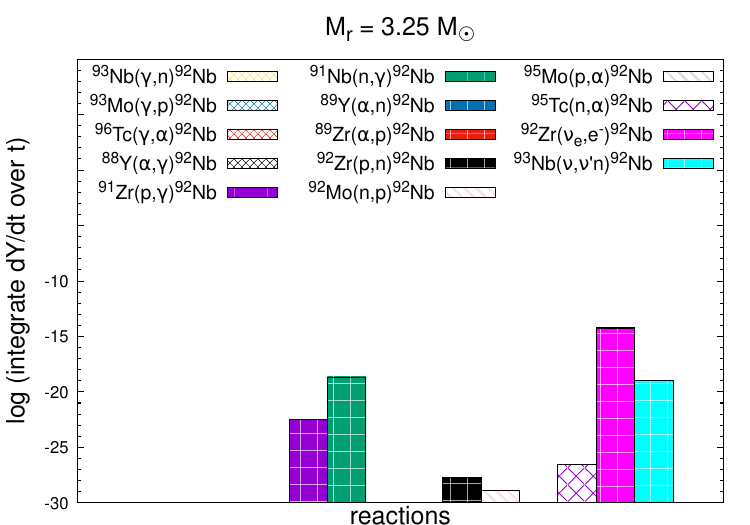}
        \includegraphics[width=8.5cm]{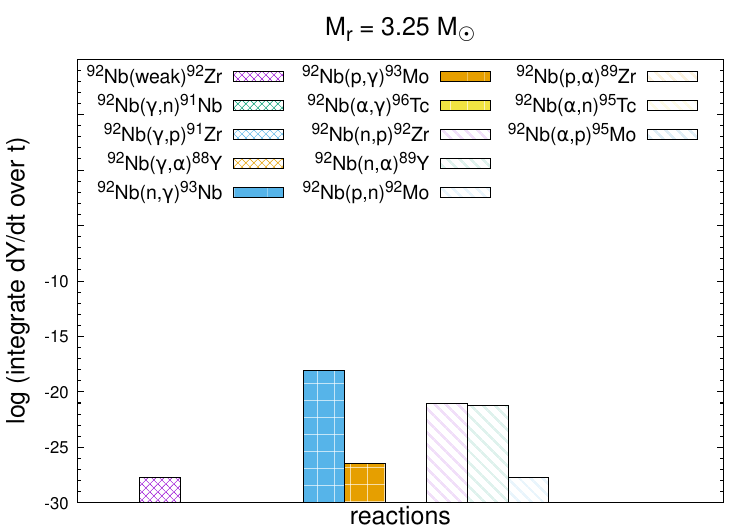}
        \includegraphics[width=8.5cm]{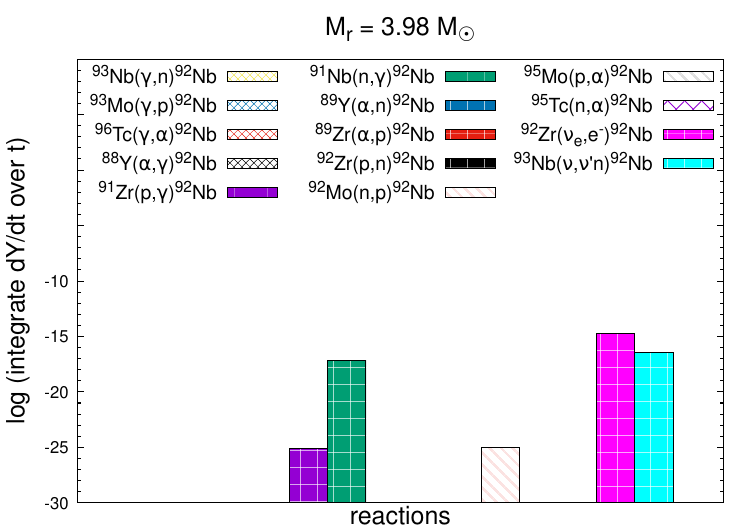}
        \includegraphics[width=8.5cm]{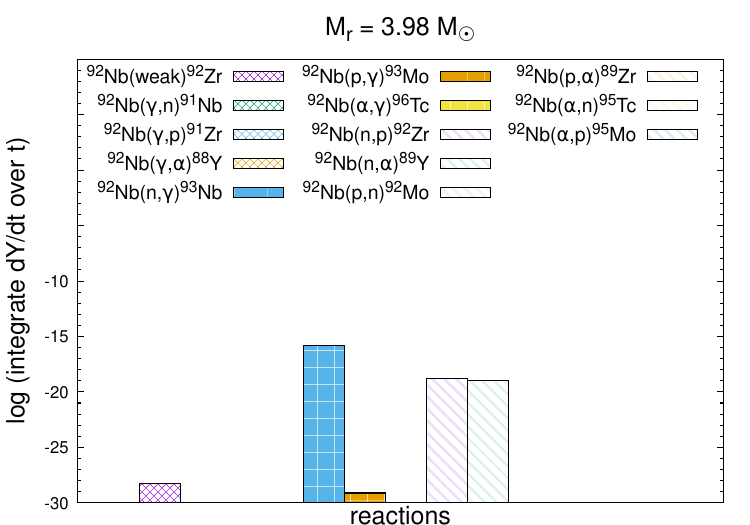}
        \includegraphics[width=8.5cm]{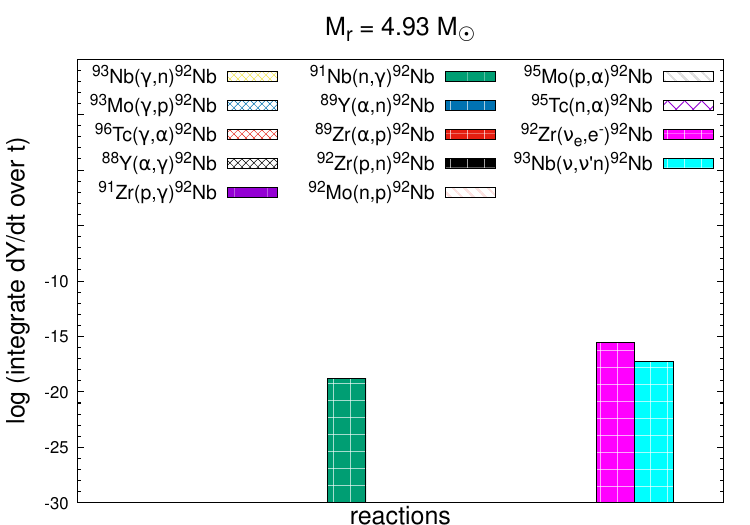}
        \includegraphics[width=8.5cm]{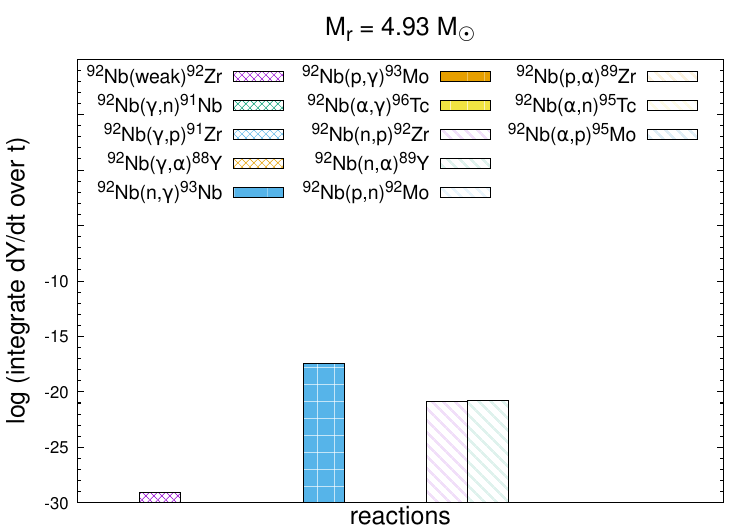}
        }
        \caption{Same as Figure \ref{fig_rea_7li}, but for $^{92}$Nb.}
        \label{fig_rea_92nb}
    \end{figure*}

\begin{figure*}[h!]
        \centering
        {
        \includegraphics[width=8.5cm]{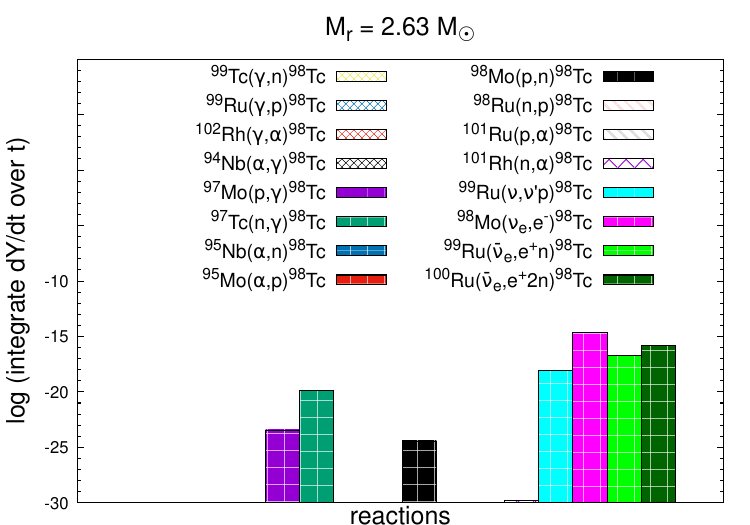}
        \includegraphics[width=8.5cm]{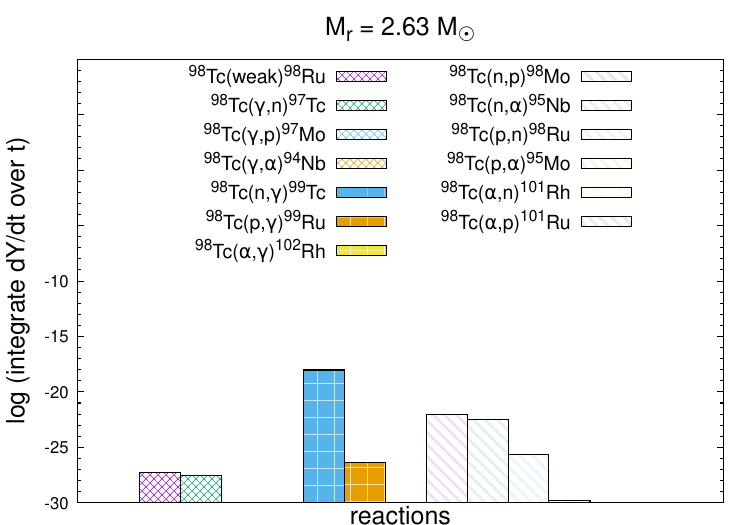}
        \includegraphics[width=8.5cm]{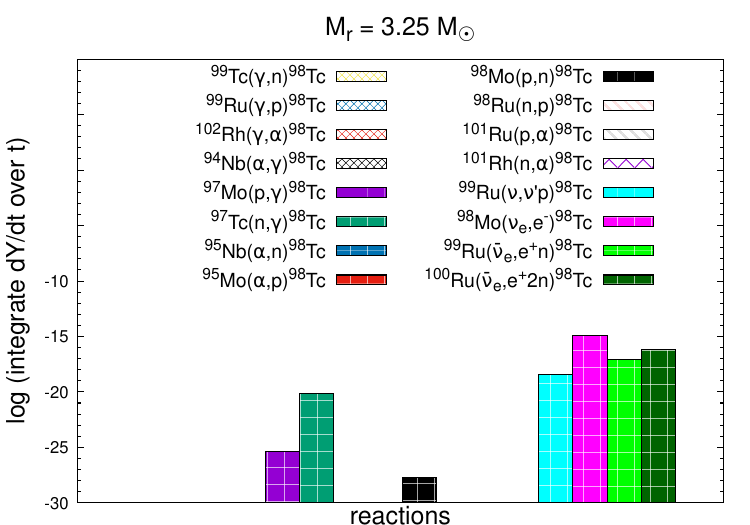}
        \includegraphics[width=8.5cm]{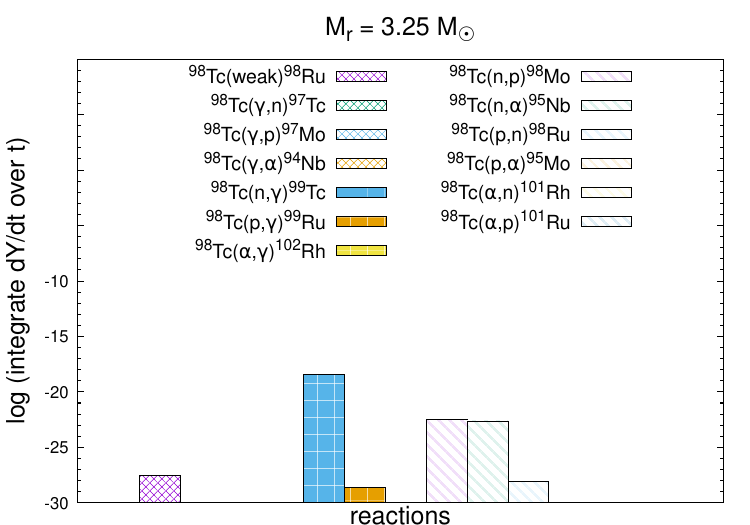}
        \includegraphics[width=8.5cm]{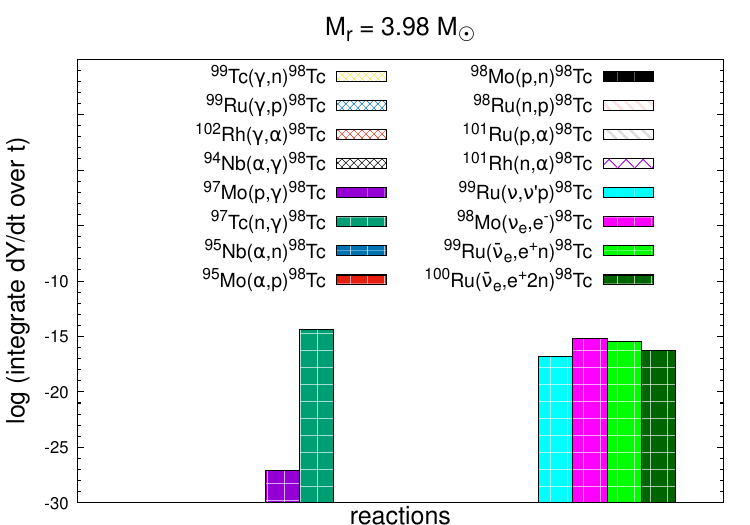}
        \includegraphics[width=8.5cm]{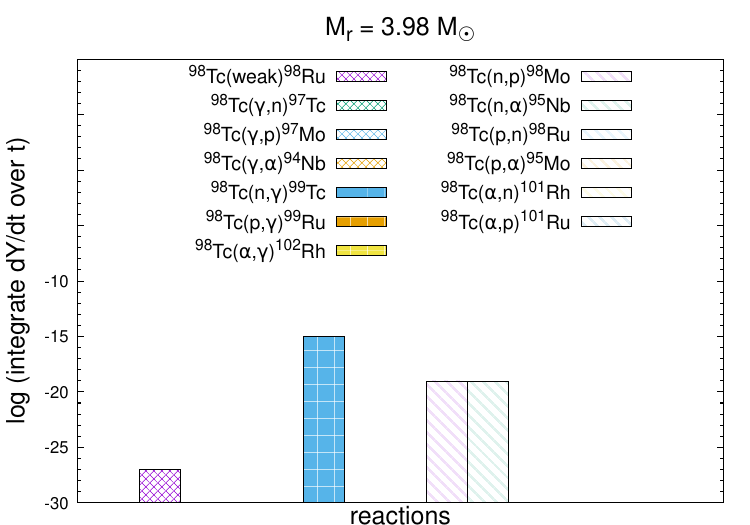}
        \includegraphics[width=8.5cm]{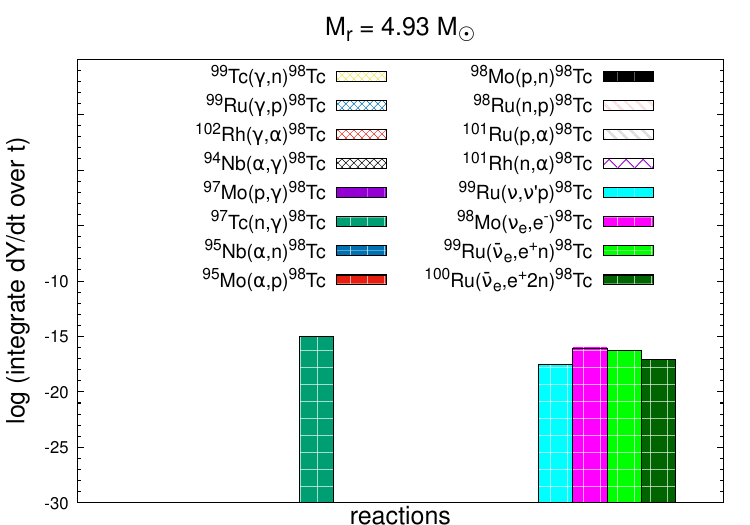}
        \includegraphics[width=8.5cm]{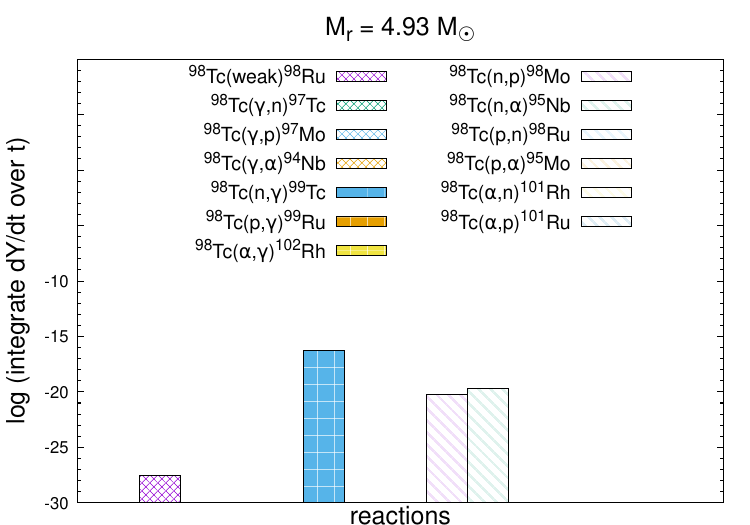}
        }
        \caption{Same as Figure \ref{fig_rea_7li}, but for $^{98}$Tc.}
        \label{fig_rea_98tc}
    \end{figure*}

\begin{figure*}[h!]
        \centering
        {
        \includegraphics[width=8.5cm]{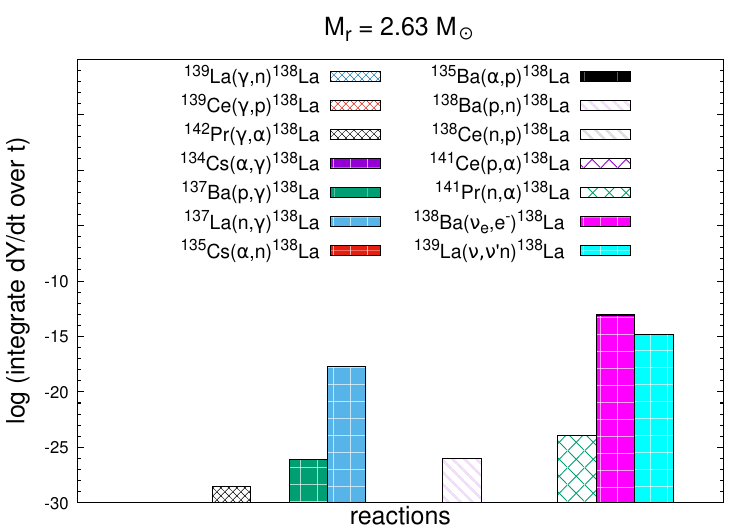}
        \includegraphics[width=8.5cm]{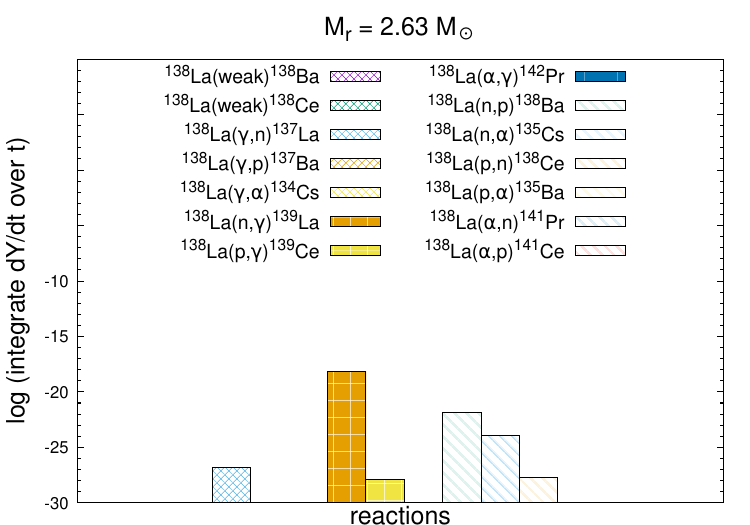}
        \includegraphics[width=8.5cm]{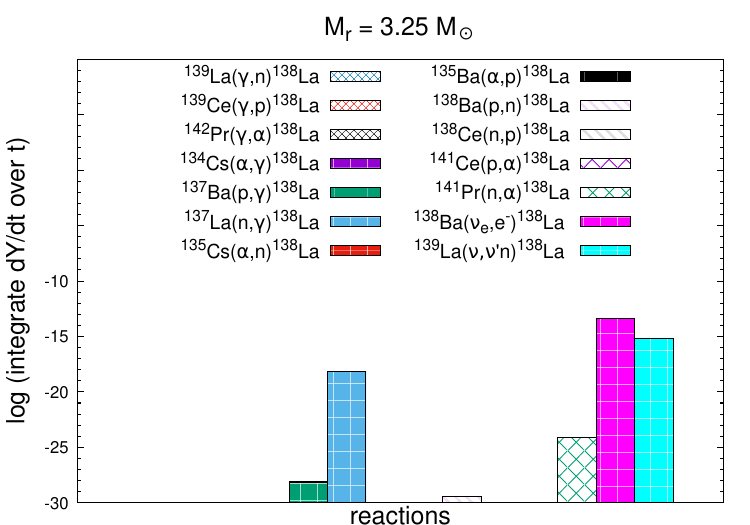}
        \includegraphics[width=8.5cm]{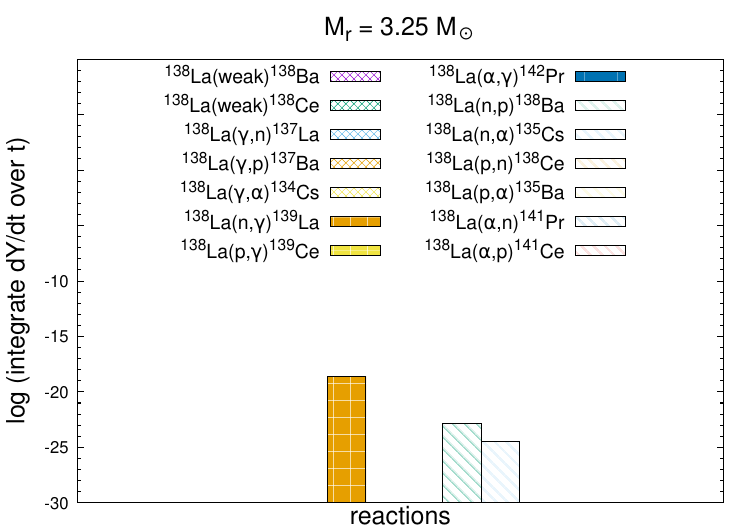}
        \includegraphics[width=8.5cm]{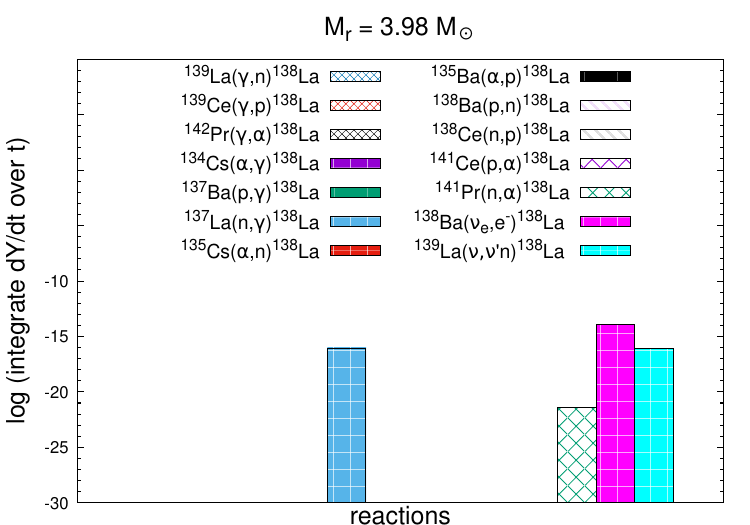}
        \includegraphics[width=8.5cm]{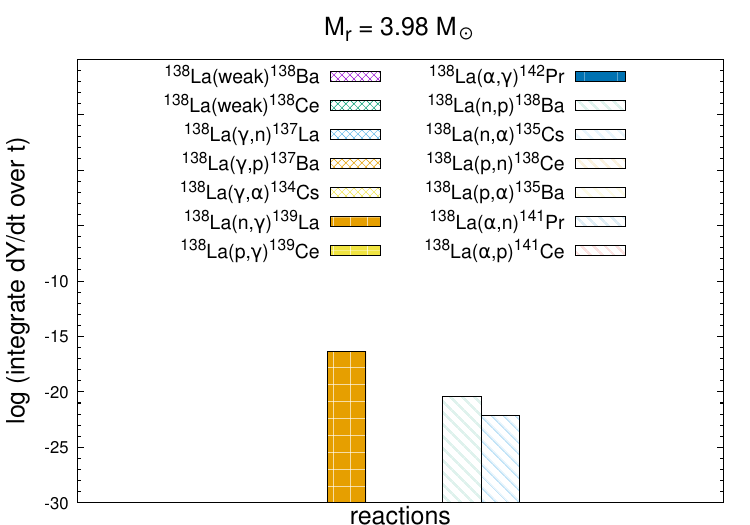}
        \includegraphics[width=8.5cm]{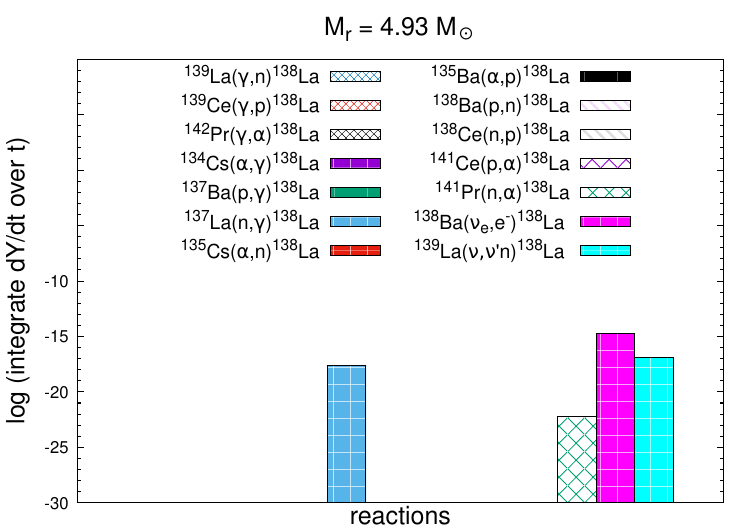}
        \includegraphics[width=8.5cm]{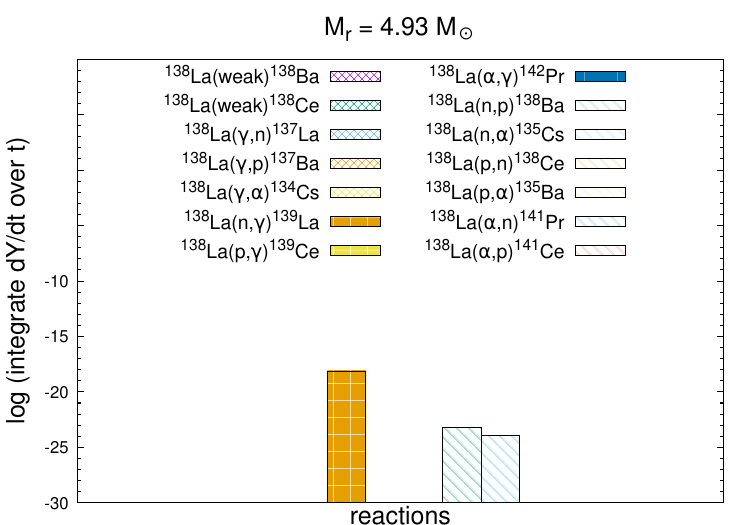}
        }
        \caption{Same as Figure \ref{fig_rea_7li}, but for $^{138}$La.}
        \label{fig_rea_138la}
    \end{figure*}

\begin{figure*}[h!]
        \centering
        {
        \includegraphics[width=8.5cm]{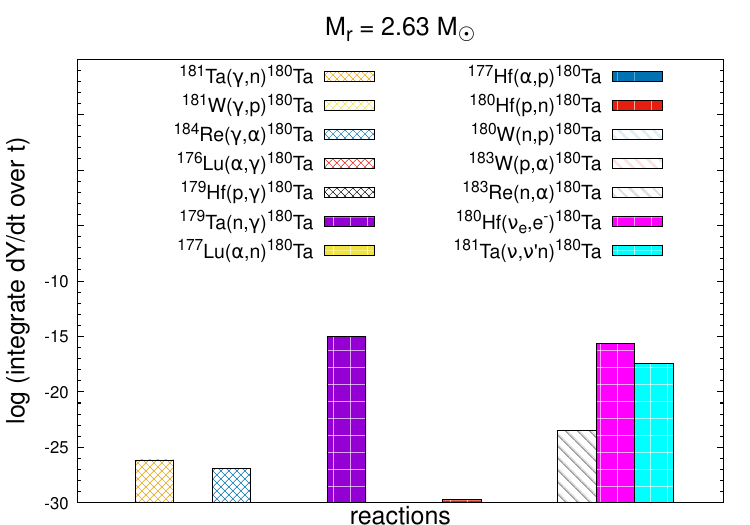}
        \includegraphics[width=8.5cm]{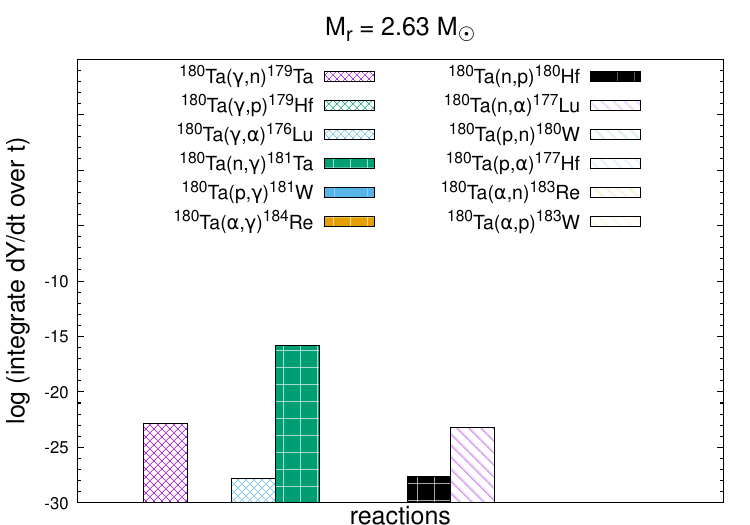}
        \includegraphics[width=8.5cm]{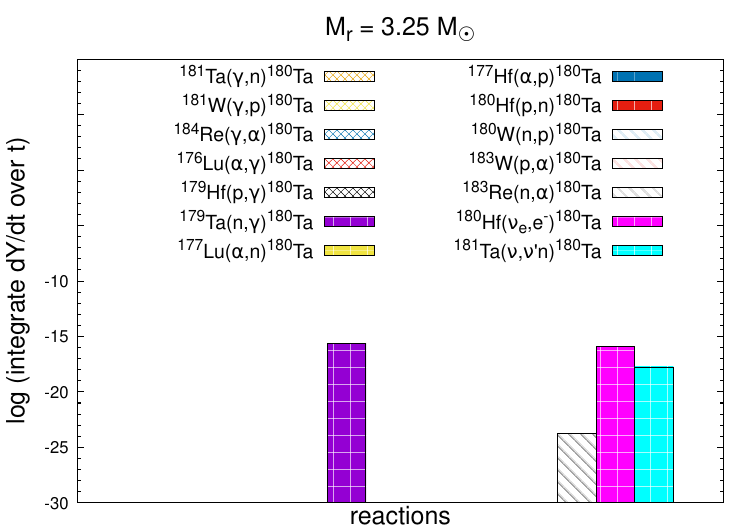}
        \includegraphics[width=8.5cm]{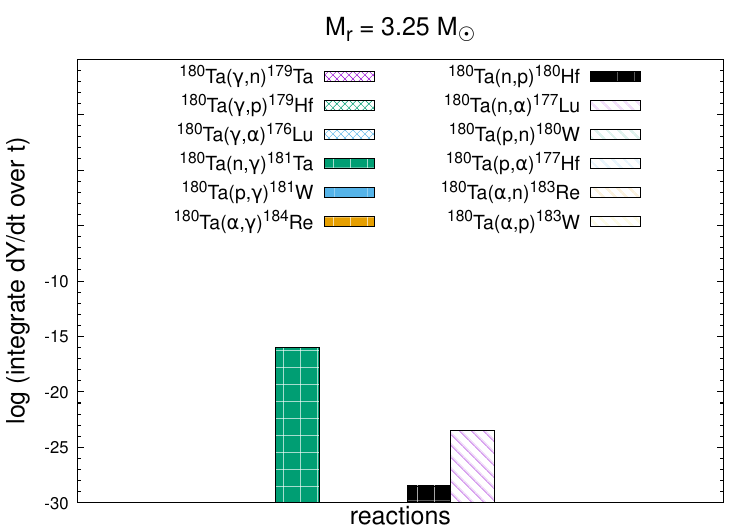}
        \includegraphics[width=8.5cm]{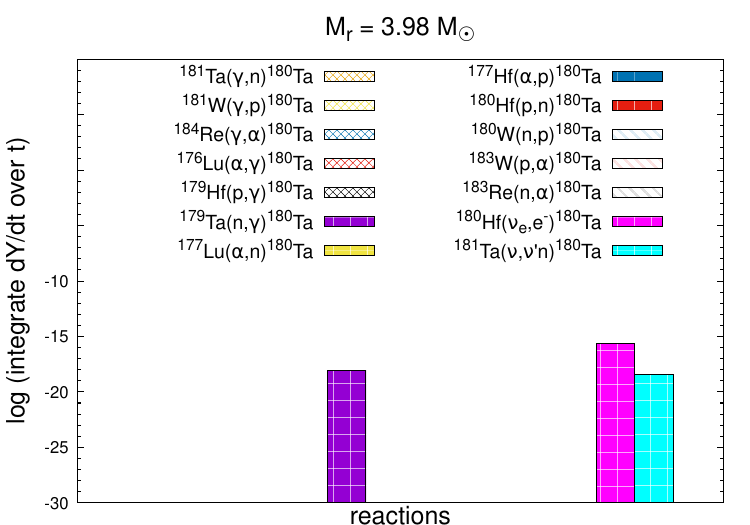}
        \includegraphics[width=8.5cm]{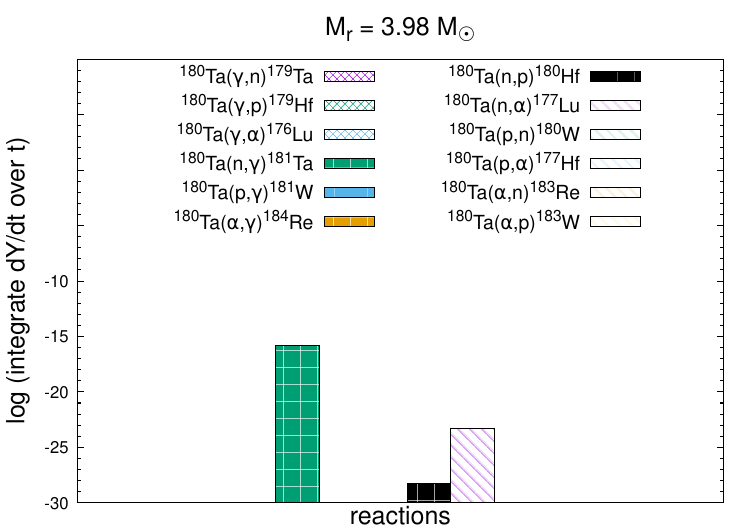}
        \includegraphics[width=8.5cm]{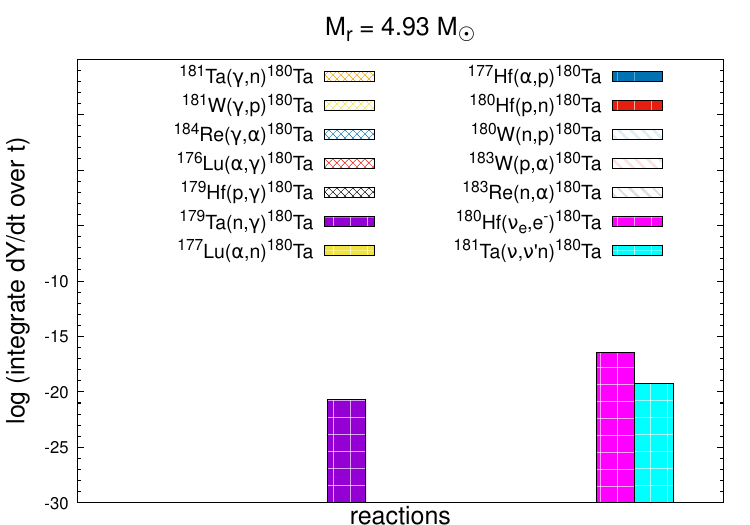}
        \includegraphics[width=8.5cm]{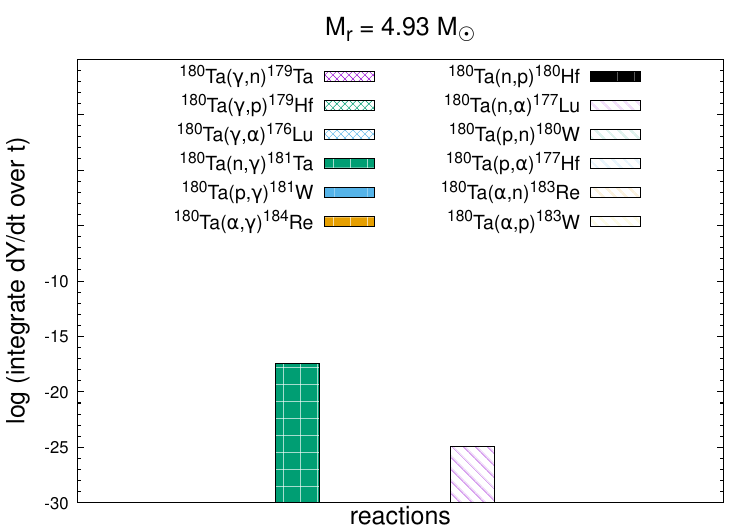}
        }
        \caption{Same as Figure \ref{fig_rea_7li}, but for $^{180}$Ta.}
        \label{fig_rea_180ta}
    \end{figure*}

Also we visualize all the main reactions relevant to these elements as histograms in Figures \ref{fig_rea_7li}, \ref{fig_rea_7be}, \ref{fig_rea_11b}, \ref{fig_rea_11c}, \ref{fig_rea_92nb}, \ref{fig_rea_98tc}, \ref{fig_rea_138la}, and \ref{fig_rea_180ta}, respectively. The left and right sides of the figures are the production and destruction reactions, respectively, for the element synthesis. These histograms could be very useful for grasping the contributions of the main nuclear reactions as well as that of the neutrino-induced reactions.
\begin{table*}[h!b]
	\centering 		
 	\begin{tabular}{r|ccccc}
	\hline
	\hline
    \multicolumn{1}{c|}{} &
    \multicolumn{4}{c}{The main production reactions in the NH case} \\
    $M_r/M_\odot$   &  2.63  & 3.25 & 3.98 & 4.93  \\
    \hline
    {$^{7}$Li} & {$^{12}$C + $\bar{\nu}_e$} & {$^{12}$C + $\bar{\nu}_e$} & {$^{4}$He$(t,\gamma)^{7}$Li} & {$^{4}$He$(t,\gamma)^{7}$Li} \\
    \multicolumn{1}{c|}{} & {$^{12}$C + $\nu$} & {$^{12}$C + $\nu$} & {$^{7}$Be$(n,p)^{7}$Li} & {} \\
    \hline
    {$^{7}$Be} & {$^{12}$C + $\nu$} & {$^{12}$C + $\nu$} & {$^{3}$He$(\alpha,\gamma)^{7}$Be} & {$^{3}$He$(\alpha,\gamma)^{7}$Be} \\
    \multicolumn{1}{c|}{} & {$^{12}$C + $\nu_e$} & {$^{12}$C + $\nu_e$} & {} & {} \\
    \multicolumn{1}{c|}{} & {$^{10}$B$(p,\alpha)^{7}$Be} & {$^{10}$B$(p,\alpha)$Be} & {} & {} \\
    \hline
    {$^{11}$B} & {$^{12}$C + $\nu$} & {$^{12}$C + $\nu$} & {$^{7}$Li$(\alpha,\gamma)^{11}$B} & {$^{7}$Li$(\alpha,\gamma)^{11}$B} \\
    \multicolumn{1}{c|}{} & {$^{12}$C + $\bar{\nu}_e$} & {$^{12}$C + $\bar{\nu}_e$} & {$^{11}$C$(n,p)^{11}$B} & {$^{12}$C + $\nu$} \\
    \multicolumn{1}{c|}{} & {} & {} & {$^{12}$C + $\nu$} & {} \\
    \hline
    {$^{11}$C} & {$^{12}$C + $\nu$} & {$^{12}$C + $\nu$} & {$^{7}$Be$(\alpha,\gamma)^{11}$C} & {$^{12}$C + $\nu_e$} \\
    \multicolumn{1}{c|}{} & {$^{12}$N$(\gamma,p)^{11}$C} & {$^{12}$N$(\gamma,p)^{11}$C} & {$^{12}$C + $\nu_e$} & {$^{12}$C + $\nu$} \\
    \multicolumn{1}{c|}{} & {$^{12}$C + $\nu_e$} & {$^{12}$C + $\nu_e$} & {$^{12}$C + $\nu$} & {} \\
    \hline
    {$^{92}$Nb} & {$^{92}$Zr$(\nu_e,e^-)^{92}$Nb} & {$^{92}$Zr$(\nu_e,e^-)^{92}$Nb} & {$^{92}$Zr$(\nu_e,e^-)^{92}$Nb} & {$^{92}$Zr$(\nu_e,e^-)^{92}$Nb} \\
    \hline
    {$^{98}$Tc} & {$^{98}$Mo$(\nu_e,e^-)^{98}$Tc} & {$^{98}$Mo$(\nu_e,e^-)^{98}$Tc} & {$^{97}$Tc$(n,\gamma)^{98}$Tc} & {$^{97}$Tc$(n,\gamma)^{98}$Tc} \\
    \multicolumn{1}{c|}{} & {$^{100}$Ru$(\bar{\nu}_e,e^+2n)^{98}$Tc} & {$^{100}$Ru$(\bar{\nu}_e,e^+2n)^{98}$Tc} & {$^{98}$Mo$(\nu_e,e^-)^{98}$Tc} & {$^{98}$Mo$(\nu_e,e^-)^{98}$Tc} \\
    \multicolumn{1}{c|}{} & {} & {} & {$^{99}$Ru$(\bar{\nu}_e,e^+n)^{98}$Tc} & {$^{99}$Ru$(\bar{\nu}_e,e^+n)^{98}$Tc} \\
    \hline
    {$^{138}$La} & {$^{138}$Ba$(\nu_e,e^-)^{138}$La} & {$^{138}$Ba$(\nu_e,e^-)^{138}$La} & {$^{138}$Ba$(\nu_e,e^-)^{138}$La} & {$^{138}$Ba$(\nu_e,e^-)^{138}$La} \\
    \hline
    {$^{180}$Ta} & {$^{179}$Ta$(n,\gamma)^{180}$Ta} & {$^{179}$Ta$(n,\gamma)^{180}$Ta} & {$^{180}$Hf$(\nu_e,e^-)^{180}$Ta} & {$^{180}$Hf$(\nu_e,e^-)^{180}$Ta} \\
    \multicolumn{1}{c|}{} & {$^{180}$Hf$(\nu_e,e^-)^{180}$Ta} & {$^{180}$Hf$(\nu_e,e^-)^{180}$Ta} & {} & {} \\
	\hline
	\hline
	\end{tabular}
    \caption{Main nuclear reactions at each mass coordinate for the nuclei considered in this work. They are deduced from the integration of respective reaction rates over time to $t \sim 50 $ s. Detailed reactions of $^{12}$C + $\nu_e$ for $^{11}$C and $^{7}$Be are shown at Table 3.}
    \label{tab_main_reactions}
\end{table*}

\clearpage

\section{Production factors normalized by different stable nuclei} \label{sec:app_e}
Here we present the production factors of $^{7}$Li, $^{11}$B, $^{138}$La and $^{180}$Ta normalized to $^{16}$O, $^{24}$Mg and $^{28}$Si, respectively. See the details in Section \ref{sec:5.5}.
    \begin{table*}[t]
        \centering
        \caption{The production factor normalized to $^{16}$O}
        \begin{tabular}{r|ccccc}
        \hline
        \multicolumn{1}{c|}{} &
        \multicolumn{5}{c}{The production factor $10^{[{\rm i}/^{16}{\rm O}]}$ in the region $M_r =1.6$--6 $M_\odot$ } \\
        \cline{2-6}
        \multicolumn{1}{c|}{} &
        \multicolumn{1}{c|}{MH} & i =  {$^{7}$Li} & {$^{11}$B} & {$^{138}$La} & {$^{180}$Ta}   \\
        \hline
        \multicolumn{1}{c|}{FD EQ} &
        \multicolumn{1}{c|}{NH} &
            {2.90$\times 10^{-1}$} & {7.72$\times 10^{-1}$} & {9.9$\times 10^{-2}$} & {3.037} \\
        \multicolumn{1}{c|}{({\it HKC18})} &
        \multicolumn{1}{c|}{IH} &
            {1.38$\times 10^{-1}$} & {8.46$\times 10^{-1}$} & {9.6$\times 10^{-2}$} & {3.028} \\
        \hline
        \multicolumn{1}{c|}{FD EQ} &
        \multicolumn{1}{c|}{NH} &
            {1.42$\times 10^{-1}$} & {4.33$\times 10^{-1}$} & {1.02$\times 10^{-1}$} & {3.379} \\
        \multicolumn{1}{c|}{({\it KCK19})} &
        \multicolumn{1}{c|}{IH} &
            {8.4$\times 10^{-2}$} & {3.73$\times 10^{-1}$} & {0.93$\times 10^{-1}$} & {3.366} \\
        \hline
        \multicolumn{1}{c|}{FD EQ Shock} &
        \multicolumn{1}{c|}{NH} &
            {1.20$\times 10^{-1}$} & {3.98$\times 10^{-1}$} & {1.13$\times 10^{-1}$} & {3.382} \\
        \multicolumn{1}{c|}{({\it KCK19})} &
        \multicolumn{1}{c|}{IH} &
            {8.5$\times 10^{-2}$} & {3.62$\times 10^{-1}$} & {9.4$\times 10^{-2}$} & {3.366} \\
        \hline
        \multicolumn{1}{c|}{SI EQ\footnote{Same as FD EQ ({\it KCK19}) NH result}} &
        \multicolumn{1}{c|}{NH} &
            {1.42$\times 10^{-1}$} & {4.33$\times 10^{-1}$} & {1.02$\times 10^{-1}$} & {3.379} \\
        \multicolumn{1}{c|}{({\it KCK19})} &
        \multicolumn{1}{c|}{IH} &
           {1.28$\times 10^{-1}$} & {6.29$\times 10^{-1}$} & {3.67$\times 10^{-1}$} & {3.493} \\
        \hline
        \multicolumn{1}{c|}{SI NEQ} &
        \multicolumn{1}{c|}{NH} &
            {1.18$\times 10^{-1}$} & {8.60$\times 10^{-1}$} & {4.11$\times 10^{-1}$} & {4.293} \\
        \multicolumn{1}{c|}{({\it KCK19})} &
        \multicolumn{1}{c|}{IH} &
            {1.06$\times 10^{-1}$} & {8.32$\times 10^{-1}$} & {3.06$\times 10^{-1}$} & {4.250} \\
        \hline
        \multicolumn{1}{c|}{FD NEQ} &
        \multicolumn{1}{c|}{NH} &
            {1.00$\times 10^{-1}$} & {9.97$\times 10^{-1}$} & {6.25$\times 10^{-1}$} & {4.386} \\
        \multicolumn{1}{c|}{({\it KCK19})} &
        \multicolumn{1}{c|}{IH} &
		  {1.41$\times 10^{-1}$} & {9.83$\times 10^{-1}$} & {6.47$\times 10^{-1}$} & {4.412} \\
        \hline
        \multicolumn{1}{c|}{SI NEQ \cite{2020ApJ...891L..24K}} &
        \multicolumn{1}{c|}{NH} &
            {2.33$\times 10^{-1}$} & {1.564}              & {4.18$\times 10^{-1}$} & {4.890} \\
        \multicolumn{1}{c|}{({\it HKC18})} &
        \multicolumn{1}{c|}{IH} &
            {1.94$\times 10^{-1}$} & {1.603}              & {3.04$\times 10^{-1}$} & {4.872} \\
        \hline
        \multicolumn{1}{c|}{FD NEQ \cite{2020ApJ...891L..24K}} &
        \multicolumn{1}{c|}{NH} &
            {1.99$\times 10^{-1}$} & {1.975}              & {6.61$\times 10^{-1}$} & {4.840} \\
        \multicolumn{1}{c|}{({\it HKC18})} &
        \multicolumn{1}{c|}{IH} &
            {3.22$\times 10^{-1}$} & {1.636}              & {6.71$\times 10^{-1}$} & {4.854} \\
        \hline
        \end{tabular} \label{PF_16O}
    \end{table*}
 \begin{table*}[t]
        \centering
        \caption{The production factor normalized to $^{24}$Mg}
        \begin{tabular}{r|ccccc}
        \hline
        \multicolumn{1}{c|}{} &
        \multicolumn{5}{c}{The production factor $10^{[{\rm i}/^{24}{\rm Mg}]}$ in the region $M_r =1.6$--6 $M_\odot$} \\
        \cline{2-6}
        \multicolumn{1}{c|}{} &
        \multicolumn{1}{c|}{MH} & i =  {$^{7}$Li} & {$^{11}$B} & {$^{138}$La} & {$^{180}$Ta}   \\
        \hline
        \multicolumn{1}{c|}{FD EQ} &
        \multicolumn{1}{c|}{NH} &
			{4.20$\times 10^{-1}$} & {1.116} & {1.44$\times 10^{-1}$} & {4.392} \\
        \multicolumn{1}{c|}{({\it HKC18})} &
        \multicolumn{1}{c|}{IH} &
			{2.00$\times 10^{-1}$} & {1.224} & {1.38$\times 10^{-1}$} & {4.379} \\
        \hline
        \multicolumn{1}{c|}{FD EQ} &
        \multicolumn{1}{c|}{NH} &
			{1.92$\times 10^{-1}$} & {5.84$\times 10^{-1}$} & {1.37$\times 10^{-1}$} & {4.560} \\
        \multicolumn{1}{c|}{({\it KCK19})} &
        \multicolumn{1}{c|}{IH} &
        		{1.14$\times 10^{-1}$} & {5.04$\times 10^{-1}$} & {1.26$\times 10^{-1}$} & {4.542} \\
        \hline
        \multicolumn{1}{c|}{FD EQ Shock} &
        \multicolumn{1}{c|}{NH} &
        		{1.61$\times 10^{-1}$} & {5.37$\times 10^{-1}$} & {1.52$\times 10^{-1}$} & {4.565} \\
        \multicolumn{1}{c|}{({\it KCK19})} &
        \multicolumn{1}{c|}{IH} &
        		{1.15$\times 10^{-1}$} & {4.88$\times 10^{-1}$} & {1.27$\times 10^{-1}$} & {4.542} \\
        \hline
        \multicolumn{1}{c|}{SI EQ\footnote{Same as FD EQ ({\it KCK19}) NH result}} &
        \multicolumn{1}{c|}{NH} &
        		{1.92$\times 10^{-1}$} & {5.84$\times 10^{-1}$} & {1.37$\times 10^{-1}$} & {4.560} \\
        \multicolumn{1}{c|}{({\it KCK19})} &
        \multicolumn{1}{c|}{IH} &
        		{1.74$\times 10^{-1}$} & {8.51$\times 10^{-1}$} & {4.97$\times 10^{-1}$} & {4.726} \\
        \hline
        \multicolumn{1}{c|}{SI NEQ} &
        \multicolumn{1}{c|}{NH} &
        		{1.60$\times 10^{-1}$} & {1.163} & {5.56$\times 10^{-1}$} & {5.810} \\
        \multicolumn{1}{c|}{({\it KCK19})} &
        \multicolumn{1}{c|}{IH} &
       	 	{1.44$\times 10^{-1}$} & {1.127} & {4.14$\times 10^{-1}$} & {5.751} \\
        \hline
        \multicolumn{1}{c|}{FD NEQ} &
        \multicolumn{1}{c|}{NH} &
        		{1.36$\times 10^{-1}$} & {1.349} & {8.46$\times 10^{-1}$} & {5.935} \\
        \multicolumn{1}{c|}{({\it KCK19})} &
        \multicolumn{1}{c|}{IH} &
        		{1.91$\times 10^{-1}$} & {1.330} & {8.76$\times 10^{-1}$} & {5.971} \\
        \hline
        \multicolumn{1}{c|}{SI NEQ \cite{2020ApJ...891L..24K}} &
        \multicolumn{1}{c|}{NH} &
        		{3.54$\times 10^{-1}$} & {2.378} & {6.35$\times 10^{-1}$} & {7.437} \\
        \multicolumn{1}{c|}{({\it HKC18})} &
        \multicolumn{1}{c|}{IH} &
        		{2.96$\times 10^{-1}$} & {2.437} & {4.63$\times 10^{-1}$} & {7.410} \\
        \hline
        \multicolumn{1}{c|}{FD NEQ \cite{2020ApJ...891L..24K}} &
        \multicolumn{1}{c|}{NH} &
            {3.02$\times 10^{-1}$} & {3.004}              & {1.006} & {7.361} \\
        \multicolumn{1}{c|}{({\it HKC18})} &
        \multicolumn{1}{c|}{IH} &
            {0.49$\times 10^{-1}$} & {2.489}              & {1.020} & {7.383} \\
        \hline
        \end{tabular} \label{PF_24Μg}
    \end{table*}

    \begin{table*}[t]
        \centering
        \caption{The production factor normalized to $^{28}$Si}
        \begin{tabular}{r|ccccc}
        \hline
        \multicolumn{1}{c|} {} &
        \multicolumn{5}{c} {The production factor $10^{[{\rm i}/^{28}{\rm Si}]}$ in the region $M_r =1.6$--6 $M_\odot$} \\
        \cline{2-6}
        \multicolumn{1}{c|} {} &
        \multicolumn{1}{c|} {MH} & i = {$^{7}$Li} & {$^{11}$B} & {$^{138}$La} & {$^{180}$Ta}   \\
        \hline
        \multicolumn{1}{c|}{FD EQ} &
        \multicolumn{1}{c|}{NH} &
				{3.49$\times 10^{-1}$} & {0.929$\times 10^{-1}$} & {1.20$\times 10^{-1}$} & {3.657} \\
        \multicolumn{1}{c|}{({\it HKC18})} &
        \multicolumn{1}{c|}{IH} &
				{1.66$\times 10^{-1}$} & {1.019} & {1.15$\times 10^{-1}$} & {3.647} \\
        \hline
        \multicolumn{1}{c|}{FD EQ} &
        \multicolumn{1}{c|}{NH} &
				{5.53$\times 10^{-1}$} & {1.681} & {3.96$\times 10^{-1}$} & {13.125} \\
        \multicolumn{1}{c|}{({\it KCK19})} &
        \multicolumn{1}{c|}{IH} &
				{3.28$\times 10^{-1}$} & {1.449} & {3.62$\times 10^{-1}$} & {13.074} \\
        \hline
        \multicolumn{1}{c|}{FD EQ Shock} &
        \multicolumn{1}{c|}{NH} &
				{4.65$\times 10^{-1}$} & {1.546} & {4.38$\times 10^{-1}$} & {13.138} \\
        \multicolumn{1}{c|}{({\it KCK19})} &
        \multicolumn{1}{c|}{IH} &
				{3.32$\times 10^{-1}$} & {1.405} & {3.67$\times 10^{-1}$} & {13.074} \\
        \hline
        \multicolumn{1}{c|} {SI EQ\footnote{Same as FD EQ ({\it KCK19}) NH result}} &
        \multicolumn{1}{c|} {NH} &
				{5.53$\times 10^{-1}$} & {1.681} & {3.96$\times 10^{-1}$} & {13.125} \\
        \multicolumn{1}{c|}{({\it KCK19})} &
        \multicolumn{1}{c|}{IH} &
				{5.00$\times 10^{-1}$} & {2.448} & {1.429} & {13.603} \\
        \hline
        \multicolumn{1}{c|}{SI NEQ} &
        \multicolumn{1}{c|}{NH} &
				{4.59$\times 10^{-1}$} & {3.348} & {1.599} & {16.721} \\
        \multicolumn{1}{c|}{({\it KCK19})} &
        \multicolumn{1}{c|}{IH} &
				{4.15$\times 10^{-1}$} & {3.242} & {1.191} & {16.553} \\
        \hline
        \multicolumn{1}{c|}{FD NEQ} &
        \multicolumn{1}{c|}{NH} &
				{3.91$\times 10^{-1}$} & {3.883} & {2.436} & {17.082} \\
        \multicolumn{1}{c|}{({\it KCK19})} &
        \multicolumn{1}{c|}{IH} &
				{5.48$\times 10^{-1}$} & {3.827} & {2.520} & {17.186} \\
        \hline
        \multicolumn{1}{c|}{SI NEQ \cite{2020ApJ...891L..24K}} &
        \multicolumn{1}{c|}{NH} &
				{2.91$\times 10^{-1}$} & {1.952} & {5.21$\times 10^{-1}$} & {6.103 } \\
        \multicolumn{1}{c|}{({\it HKC18})} &
        \multicolumn{1}{c|}{IH} &
				{2.43$\times 10^{-1}$} & {2.000} & {3.80$\times 10^{-1}$} & {6.080} \\
        \hline
        \multicolumn{1}{c|}{FD NEQ \cite{2020ApJ...891L..24K}} &
        \multicolumn{1}{c|}{NH} &
                {2.48$\times 10^{-1}$} & {2.465} & {8.25$\times 10^{-1}$} & {6.404} \\
        \multicolumn{1}{c|}{({\it HKC18})} &
        \multicolumn{1}{c|}{IH} &
                {4.02$\times 10^{-1}$} & {2.042} & {8.37$\times 10^{-1}$} & {6.058} \\
        \hline
        \end{tabular} \label{PF_28Si}
    \end{table*}


\clearpage

\end{document}